\pdfoutput=1 
\documentclass[11pt,a4paper]{article}

\usepackage{booktabs} 
\usepackage[utf8]{inputenc}
\usepackage[T1]{fontenc}
\usepackage{amsmath,amssymb,amsfonts,amsthm} 
\usepackage{graphicx}                         
\usepackage[section]{placeins}                
\usepackage{xcolor}
\usepackage{geometry}                         
\usepackage{authblk} 

\title{\textbf{Bridging the Gap between Extreme Environments and Precision Measurements: Recent Progress in Megagauss Physics}}

\author{Shojiro Takeyama}

\affil{The International MegaGauss Science Laboratory, The Institute for Solid State Physics, The University of Tokyo, 5-1-5 Kashiwanoha, Kashiwa, Chiba 277-8581, Japan}

\usepackage{hyperref}
\usepackage[all]{hypcap} 

\hypersetup{
    colorlinks=true,
    linkcolor=blue,
    citecolor=blue,
    urlcolor=blue
}
\usepackage{upgreek}
\usepackage{amsmath}
\usepackage{amssymb}
\usepackage{bm}
\usepackage{siunitx}
\usepackage{tabularx}
\begin{document}

\maketitle

\let\thefootnote\relax\footnotetext{Correspondence: takeyama@issp.u-tokyo.ac.jp}
\let\thefootnote\relax\footnotetext{Current address: 470-51 Yotsukaido-Shi, Chiba-Ken, Japan}

\begin{abstract}
Ultrastrong magnetic fields, ranging from 100~T to 1,000~T, are generated exclusively by destructive pulsed magnets. While various generation methods exist, this review focuses on the Single-Turn Coil (STC) and Electromagnetic Flux Compression (EMFC) techniques, which provide optimal environments for high-precision measurements in materials science. First, we present recent technological breakthroughs in the EMFC method that have successfully achieved fields exceeding 1,000~T. We then describe specialized measurement infrastructures for magneto-optics, magnetization, and magneto-transport, highlighting the development of miniaturized all-plastic cryostats and custom sample holders designed for the dual extremes of cryogenic temperatures and megagauss fields. Representative physical phenomena revealed through these techniques are discussed, including quantum phase transitions in frustrated magnets, Aharonov--Bohm effects in carbon nanotubes, and semiconductor-to-metal transitions in strongly correlated systems. Furthermore, we address emerging measurement platforms such as magnetostriction, specific heat, and ultrasound velocity. Throughout this review, we emphasize the instrumentation and experimental refinements that ensure reliable data acquisition in the ultrastrong pulsed field regime.
\end{abstract}

\vspace{1em}

\clearpage 

\section{Introduction}

Generating an ultra-strong magnetic field is, in itself, of little significance. In the cosmos, particularly near neutron stars, environments with magnetic fields reaching are abundant$10^{8}$~T \cite{Longair1994} . However, such fields remain useless to us because they are beyond human control. Just as fire was nothing more than a menace until humanity learned to harness it, a tool can only truly be called ours once it is mastered. Therefore, a ``usable'' magnetic field must be one that humans can freely control and manipulate at will. The ultra-strong magnetic fields developed here are precisely that: they are reproducibly controlled, enabling high-precision physical measurements that pave the way for acquiring new scientific insights. However, the precise control of magnetic fields is by no means an easy task, a reality reflected in the words of Soshin Chikazumi: ``Nature does not favor strong magnetic fields ''\cite{Miura1984}. From a physical standpoint, this inherent difficulty stems from the Maxwell equation $\nabla \cdot \mathbf{B} = 0$, which is intimately linked to the fact that nature hides the existence of magnetic monopoles.

Magnetic fields, as a direct consequence of the physical law $\nabla \cdot \mathbf{B} = 0$, possess the unique capability to directly intervene in the behavior of electrons that govern the properties of matter. Since magnetic flux lines are required to be continuous and form closed loops without termination, they inevitably ``penetrate'' the electronic system, fundamentally transforming the electronic state and giving rise to entirely new phases of matter.
Prominent examples of this transformation include the integer and fractional quantum Hall effects, the emergence of exotic superconducting states, and the Aharonov--Bohm (AB) effect observed in carbon nanotubes. To illustrate the extreme scale of this interaction, generating a single magnetic flux quantum through a carbon nanotube with a diameter of 10~nm would require an ultra-high field of approximately 5600~T~\cite{Ando2004}. Furthermore, the influence of magnetic fields extends beyond individual electronic states; their presence has been indispensable in manifesting universal fractal structures within condensed matter. By directly addressing both the charge and spin degrees of freedom, strong magnetic fields unlock inherent functionalities and hidden potentials within materials, paving the way for the discovery of unprecedented quantum phenomena (for details; see also Ref.~\cite{Berthier2002}).
.

Physical precision measurements in ultrastrong magnetic fields exceeding 100~T require both a minimum sample volume and a sufficient pulse duration (at least on the order of microseconds). However, such fields are typically generated by destructive magnets with extremely short pulse widths (5--50~$\upmu$s). To address this, the single-turn coil (STC) technique, powered by a fast-discharging capacitor bank with a relatively low energy of less than 0.2~MJ, can provide a well-controlled environment for fields up to 300~T. Generating 1,000~T-class magnetic fields via flux compression generally demands massive energy sources on the scale of several to dozens of MJ, supplied by either capacitor banks or chemical explosives. In contrast, recent advancements in electromagnetic flux compression (EMFC) have enabled a record-breaking 1,200~T field with an injection energy of 3.2~MJ \cite{Takeyama1000T, Takeyama1200T}. (This value is relatively low compared to what is required for chemical explosion methods).

These techniques have advanced to a stage where highly reliable precision measurements are now feasible. Various measurement methodologies have been developed and successfully applied to condensed matter physics, even at cryogenic temperatures, within the extreme environments generated by ultrastrong magnetic fields. These unique techniques, specifically tailored for destructive magnets, continue to be instrumental in exploring novel physical phenomena that only manifest in the extreme quantum limit. 
While existing reviews provide a comprehensive history of strong and ultrastrong magnetic fields \cite{HMF-ST2003, MiuraHerachPulseStrong1985, Portugall1998, Portugall1999}, this work aims to shed light on more recent developments. Specifically, we review the scientific and technological breakthroughs in the ultrastrong field regime over the past twenty years.

While various methods generate ultra-high magnetic fields, this review focuses on techniques that permit precision measurements of physical properties. Consequently, ultra-short pulse methods---such as laser-driven and plasma focus techniques---are excluded.
The rationale lies in the extreme scaling of time and space. In these systems, fields are confined to nanosecond order ($\sim 10^{-9}$~s) and sub-millimeter volumes. To put this into perspective: if the 5--50~$\upmu$s pulse of STC or EMFC system represents one ``hour'' of experimental observation, the nanosecond window shrinks to a mere fraction of a second.
Although STC and EMFC measurements often occur under adiabatic conditions where hysteresis is unavoidable, it is essential to maximize the temporal window to approach ``quasi-steady-state'' conditions as closely as possible. Such duration is vital for the relaxation of collective excitations and ensuring adequate signal-to-noise ratios under a single-shot measurement. For the experimentalist, a nanosecond-scale field is a transient event rather than a viable environment for precision physics.

This review provides a broad overview of destructive ultrastrong magnetic fields and the associated measurement technologies, covering diverse materials such as semiconductors, magnetic materials, and superconductors. By establishing a reliable and high-precision megagauss environment, we can survey the vast landscape of condensed matter physics from this unique vantage point, tackling cutting-edge problems across various disciplines. While experiments in ultrastrong magnetic fields require large-scale facilities, this article demonstrates that the ultimate key to successful measurement lies in the meticulous, ``hands-on'' craftsmanship performed at the laboratory bench. Going beyond conventional reporting, this article aims to reveal the practical ``know-how'' and experimental intricacies developed through years of trial and error. Often overlooked in formal publications, these essential techniques are, in fact, the decisive factors in achieving reliable results under extreme conditions. 
To effectively convey these practical aspects, this review includes a large number of figures and photographs with detailed captions, allowing readers to grasp the essence of the topics by browsing through the visual materials and their descriptions.
In this review, a large number of figures and photographs are included with detailed captions, allowing readers to grasp the essence of the topics by browsing through the visual materials and their descriptions.

To logically bridge the gap between engineering breakthroughs and fundamental science, this review is structured to first establish the foundation of field generation, proceed through the essential measurement and cryogenic infrastructures, and~ultimately culminate in specific solid-state physics applications. To~reflect this logical approach, the~article is organized as follows.
Section~\ref{sec:GenerationFields} briefly introduces the STC and chemical explosive-driven flux compression techniques, followed by a more detailed discussion of recent progress in EMFC that has enabled magnetic fields exceeding 1000~T. Section~\ref{sec:B_fieldMeasur} highlights state-of-the-art precision measurement techniques for solid-state physics, specifically as applied to STC and EMFC methods. This section also describes the accuracy of magnetic field measurements using pickup probes within destructive, short-pulse magnets. Finally, \mbox{Section~\ref{sec:cryo}} presents miniature all-plastic cryostats designed to achieve cryogenic temperatures with sufficient stability in ultrastrong magnetic fields, tailored to various experimental requirements under extreme pulsed~fields.

The following sections are devoted to instrumentation for precise solid-state physics measurements and their specific applications to materials science research. Section~\ref{sec:MagMeasure} describes substantial improvements in magnetization measurements achieved through both the induction pickup coil and Faraday rotation methods. These techniques have revealed the full magnetization processes of spin-frustrated materials and contributed to the discovery of diverse field-induced quantum magnetic phases. In~Section~\ref{sec:StreakSpec} magneto-optical absorption streak spectroscopy is shown to be a powerful tool for investigating magnetic quantum phases by probing intra-atomic $d$--$d$ optical transitions and associated transitions. Furthermore, infrared absorption streak spectroscopy has successfully unmasked the A-B effect in single-wall carbon nanotubes, resolving it from the complex band-edge exciton behavior under ultrastrong magnetic~fields.

Section~\ref{sec:InfraredCR} describes the use of cyclotron resonance spectroscopy in ultrastrong magnetic fields, utilizing infrared and near-infrared lasers. This approach has served as a powerful tool for investigating low-mobility electronic states in ferromagnetic semiconductors and has provided new insights into the relativistic Dirac electron systems found in graphene. In~Section~\ref{sec:RF}, we discuss the development of the radio-frequency (RF) self-resonant coil technique for magneto-conductivity measurements. This method was applied to cuprate high-temperature superconductors and extended to study field-induced semiconductor-to-metal transitions in the correlated narrow-gap semiconductor FeSi under ultrastrong magnetic fields. 
Finally, Section~\ref{sec:outlook} addresses the remaining challenges and solutions for improving the ``field quality''---which implies high spatial magnetic field homogeneity, a slowly evolving temporal profile, low high-frequency ripple and noise, and excellent shot-to-shot reproducibility, all of which are essential for high-quality measurements---in field generation, and~subsequently introduces several ongoing precision measurement techniques along with their limitations.

\section{Techniques for Generating Ultrastrong Magnetic Fields: Destructive Pulsed Magnets}
\label{sec:GenerationFields}
\subsection{Single-Turn Coil Technique}
\label{STC}

The single-turn coil technique provides a straightforward, convenient approach to generating 100--300~T magnetic fields with microsecond-order lifetimes~\cite{HMF-ST2003, MiuraHerachPulseStrong1985, Portugall1998, Portugall1999, NakaoSTC}.
A mega-ampere current (1--4~MA) injected into a single-turn small coil---typically constructed from 3~mm thick copper plate---generates intense magnetic fields with high homogeneity. 
These fields, produced within a 3--8~mm bore (0.02--4~cm$^{3}$), enable precise and sophisticated solid-state physics experiments.
The coil is destroyed by a combination of rapid magnetic expansion and Joule heating. 
Since the destruction is directed exclusively outward, the~measurement probes and samples inside remain intact. 
This allows for highly reliable, repeatable measurements under identical experimental conditions, requiring only a new coil for each~run.

\begin{figure}[htbp]
\centering
\includegraphics[width=0.5\columnwidth]{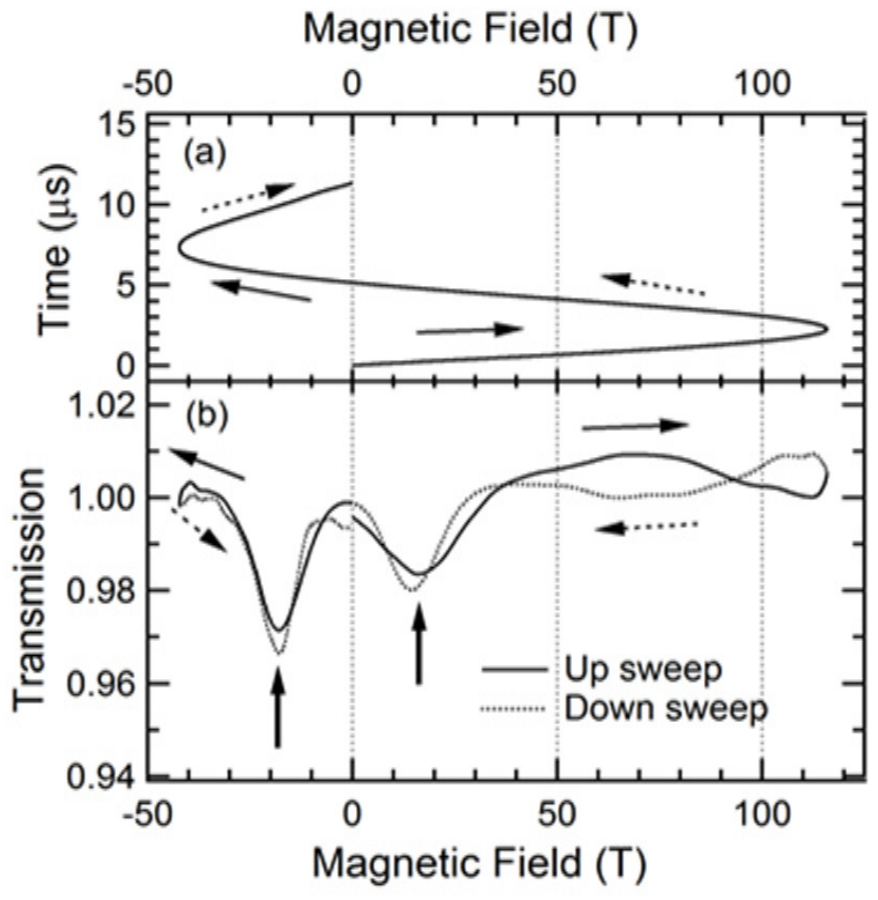}
 \caption{\label{fig:waveformSTC}
 (a) Typical temporal evolution of a pulsed magnetic field applied by the single-turn coil method. 
 Solid and dashed arrows indicate signals in an up- and a down-sweep of a pulsed magnetic field, respectively. 
 (b) Cyclotron resonance transmission of atomic monolayer graphene plotted with respect to magnetic field corresponding to that in (a). A CO$_{2}$ gas laser (wavelength: 9.46~$\upmu$m) is used as an irradiation light source. Upper arrows indicate the position of the magnetic field at a resonance absorption peak of the cyclotron resonance. 
 The resonance absorption peak was also recorded in a reversed direction of a pulsed magnetic field at $B = -20$~T. 
 [Figure originally created by the author and also presented in the doctoral thesis of H. Saito (The University of Tokyo, 2014) \cite{saito}].}
\end{figure}

Driven by a high-current discharge from a fast capacitor bank, the~pulsed magnetic field persists until the coil undergoes catastrophic, outward explosive destruction. 
This oscillating field profile facilitates comprehensive measurements---including up-sweep, down-sweep, and~reversed-field regimes---all within a single shot, as~demonstrated in Figure~\ref{fig:waveformSTC}. 
Furthermore, the~ultra-high sweep rate of 50--100~T/$\upmu$s is advantageous for the sensitive detection of relaxation phenomena, particularly those involving~hysteresis.
\vspace{6pt}

%

\begin{figure}[htbp]
\centering
\includegraphics[width=0.55\columnwidth]{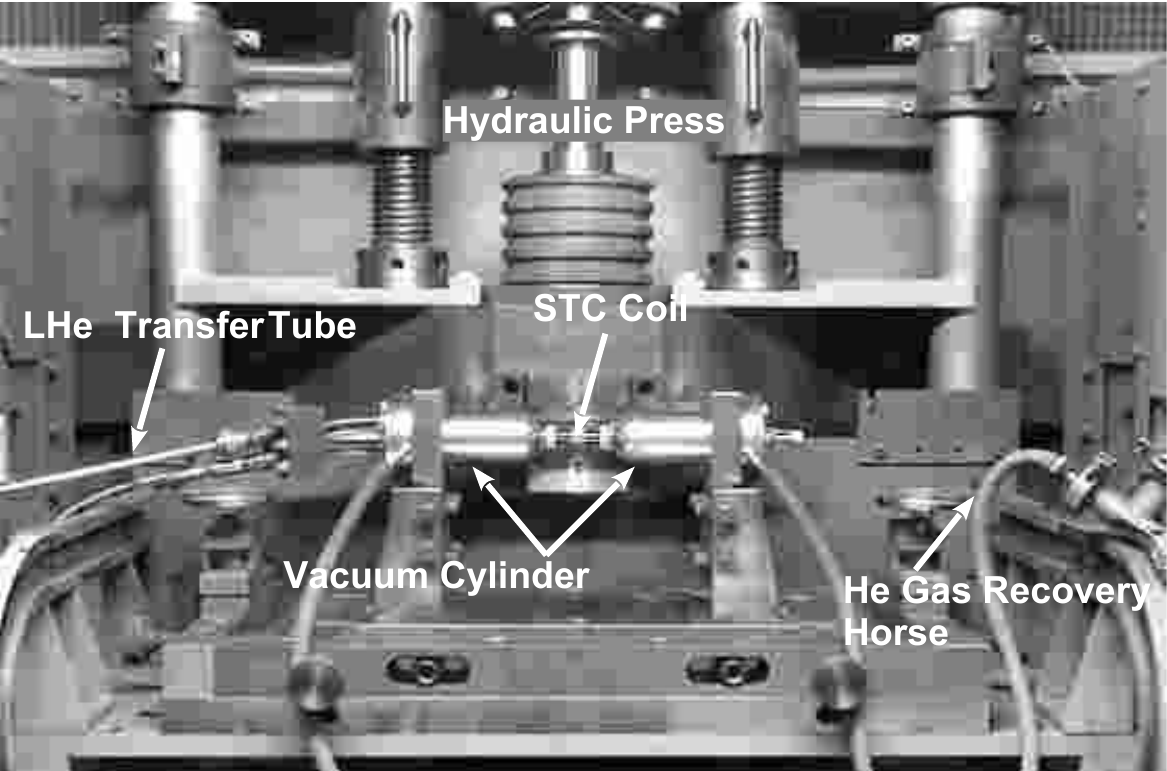}
 \caption{\label{fig:HSTC}
Photographic view of the horizontal single-turn coil megagauss generator. A single-turn copper coil is firmly clamped by a hydraulic press with a force of approximately 20 tons from above and secured with bolts on both sides. Vacuum-sealed metal cylinders extend from both sides toward the coil, supporting a miniature all-plastic cryostat positioned at the coil's center. Liquid helium is supplied to the cryostat via a transfer tube from the left, while the exhaust helium gas is recovered through a rubber hose. 
[Photograph taken by the author at ISSP, The University of Tokyo].
}
\end{figure}
\begin{figure}[htbp]
\centering 
\includegraphics[width=0.6\columnwidth]{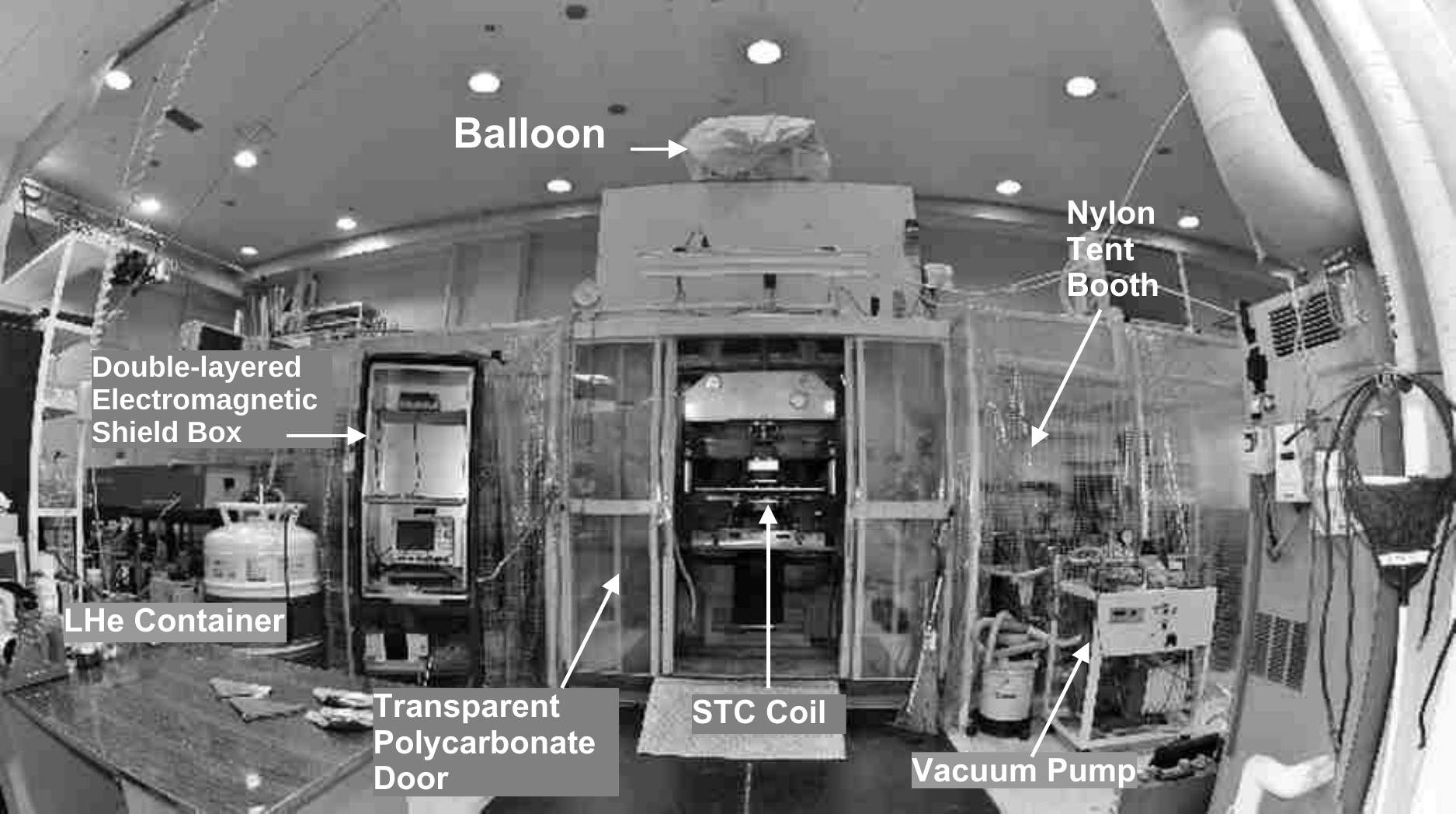}
 \caption{\label{fig:panoramaHSTC}
Horizontal single-turn coil system surrounded by measurement instruments.
The single-turn coil is mounted at the center of an explosion-proof iron chamber (center of the photo). A shockwave-absorbing balloon ($1\,\text{m} \times 1\,\text{m} \times 2\,\text{m}$, made of Kevlar synthetic fiber) is installed on the chamber's ceiling. The front access sliding doors consist of double-layered, 5-mm-thick transparent polycarbonate plates, which provide sufficient blast resistance while minimizing shockwave back-reflection toward the sample holder. These transparent doors also allow for immediate visual assessment in case of an emergency following a shot.
Lasers are housed in a nylon tent booth on the left, which protects them from explosion-generated dust. To ensure signal integrity, noise-sensitive instruments are kept in a double-layered electromagnetic shield box (iron outer wall, aluminum inner wall) located next to the liquid helium container. Light transmitted through the sample is directed into the left booth, where mirrors and detectors are aligned on an optical bench.
[Photograph taken by the author at ISSP, The University of Tokyo].}
\end{figure}
\begin{figure}[htbp]
\centering
\includegraphics[width=0.55\columnwidth]{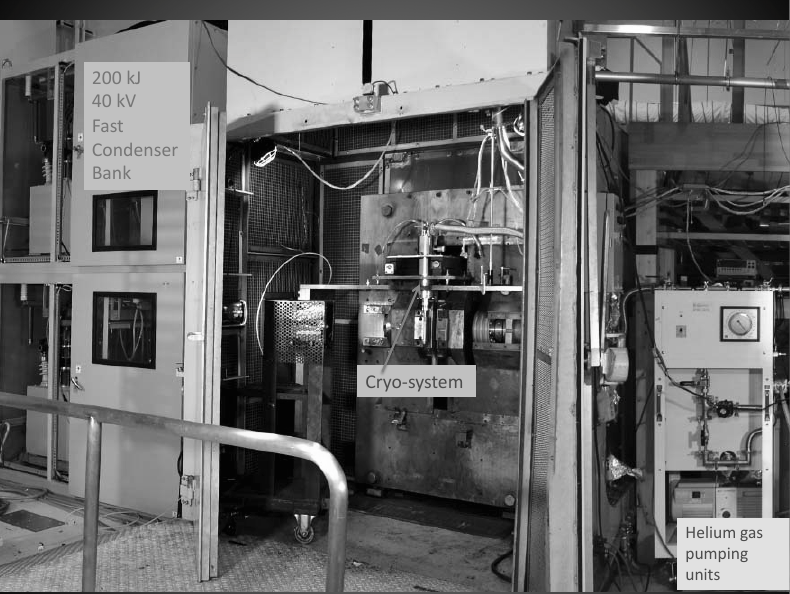}
 \caption{\label{fig:VSTC}
Overview of the vertical single-turn coil (V-STC) megagauss generator and surrounding diagnostic instruments. 
To the left of the explosion-proof chamber are the fast-discharge capacitor bank units equipped with air-gap switches. 
The sample temperature is controlled from room temperature down to 1.6~K using a helium gas pumping unit (visible to the right of the chamber) that regulates the pressure within the cryostat. 
The cryostat, inserted into the STC coil, is visible at the center (details are described in Section~\ref{sec:cryoV}). 
[Photograph taken by the author at ISSP, The University of Tokyo]}
\end{figure}
\begin{figure}[tbp]
\centering
\includegraphics[width=0.3\columnwidth]{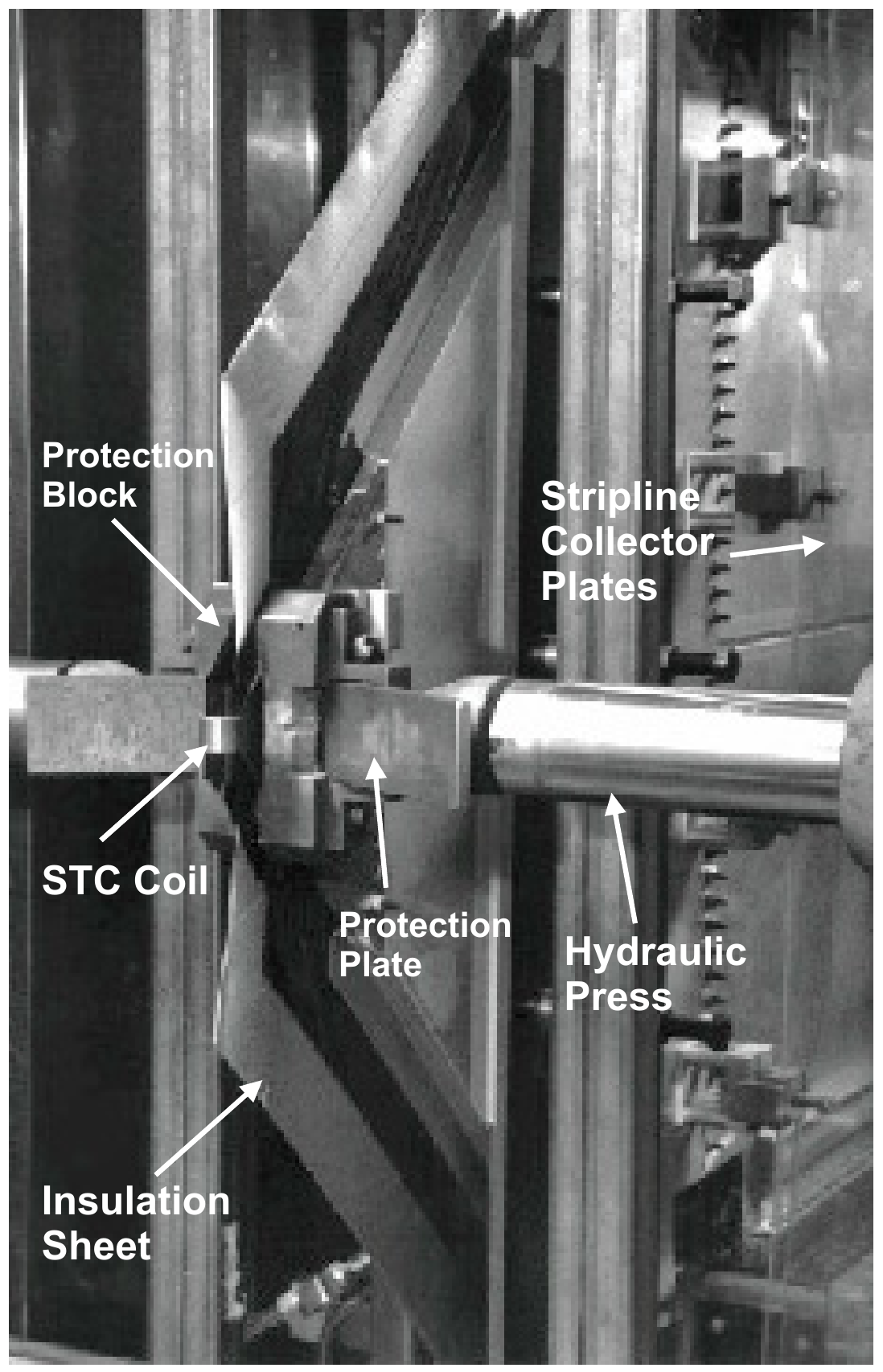}
 \caption{\label{fig:vstcPortugall}
Vertical single-turn coil megagauss generator installed at Humboldt University (now relocated to the Laboratoire National des Champs Magn\'{e}tiques Intenses (LNCMI) in Toulouse).
The single-turn coil is clamped between two rails from the right-hand side, where the hydraulic press exerts pressure on one of the opposing rails. Behind the piston bar, the stripline collector plates are partially visible; these are directly connected to the capacitor banks via rail-gap switches. Protruding steel plates on both sides of the coil serve as fragment-catching shields. 
[Reproduced from Ref.~\cite{Portugall1999} with permission from IOP Publishing  (1999).]}
 \end{figure}

\vspace{6pt}
Single-turn coil megagauss generators come in two types based on the orientation of the coil axis (and magnetic field): horizontal and vertical. 
While the horizontal configuration facilitates optical diagnostics with complex laser and mirror setups (Figures~\ref{fig:HSTC} and \ref{fig:panoramaHSTC}), the vertical type (Figures~\ref{fig:VSTC} and \ref{fig:vstcPortugall}) is better suited for cryogenic experiments. 
It allows for direct immersion into liquid helium within a cryostat placed inside the coil, enabling measurements at temperatures from 4.2~K down to 1.6~K via liquid pumping (see Figure~\ref{fig:panoramaHSTC}).

The coil and sample holder setups for both systems are detailed in Section~\ref{sec:cryo}.
The vertical single-turn coil megagauss generator, originally installed at Humboldt University and now relocated to Toulouse (Figure~\ref{fig:vstcPortugall}), features a unique design. It is constructed with a lightweight, low-inductance discharge circuit based on a stripline architecture and equipped with rail-gap switches. 
This system was specifically engineered to minimize electromagnetic discharge noise, distinguishing it from the configuration in Figure~\ref{fig:VSTC}, which utilizes conventional air-gap switches and long high-voltage coaxial cables. Detailed performance data are available in Refs.~\cite{Portugall1998, Portugall1999}.

\vspace{12pt}
The single-turn coil megagauss generator at the Los Alamos National High Magnetic Field Laboratory \cite{Mielke_stc} features a low-inductance, 60~kV capacitor bank, with an overall design similar to the horizontal system at the University of Tokyo. A unique feature of this installation is the placement of the capacitor banks and gap switches directly above the ceiling of the explosion-proof chamber. This configuration minimizes the length of the power cables connecting the switches to the collector plate adjacent to the load coil. Minimizing cable length is critical for reducing both power-source inductance and the ringing-type discharge noise inherent to single-turn coil systems.

\subsection{Magnetic Flux Compression Techniques}

To generate ultrastrong magnetic fields, magnetic flux initially established in a large volume is rapidly compressed into a smaller space---a method known as the magnetic flux compression technique. A metal cylinder, referred to as a ``liner,'' is used to compress the initial magnetic flux (the seed field). Ultrastrong fields are achieved through high-speed liner implosion, typically at velocities of 2--5~km/s. 
The peak magnetic field intensity is roughly proportional to the final implosion speed. This relationship follows the ``Hugoniot curve'' trend, where the particle velocity of the metal liner relates to the pressure generated by the shock wave. The fundamental principles of flux compression are detailed in Ref.~\cite{HMF-ST2003}.
\subsubsection{Chemical Explosive-Driven Flux Compression}

One effective method for achieving such high implosion velocities is to utilize chemical explosives, a technique known as explosive-driven magnetic flux compression. Fowler {\it et al.} were the first to report the generation of magnetic fields exceeding 1,000~T using this method \cite{Fowler}.
Subsequently, magnetic fields in the range of 1,000--1,500~T were achieved by employing the ``cascades'' method, with a record value of 2,800~T documented \cite{Bykov2001} using a massive quantity of TNT explosives (170~kg, equivalent to 680~MJ). However, the scale of the explosion required for such field generation results in the total destruction of all equipment within a several-meter radius. This necessitates conducting experiments in outdoor settings, which poses significant impediments to sophisticated, reproducible, and high-precision physical measurements. Consequently, the frequency of experiments is extremely limited, and verifying the reproducibility of the obtained data remains a formidable challenge.

Beginning in the late 1990s, significant progress was made in utilizing explosive-driven flux compression for advanced solid-state research. A notable example is the ``Kapitza series experiment,'' where Kudasov et al. \cite{Kudasov1998, Kudasov1999} implemented high-frequency conductivity and susceptibility diagnostics using a liquid helium cryostat integrated with an MK-1 generator. Their experiments on FeSi in fields reaching 450~T identified a gradual conductivity increase starting at 250~T and a metamagnetic-like transition near 360~T. These results, which diverged from the spin-density fluctuation model \cite{Takahashi1997}, highlighted the need for more systematic temperature-dependent studies and motivated the pursuit of even higher fields, up to 1,000~T.

Around the same time, the ``Dirac series experiment'' was launched as an international collaboration involving research groups from Australia, Japan, Russia, and the USA. This project aimed to explore solid-state physics under the extreme conditions provided by 1,000~T-class magnetic fields. Key results are summarized in Refs.~\cite{Dzurak1998a, Dzurak1998b, Kane1997}.
To perform high-frequency AC-conductivity transport measurements, a specialized sample mounting device using a coplanar transmission line was developed. This design was essential to mitigate sample heating caused by the massive eddy currents induced by the rapid magnetic flux change ($dB/dt$), which reaches orders of $10^{8}$--$10^{9}$~T/s during microsecond pulses generated by explosive-driven flux compression. This technique was applied to study the in-plane upper critical field ($H_{c2}$) of high-$T_{c}$ cuprate superconductors, specifically YBa$_{2}$Cu$_{3}$O$_{7-\delta}$, and to search for novel superconducting phases in two-dimensional, low-density electron gases within GaAs/Al$_{x}$Ga$_{1-x}$As parabolic quantum wells. While the group successfully obtained meaningful data up to 300~T, signals above this range were obscured by massive electromagnetic noise originating from the fusing of the third generator coil (the third cascade).

Optical measurements are generally considered robust against the electromagnetic noise associated with current discharges or the fusion of an imploding metal liner. Puhlmann and co-workers performed far-infrared spectroscopic measurements using a CO$_{2}$ laser on p-type cubic GaN thin films in magnetic fields up to 700~T, generated by a three-cascade MC-1 megagauss generator~\cite{Puhlmann2001}. Two optical transmission minima, attributed to inter-valence band transitions of holes, were resolved with a relatively high signal-to-noise ratio in fields up to 300~T---until the fusion of the second coil occurred. Although they reported a third transmission minimum around 410~T, its reliability appears questionable. Not only is there no corresponding transition in their calculated valence-band Landau levels, but the timing of this minimum also coincides exactly with the fusion of the second coil.

These observations suggest that the practical upper limit for reliable physical measurements, or even precise field determination, using chemical explosive-driven flux compression is likely restricted to the 500--600~T range, which is achievable with a single-stage liner configuration. This range represents a realistic boundary for high-fidelity data acquisition before coil failure, as implied by the analysis of, for example, the work by Dzurak et al. Refs.~\cite{Kane1997,Dzurak1998b}.


\subsubsection{Operating Principle of the Electromagnetic Flux~Compression}
\label{sec:emfc_principle}

The technique (EMFC) is a more controlled alternative and better suited for solid-state physics measurements in fields exceeding 300~T, whereas the single-turn coil method is more convenient for fields below this threshold. Unlike the single-turn coil method, the complete destruction of the magnet and the sample is unavoidable after each shot. However, the explosive force is confined to a relatively small area within an indoor explosion-proof chamber. This allows various experiments to be conducted repeatedly in a laboratory environment, with comprehensive protection against the electromagnetic noise associated with massive current discharges. A peak current on the order of mega-amperes is injected into the coil from electrical energy stored in capacitor banks.

The technique, known as the $\theta$-pinch, was originally developed by Cnare \cite{Cnare} and subsequently refined at the Megagauss Laboratory of the Institute for Solid State Physics (ISSP), the University of Tokyo, where a prototype megagauss generator was constructed in 1972. Using a 285~kJ (30~kV) capacitor bank, pulsed fields exceeding 100~T were successfully generated \cite{Miuraetal1975}. In 1980, the laboratory began installing upgraded equipment, including a 5~MJ (6.25~mF, 40~kV) main capacitor bank for the primary coil and a 1.5~MJ (30~mF, 10~kV) sub-bank for the seed fields. In the early 2000s, a peak magnetic field ($B_{\text{max}}$) of approximately 600~T was achieved by integrating a ``feed-gap compensator'' into the primary coil \cite{Matsuda2002}. Concurrently, a group at Loughborough University reported generating 300~T using a compact, fast capacitor bank system (24.5~kV, 63.2~kJ). A notable feature of their work was the successful recording of both the rising and falling sweeps of the magnetic field pulse, achieved by introducing an aluminum powder cascade \cite{novac2004}.

\begin{figure}[htb]
\centering
\includegraphics[width=0.8\columnwidth]{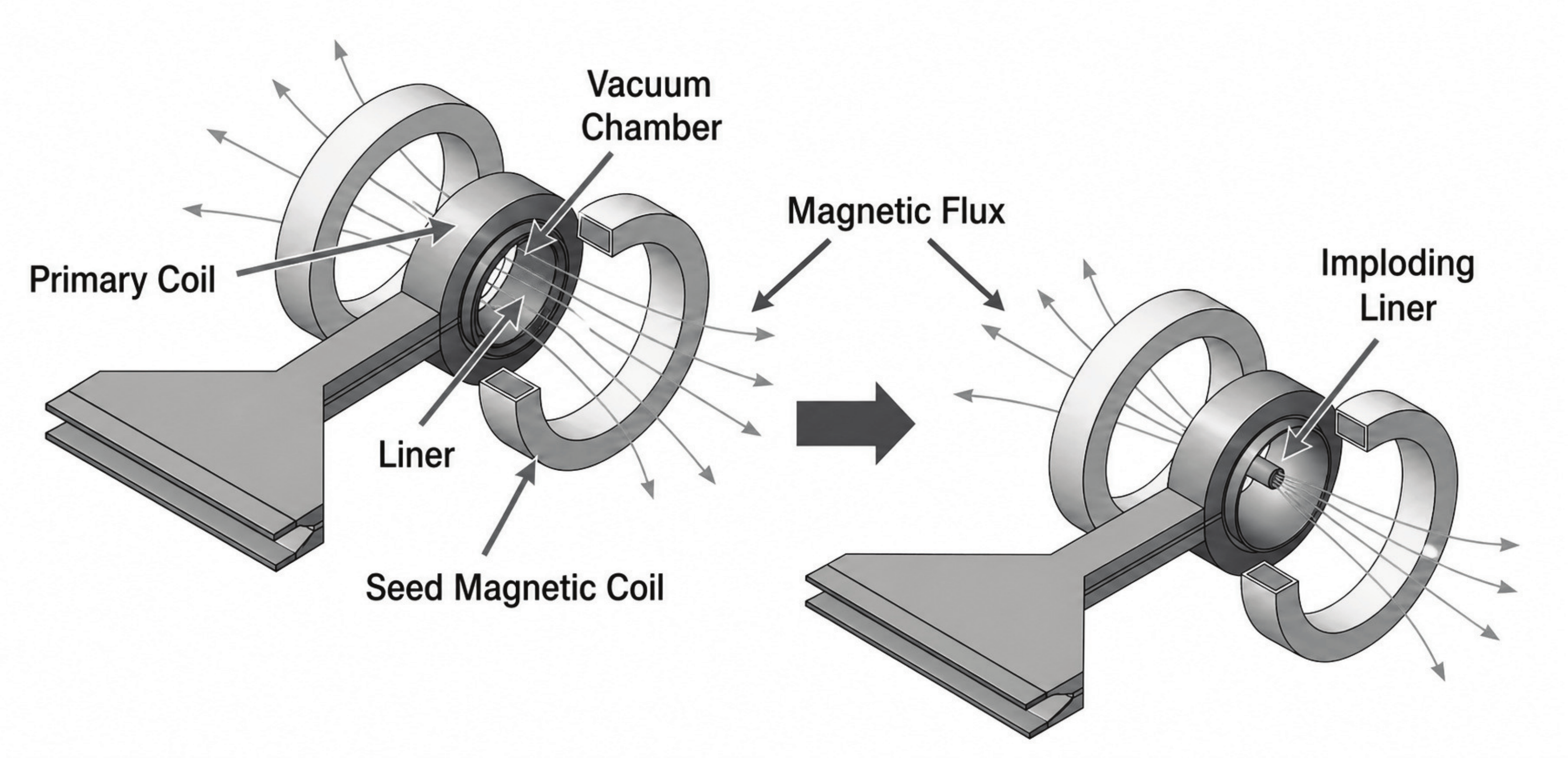}
\caption{\label{fig:Princip}
Generation of ultrastrong magnetic fields via EMFC.
The process begins with an initial magnetic flux (seed field of 3--4~T) generated by a pair of seed coils positioned on either side of the primary coil. This flux is then compressed by an imploding liner, which is accelerated by the Lorentz force (Maxwell stress) generated between the liner and the primary coil upon injecting a pulsed current of 3--4~MA.}
\end{figure}
As shown in Figure~\ref{fig:Princip}, the principal idea of generating very high magnetic fields is the use of the electromagnetic effect to compress the initial magnetic flux (the seed field).
In the $\theta$-pinch technique, an initial magnetic flux of several Tesla is generated by a pair of non-destructive pulse magnets positioned on both sides of the primary coil. A thin metal cylinder, known as the ``liner,'' is placed coaxially inside the single-turn primary coil, separated by a thin insulating tube (typically made of Bakelite). When a massive pulsed current is injected into the primary coil, a counter-flowing eddy current is electromagnetically induced in the liner (usually copper). Due to the opposing directions of these currents, the liner is accelerated toward the center by Maxwell stress (electromagnetic force), thereby compressing the initial magnetic flux (the seed field).

\subsubsection{Copper-Lined Main Coil}
\label{sec:clc}

Regarding the primary coil, the choice of material presents a significant challenge. Steel (SS400) is conventionally used for its cost-effectiveness and structural strength. However, its relatively low electrical conductivity results in a large skin depth, even for high-frequency discharging currents. Furthermore, the high electrical resistance of steel limits efficient current delivery from the power source and increases energy loss through Joule heating. This large skin depth also increases the effective distance between the primary coil and the liner, thereby degrading the electromagnetic coupling.
Since the driving coil is a single-turn design, a feed gap is inevitable. This gap disrupts the symmetry of the liner implosion, ultimately suppressing the peak magnetic field. While reducing the gap size is limited by the required thickness of high-voltage insulation, a ``feed-gap compensator'' was developed to address this issue. This device consists of six insulated copper blocks placed inside the primary coil to mitigate the gap's influence \cite{Matsuda2002}. The implementation of this compensator significantly improved the spatial symmetry of the imploding liner, leading to the achievement of a peak field of 600~T.
However, the fabrication of such a primary coil, including the integration of the feed-gap compensator, is labor-intensive and involves significant manufacturing costs. 

\begin{figure}[tbp]
\centering
\includegraphics[width=0.6\columnwidth]{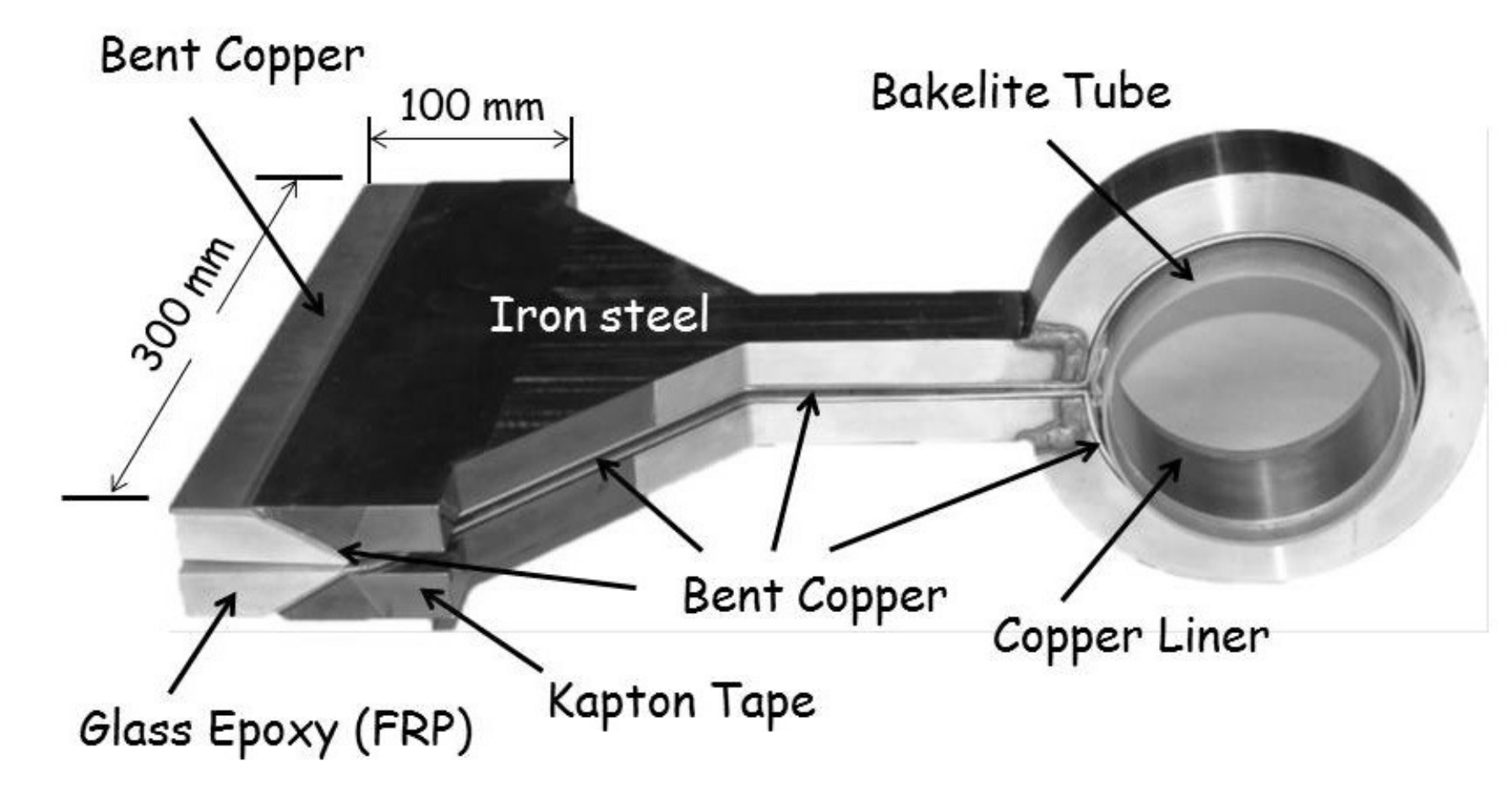}
\caption{\label{fig:clcoil}
Photograph of the CL coil. 
The copper current-feed plate is precision-machined to match the internal profile of the steel primary outer coil (the steel part), into which it is laterally inserted to ensure seamless mechanical and electrical contact.
The inner part of the primary coil consists of a single bent copper plate (2~mm thickness), which is precisely lined to the outer iron coil. The liner and measurement probes, along with the sample, are inserted into the vacuum chamber (constructed from a Bakelite tube, not shown in the photograph) and are fitted to the CL coil with high precision. Multilayered insulation sheets (comprising polyethylene and Kapton sheets, not shown in the photograph) with a thickness of 0.2--0.5~mm are inserted into the gap between the upper and lower current feeding plates. The vacuum within the chamber is typically maintained at 0.05--0.1~Pa.}
\end{figure}

To overcome the aforementioned problems, a copper-lined coil (hereafter referred to as CL coil) was developed as a high-efficiency primary coil \cite{Kojima2006, TakeCL2011}. This coil consists of a bent tough-pitch copper plate lined inside an outer steel coil (SS400) (see Figure~\ref{fig:clcoil}). The copper plate is shaped to optimize electrical contact with the clamping electrodes of the collector plate, which aggregates the high-voltage power cables from the capacitor banks. These electrodes inject current by pressing against both the copper and steel plates over an area of 300 $\times$ 100~mm$^{2}$, as illustrated in Figure~\ref{fig:clcoil}. The contact area is secured by an external hydraulic press with a force of 1,000~kN.
Since the conductivity of copper is nine times higher and its skin depth three times shallower than those of steel, the copper lining carries the vast majority of the current. Based on a dynamical analysis of current density distribution and Joule heating (detailed in the Appendix of Ref.~\cite{TakeCL2011}), a 2-mm-thick copper plate was selected to line the 25-mm-thick steel coil. In this configuration, the outer steel coil provides sufficient mass inertia to minimize the deformation of the copper lining during liner acceleration.
For a detailed description of the magnetic field pickup coil setup and its installation, refer to Section~\ref{sec:sampleset}.

\begin{figure}[tbp]
\centering
\includegraphics[width=0.6\columnwidth]{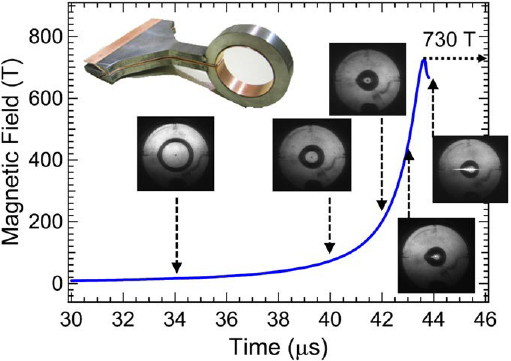}
\caption{\label{fig:700T}
Pulsed magnetic field profile reaching a peak of 730~T. The inset photograph shows the CL primary coil used in the experiment. High-speed framing images of the imploding liner are displayed at corresponding points of the magnetic field pulse. 
[Reproduced from Ref.~\cite{TakeCL2011} with permission from IOP Publishing  (2011).]}
\end{figure}

The CL coil was found to improve the geometrical symmetry of the liner's imploding motion compared to previously used coils. Reproducibility of the magnetic field generation across shots has significantly increased, leading to stable operation and enhanced controllability. With this new coil \cite{Kojima2006}, the energy transfer efficiency from the capacitor bank to the magnetic energy was nearly doubled compared to previous results \cite{Matsuda2002}. A peak field exceeding 700~T was generated within a final bore of 6.2~mm via a discharge from a 40~kV, 4~MJ capacitor bank (see Figure~\ref{fig:700T}) \cite{TakeCL2011}.

In EMFC, the measurement probe rod must be positioned precisely at the center of the imploding liner, as the inner wall of the liner closely approaches the probe near the peak field intensity. Such precision is achieved by fixing the probe rod directly to the primary coil assembly. Misalignment can cause a substantial reduction in the peak field and the suppression of the turn-around phenomenon (see Section~\ref{sec:simulation}).
\begin{figure}[tbp] 
\centering
 \centerline{\includegraphics[width=0.7\columnwidth]{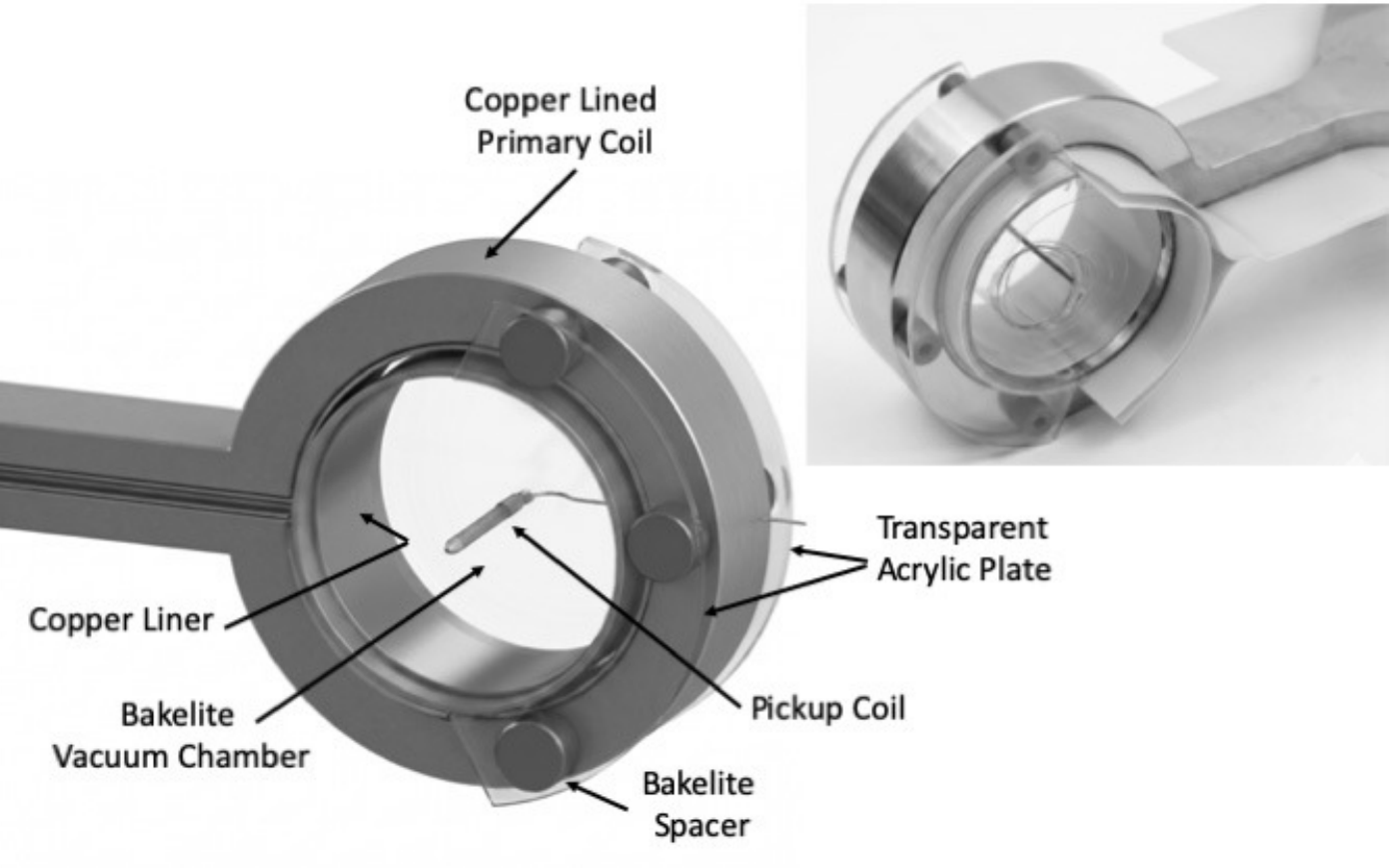} }
   \caption{
   View of the primary coil assembly and pickup coil setup. A copper liner and a Bakelite vacuum chamber are installed within the steel primary coil (outer diameter: 185 mm; inner diameter: 130 mm; length: 45 mm). White polyethylene sheets provide insulation between components. This setup is designed for synchronized high-speed framing photography and magnetic field measurements.
   }
    \label{emfccoil}
\end{figure}

As shown in Figure~\ref{emfccoil}, the liner vacuum chamber, equipped with the measurement probe, is firmly secured to the outer steel coil \cite{Takeyama2018} (previously, the chamber was mounted to the flange of the outermost seed magnetic field coil; however, since the relative positioning between the seed coil and the primary coil could vary by several millimeters in each experiment, sufficient alignment precision could not be guaranteed). This configuration allows the probe to be aligned with the center of the liner implosion using the primary coil as a fixed reference, which remains stationary during the process due to its significant mass and inertia in comparison to the thin copper liner.

Crucially, empirical data from repeated EMFC experiments have revealed that the center of the liner implosion consistently shifts by approximately 2~mm toward the feed-gap of the primary coil. This phenomenon occurs because the electromagnetic coupling between the primary coil and the liner is locally weakened by the presence of the feed-gap. To compensate for this systematic shift, the Bakelite vacuum cylinder containing the liner is intentionally offset by inserting 2-mm-thick polyethylene insulating sheets on the feed-gap side. This deliberate misalignment of the initial setup ensures that the liner converges precisely at the position of the measurement probe at the final stage of compression. Additionally, transparent acrylic flanges cover both sides of the vacuum chamber, serving as windows for framing photography to observe the liner's imploding motion. Ultimately, the combination of the high reproducibility provided by the CL coil and this precise centering technique was the key factor in successfully achieving magnetic fields up to 1,200~T.

\begin{figure}[tbp]
\centering
\includegraphics[width=0.7\columnwidth]{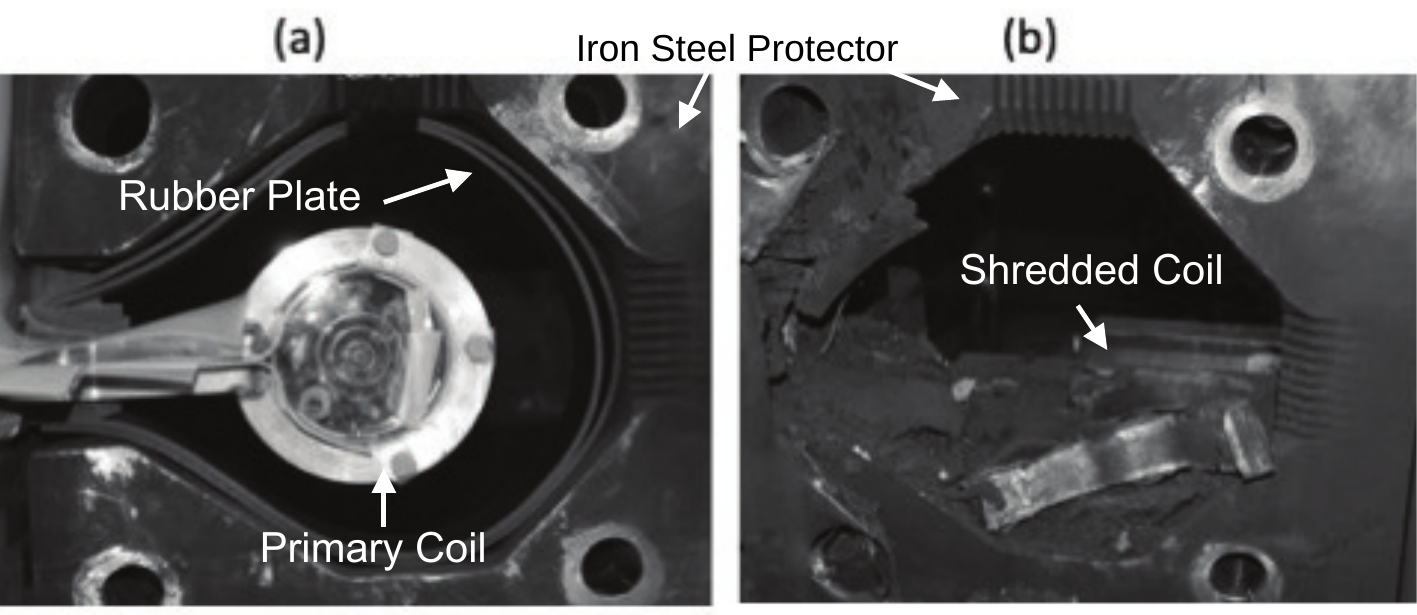}
\caption{\label{fig:coilset}
(a) The experimental configuration showing the primary coil installed and enclosed within the iron steel protector prior to the discharge. (b) The state of the assembly immediately after the ultrahigh magnetic field generation. The steel body of the primary coil has been shattered into three or four distinct segments, which are seen scattered inside the protection block. This visual evidence highlights the immense electromagnetic forces and internal pressure generated during the 4-MJ energy injection, which necessitates the use of a robust containment system.
[Photograph taken by the author at ISSP, The University of Tokyo.]}
\end{figure}

The 4-MJ energy injection into a flux compression coil triggers a massive, explosive destruction of not only the sample and the coil but also the insulation sheets and surrounding components. This energy release is equivalent to the detonation of several sticks of dynamite. 
A physical testament to these unprecedented energy densities can be seen in Video Movie \cite{ieeevideo}, which records the explosion at 1.7~MJ ($\sim$400~T). Note that for injections exceeding 3~MJ, the resulting destruction is so massive that it defies conventional imaging.
To contain this force, the entire assembly---including the sample, liner, and coils---is enclosed within a heavy steel protector, as shown in Figure~\ref{fig:coilset}. Image (a) in Figure~\ref{fig:coilset} displays the setup prior to the shot, where the seed-field coils are mounted to the steel protection block from both the front and rear. Image (b) of Figure~\ref{fig:coilset} shows the aftermath immediately following the magnetic field generation. The primary coil has been torn into fragments, and the copper liner, along with the copper plating of the primary coil (CL coil), has completely evaporated. To manage the extreme internal pressure, the protection block is specifically engineered with several venting mechanisms to safely release the blast pressure.

The primary role of the protection block extends beyond mere containment; it serves as a critical buffer to safeguard the high-voltage infrastructure located behind the coil assembly. This includes sensitive collector plates and a massive array of 480 high-voltage power cables, all of which must be isolated from the 50-kV discharge. By engineered venting of the blast pressure, the system effectively mitigates the shockwave, preventing catastrophic damage to the surrounding electrical architecture and ensuring the reproducibility of these extreme experiments.
To mitigate the shockwave and prevent secondary hazards, we replaced the traditional wooden cushioning with specialized rubber padding to wrap the coil within the protection block. While wooden blocks were historically used, their flammability posed a significant fire risk following the energetic discharge. Drawing inspiration from historical advancements in ballistic protection---specifically the composite armor techniques that utilize rubber to dissipate kinetic energy---we found that the rubber layer effectively absorbs the initial impact and prevents the localized stress concentrations that previously led to structural failure. This transition significantly enhanced the safety and stability of the experimental setup.

The generation of magnetic fields exceeding 1,000~T relies on the efficient conversion of electrical energy into the kinetic energy of the liner. Driven by the electromagnetic coupling with the primary coil, the liner reaches an implosion velocity of approximately 5~km/s, effectively compressing the magnetic flux. However, at the peak of the magnetic field, the accumulated magnetic energy is reflected back into the kinetic energy of the liner. This results in a violent outward expansion, where the liner fragments collide with the primary coil at hypersonic speeds. The structural failure of the external protection housing observed during the 1,200-T generation was a direct consequence of this immense shockwave, a physical testament to the unprecedented energy densities achieved in these experiments.

\subsubsection{The Liner Imploding Dynamics and Computer Simulation}
\label{sec:simulation}

\begin{figure}[tbp]
\centering
\includegraphics[width=0.3\columnwidth]{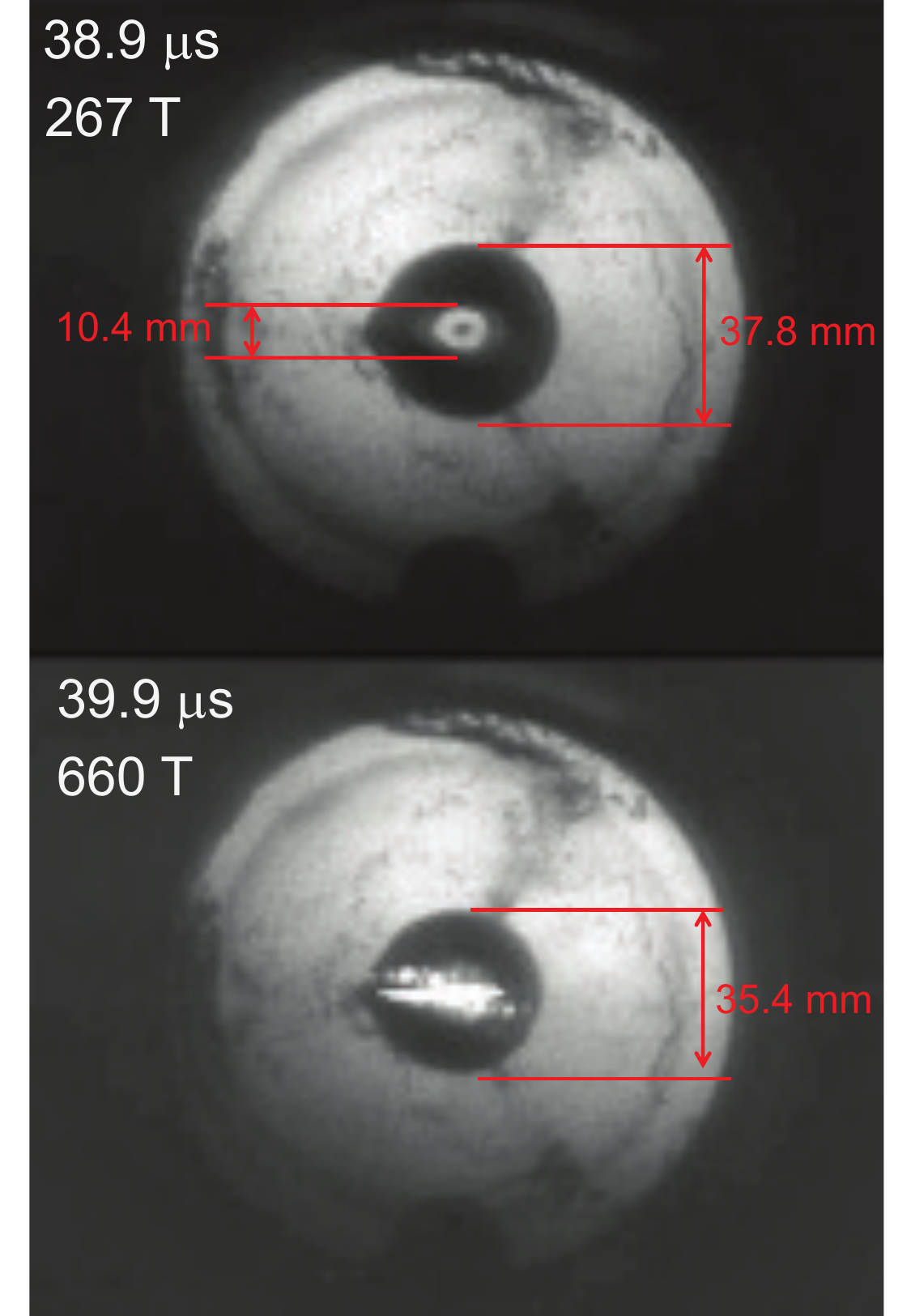}
\caption{\label{fig:framingphoto}
High-speed transmission photographs of the imploding liner during flux compression at 38.9~$\upmu$s (267~T) and 39.9~$\upmu$s (660~T, near the turnaround point). 
The dark spot at the coaxial center of the coil corresponds to the pickup coil rod for magnetic field measurement. An intense arc flash obscures the image of the liner's inner region. Experimental parameters: discharge = 40~kV/4~MJ; primary coil: length = 45~mm; liner: initial diameter = 119~mm, length = 50~mm, thickness = 1.5~mm; seed field = 3.8~T.}
\end{figure}

Magnetic field generation via EMFC is largely governed by the dynamics of the imploding metal liner and its three-dimensional, time-evolving deformation. The liner's motion is experimentally monitored using a high-speed framing camera. Examples are shown in Figures~\ref{fig:700T} and \ref{fig:framingphoto}, which present optical transmission images captured at specific points during the magnetic field rise. These images are conventionally used to analyze the velocity and radial profile of the imploding liner; however, capturing the full three-dimensional nature of the developing deformation remains a challenge.

High-speed flash X-ray photography can provide a clearer view of the liner, which is resistant to the disturbance of an extremely bright arc flash at the final stage of the implosion as seen in Figure~\ref{fig:framingphoto}, and is even capable of distinguishing liquid or plasma from the metal solid-state (see pages 275--276 in Ref.~\cite{HMF-ST2003}). However, the time sequence of the X-ray photo is difficult to record during one shot of the destructive explosive experiment.

\begin{figure}[tbp]
\centering
\includegraphics[width=0.9\columnwidth]{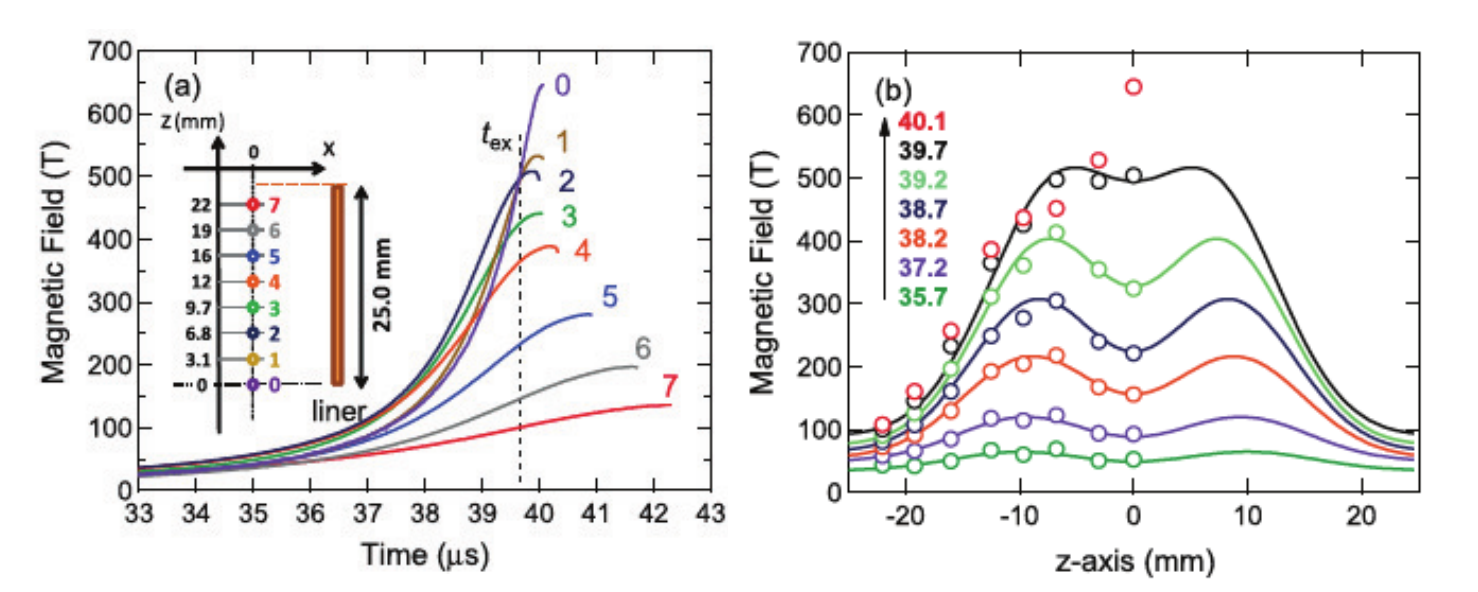}
\caption{\label{fig:FieldProfile}
(a) Magnetic field curves measured by eight calibrated pickup coils positioned along the field axis, as illustrated in the inset. The numbers 0--7 on the curves correspond to the respective positions of the pickup coils along the $x$-axis, which are arranged along the axis covering one half-side of the liner (total liner length: 50~mm). (b) Time evolution of the spatial magnetic field profiles, where bold numbers indicate the time in microseconds. 
[This figure is adapted and rearranged from Ref.~\cite{NakamuraTakeyama2014}
with permission. \copyright(2014) American Institute of Physics.]}
\end{figure}
On the other hand, computer simulations provide valuable insights for analyzing the dynamic behavior of the liner and optimizing the coil parameters for the entire condenser discharge system. A realistic simulation for EMFC was first conducted using a finite element method by Miura and Nakao \cite{MiuraNakao}. Their 2D model, which accounted for finite liner thickness as well as current and temperature distributions, was an extension of an earlier 1D fundamental model \cite{MiuraChikazumi}. Subsequently, a research group at Loughborough University in the U.K. refined this approach by allowing for the independent movement of filamentary rows, fully considering the magnetic pressure gradients within both the liner and the driving primary coil \cite{Novac2006}.

\begin{figure}[tbp]
\centering
\includegraphics[width=0.8\columnwidth]{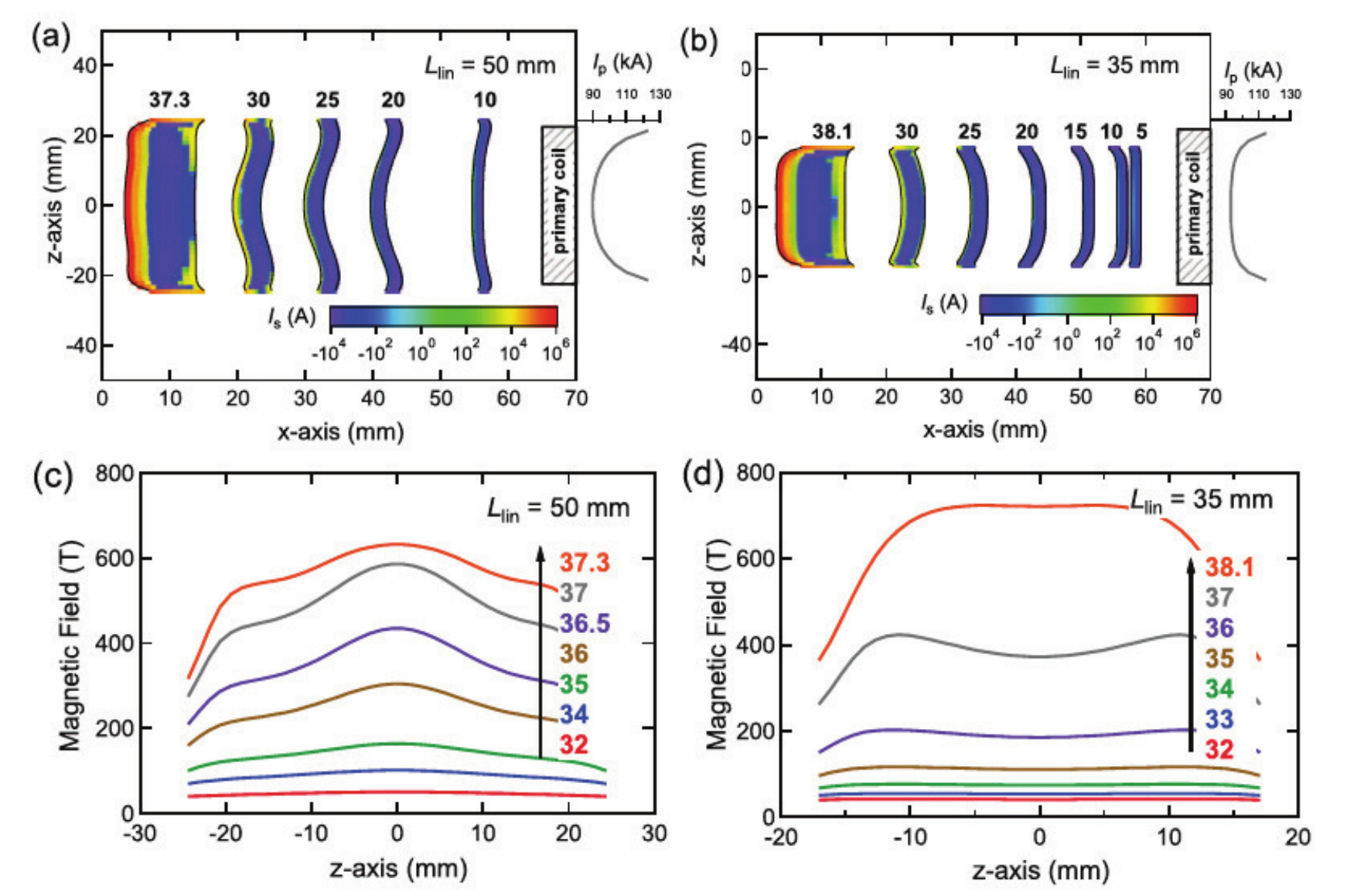}
\caption{\label{fig:SimulationFieldProf}
Simulations of liner dynamics for (\textbf{a}) a liner length $L_{\text{lin}}$ = 50~mm and (\textbf{b}) $L_{\text{lin}}$ = 35~mm, with~a constant primary coil length $L_{\text{prim}}$ = 45~mm (bold numbers indicate time in microseconds). 
The current intensity in the liner, $I_{\text{s}}$, is color-mapped onto the liner's cross-section. 
The primary coil current distribution profile $I_{\text{p}}(z)$ at time 10~$\upmu$s is shown to the right of each figure. The~time evolution of the magnetic field intensity distribution along the $z$-axis is displayed for (\textbf{c}) \mbox{$L_{\text{lin}}$ = 50~mm} ($L_{\text{lin}} > L_{\text{prim}}$) and (\textbf{d}) \mbox{$L_{\text{lin}}$ = 35~mm} ($L_{\text{lin}} < L_{\text{prim}}$), where bold numbers represent the time in microseconds, and the upward arrows indicate the direction of time evolution.
[Reproduced from Ref.~\cite{NakamuraTakeyama2014} with permission from IOP Publishing  (2014)].}
\end{figure}

Computer simulations become increasingly valuable when validated against experimental data. The temporal and spatial distributions of the magnetic field have been systematically investigated in EMFC experiments and rigorously compared with a quasi-3D simulation \cite{NakamuraTakeyama2014}. This model allows for the independent radial movement of each liner segment while assuming a constant liner length along the field axis. By explicitly modeling the CL driving coil---where the primary current is constrained within a well-defined thin copper plate---the simulation yields more accurate and realistic predictions of the experimental results.
The spatio-temporal distribution of the magnetic field generated via EMFC was investigated using eight independent pickup coils aligned along the $z$-axis. The results are presented in Figure~\ref{fig:FieldProfile} \cite{NakamuraTakeyama2014}. As shown in Figure~\ref{fig:FieldProfile}(b), the magnetic field profile initially exhibits a ``camelback'' structure with double peaks until 39.7~$\upmu$s; subsequently, these peaks abruptly merge toward the center as time elapses. Numerical simulations were performed under the same experimental conditions ($L_{\text{lin}}$ = 50~mm, $L_{\text{p}}$ = 45~mm). Figures~\ref{fig:SimulationFieldProf}(a) and (b) illustrate the time evolution of the liner deformation alongside cross-sectional current density mapping, while Figures~\ref{fig:SimulationFieldProf}(c) and (d) display the corresponding magnetic field distributions along the $z$-axis.

The experimental results shown in Figure~\ref{fig:FieldProfile}(b) exhibited a similar behavior to the simulation case for ($L_{\text{lin}} < L_{\text{prim}}$) [Figure~\ref{fig:SimulationFieldProf}(d)], despite the fact that $L_{\text{lin}} > L_{\text{prim}}$ at $t=0$ in the experiment. This suggests that the liner implodes while shrinking in the $z$-direction, and that the gradual approach of the two-hump structure observed in Figure~\ref{fig:FieldProfile}(b) results from the further development of this shrinkage.
To improve the simulation, a three-dimensional finite element method (FEM) was employed \cite{Takekoshi2015, Takekoshi2018}. The commercial FEM solver ``LS-DYNA,'' which supports structural, thermal, and electromagnetic analyses, was used.

\begin{figure}[tbp]
\centering
\includegraphics[width=0.6\columnwidth]{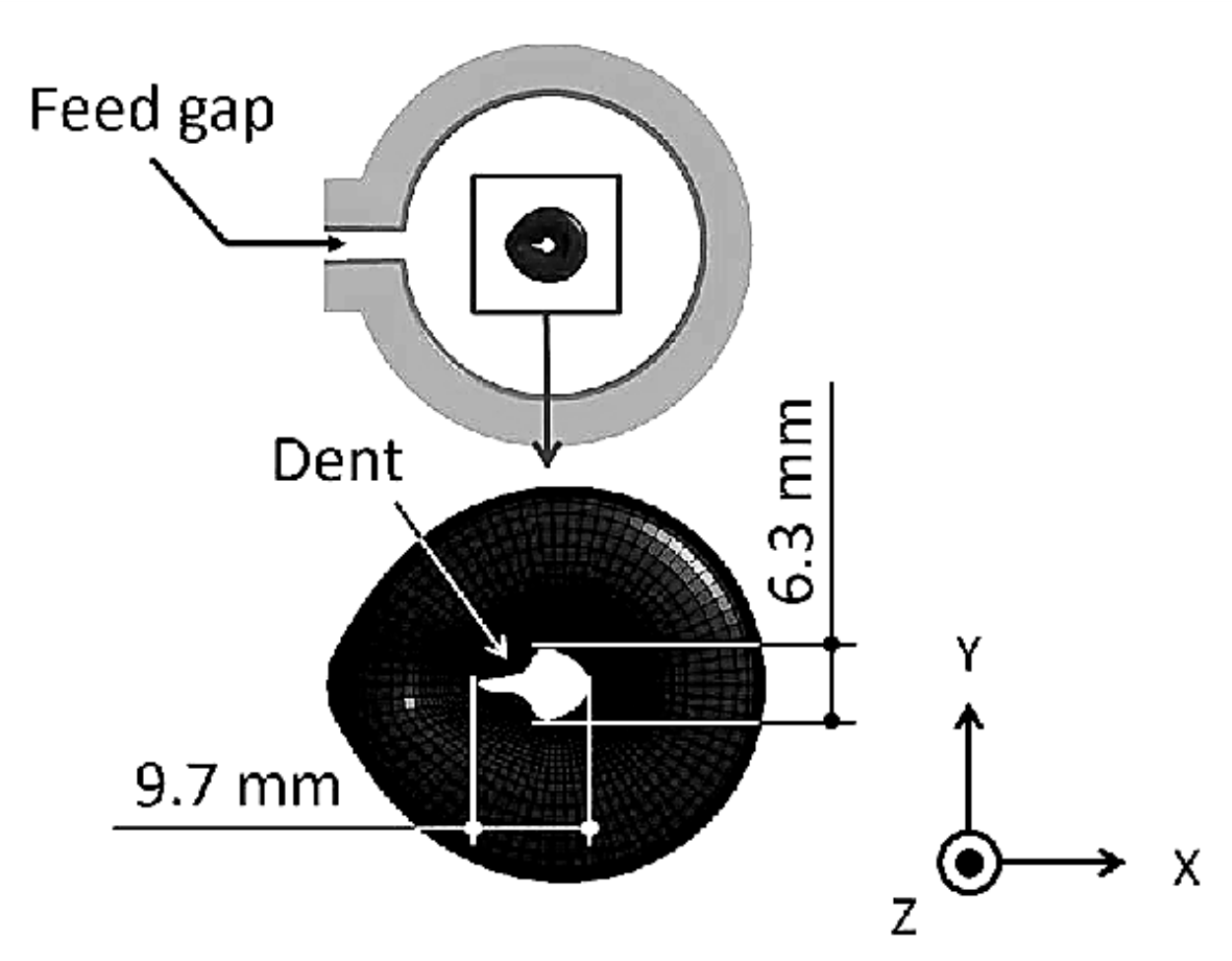}
\caption{\label{fig:TakekoshiView}
Cross-sectional view of the liner at the moment of peak magnetic field, obtained from the 3D finite element model. The calculated ``dent'' structure successfully reproduces the liner images captured by the high-speed framing camera (cf. Figures~\ref{fig:700T} and \ref{fig:framingphoto}). 
[Adapted from Ref.~\cite{Takekoshi2018}, with permission from the author, K.Takekoshi. \copyright  (2018) IEEE.]}
\end{figure}

The 3D simulation could reproduce the time evolution of spatial profiles of magnetic field intensity shown in Figure~\ref{fig:FieldProfile}(b) and also the imploding liner shape of the radial direction (a cross-sectional view) as displayed in Figure~\ref{fig:TakekoshiView}. A dent structure appears at the feed-gap side in the cross-sectional illustration. A similar structure is captured by the framing photography at the end of implosion as demonstrated in Figure~\ref{fig:framingphoto}. The dent structure induces a strong flash arc which obscures an image of the inner wall of the liner. Nevertheless, the dent starts to appear 1~$\upmu$s before the peak field as is noted in the upper picture of Figure~\ref{fig:framingphoto}. 

While the 3D simulation provides detailed structural insights, its prohibitive computational cost and inability to reproduce the turn-around phenomenon limit its practicality for guiding experiments. In contrast, the quasi-3D simulation \cite{NakamuraTakeyama2014} serves as a far more effective tool for experimental planning. It not only accurately predicts the time evolution, including the turn-around phenomenon, but also achieves 10\% accuracy in peak field estimation with significantly reduced computation time. This efficiency makes it indispensable for the rapid optimization of coil parameters and experimental conditions.

\subsubsection{Ultrafast Condenser Power Supply}

In 1,000~T-class EMFC experiments, a current on the order of mega-amperes (3--4~MA) is discharged into the primary coil from high-voltage, low-impedance capacitor bank modules, which store energies on the order of mega-joules (2--5~MJ).
Efficiently injecting such massive energy into a localized coil space at ultra-high speeds is a unique technological challenge. Due to the explosive destruction of the primary coil, the load must be clamped under high pressure within a robust protection chamber, necessitating a transmission distance of 10--30~m from the power supply. While higher charging voltages are desirable to minimize the system volume and increase discharge speeds, the practical limit is currently around 50~kV due to dielectric breakdown risks across the long transmission cables and the coil assembly.

Furthermore, high-voltage ultrafast air-gap switches are employed to handle the discharge. To minimize system impedance and accommodate the current limits of individual units, several to dozens of these switches must be connected in parallel. For such a configuration to function effectively, the synchronization of these multiple switches must be precisely controlled, typically within a jitter of less than 50~ns.
In addition to precise synchronization, preventing ``prefire''---an accidental breakdown during or after the charging process---is essential for experimental safety and reliability. Achieving a ``prefire-free'' operation at a high charging voltage of 50~kV is extremely challenging, requiring sophisticated engineering and meticulous maintenance of the switch insulation and gap spacing.
To achieve the requisite low impedance for ultrafast current injection, the system employs hundreds of high-voltage coaxial cables in parallel, integrated with high-current ultrafast switches. The primary difficulty in power supply design lies in the fundamental trade-off: increasing the capacitance provides the necessary energy for higher fields but inherently reduces the discharge speed. Consequently, optimizing the balance between total energy, charging voltage, and system impedance is critical for reaching the megagauss regime.
\begin{figure}[tbp] 
\centering
 \centerline{\includegraphics[width=0.75\columnwidth]{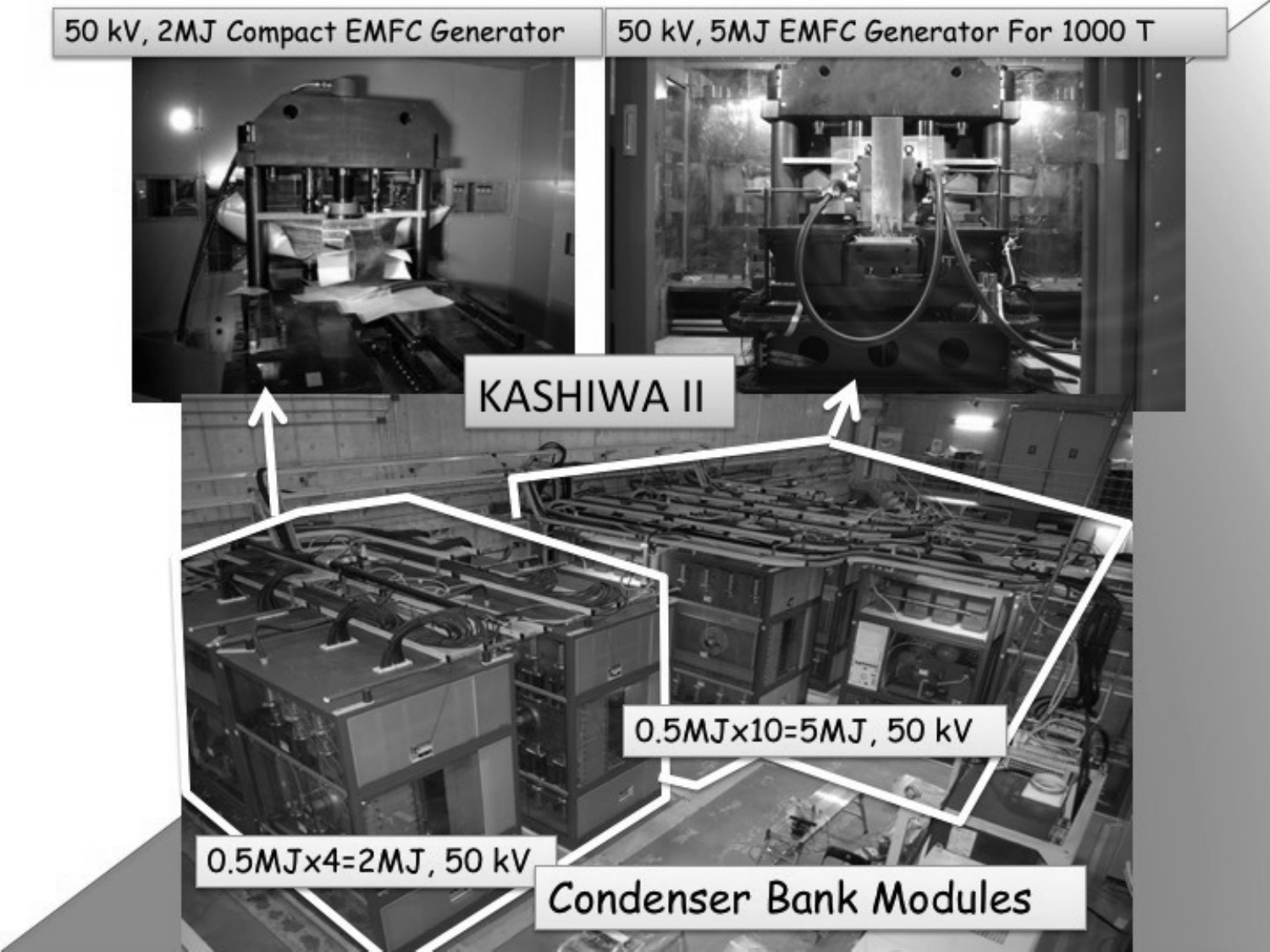} }
   \caption{
A photographic view of the capacitor module room ($11~\text{m} \times 24~\text{m} \times 12~\text{m}$ height) at ISSP, the University of Tokyo, where the 5~MJ and 2~MJ main capacitor modules are systematically arranged. In the foreground, a compressor is partially visible, which supplies dry air to the air-gap switches while precisely regulating the air pressure. 
[Reproduced from Ref.~\cite{Takeyama2018} (2018) with permission from IEEE.]}
    \label{modulephoto}
\end{figure}
A photograph of the condenser modules currently installed at ISSP, the University of Tokyo, is presented in Figure~\ref{modulephoto}. The characteristic parameters of the condenser modules, power switches, and high-voltage power cables are summarized in Table~\ref{spec_cond}.
The electric current is synchronously discharged from a system consisting of ten independent capacitor modules (0.5~MJ each), providing a total stored energy of 5~MJ. Each module integrates the capacitor units and an air-gap switch into a single unit measuring 1.8~m $\times$ 2.0~m $\times$ 2.6~m (height). These ten modules are installed in a dedicated capacitor room, which is located adjacent to the magnetic field experimental laboratory equipped with an explosion-proof chamber. 
Figure~\ref{collectorp} presents a 3D schematic illustration of the collector plates and the load-coil clamping press-gates, viewed from above.
\begin{figure}[tbp] 
 \centerline{\includegraphics[width=0.6\columnwidth]{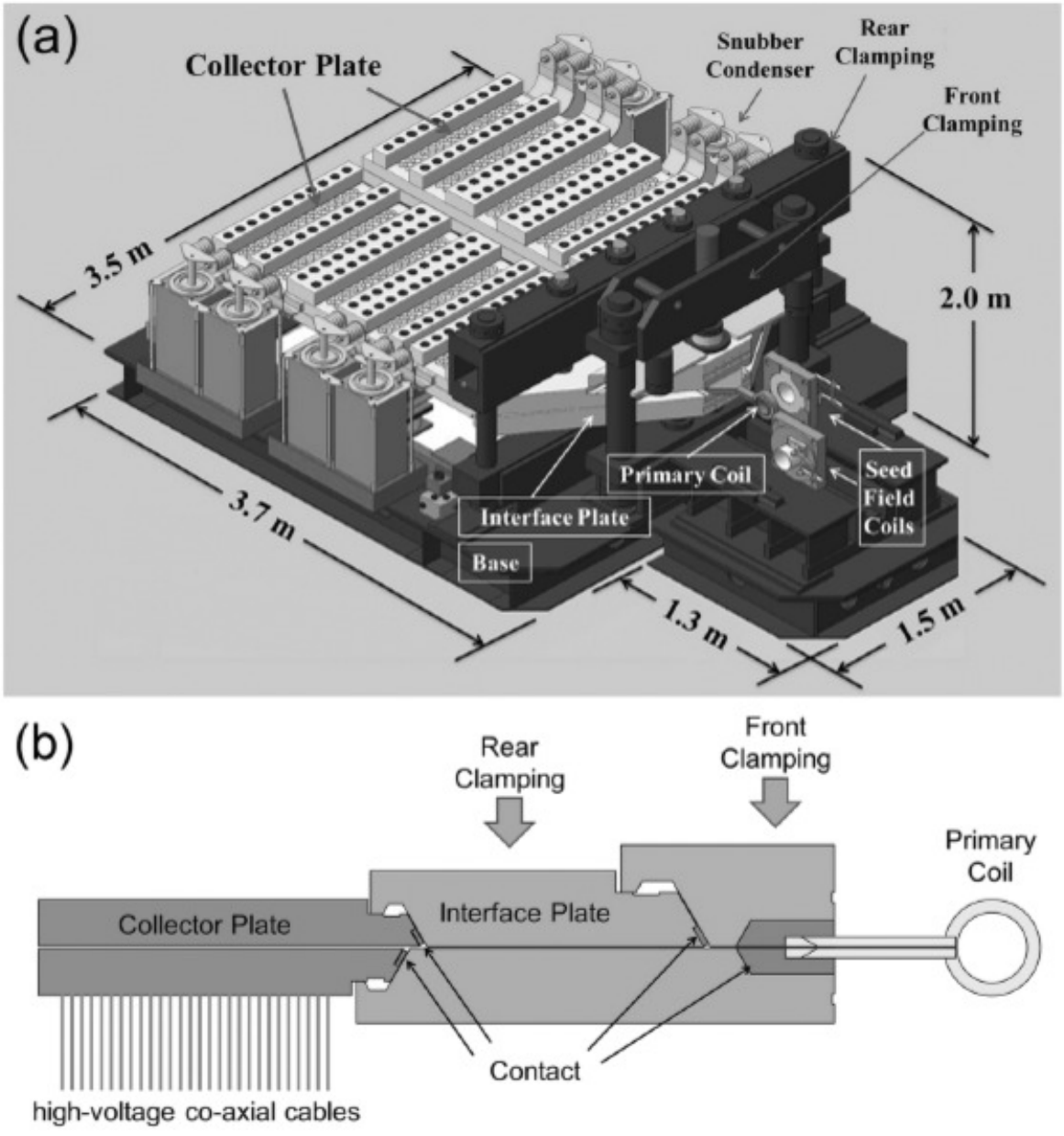} }
   \caption{
 (a) A schematic 3D view of the collector plates and the load-coil clamping press-gates. (b) Side view of the collector plate, showing the high-voltage coaxial cables and the load-coil clamping press. Multi-layered Mylar sheets (0.1~mm thick, 30 layers) are used for high-voltage insulation between the upper and lower plates. 
[Reproduced from Ref.~\cite{Takeyama1200T} (2018) with permission from AIP Publishing.]}
\label{collectorp}
\end{figure}
The discharged current is collected by two 15-cm-thick parallel collector plates fabricated from aluminum alloy (A6061P-T651) (see Figure~\ref{collectorplatebackphoto}). These plates are connected via 480 high-voltage coaxial cables extending from the gap switches of each module. 
To ensure high-precision synchronization of the current injection, all 480 cables are maintained at a uniform length of 30~m. 

As shown in Figure~\ref{collectorp}, the two parallel collector plates are coupled to a trapezoidal interface plate via a rear clamping press-gate, while the load coil is secured at the front clamping press-gate using a 100-ton hydraulic press. To ensure optimal electrical contact, a 1-mm diameter silver wire is placed across the interface between the collector plate and the trapezoidal plate. Upon tightening the rear clamping press, the silver wire is flattened to a width of approximately 3~mm, providing a robust contact area. This width is sufficient to accommodate the skin depth (approximately 2--3~mm) associated with the MHz-range high-frequency components of the injected pulsed current. Furthermore, snubber capacitors with matching resistors are connected to both sides of the collector plates. This configuration maintains impedance matching with the capacitor power sources located at a distance (10--30~m away), thereby preventing excessive voltage surges at the collector plates.

\begin{figure}[tbp] 
\centering
 \centerline{\includegraphics[width=0.5\columnwidth]{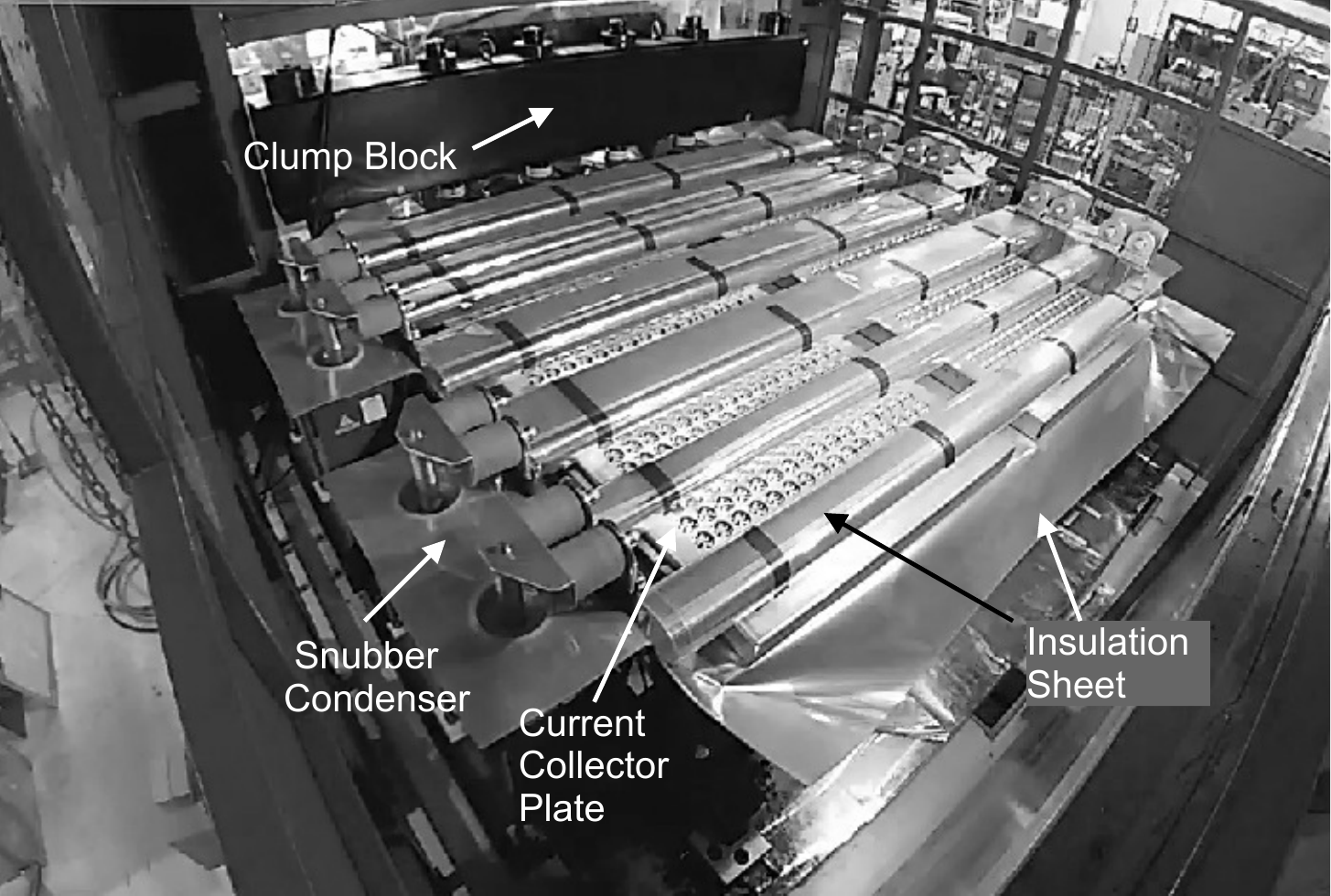} }
   \caption{
 Photograph of the collector plate situated at the rear of the explosion-proof chamber. A total of 480 high-voltage coaxial power cables (30~m in length) from the capacitor modules, located in the adjacent room, are connected to the collector plate from the bottom.
 [Photograph taken by the author at ISSP, The University of Tokyo.]}
\label{collectorplatebackphoto}
\end{figure}

\begin{table}[htbp]
\caption{Characteristic parameters of the capacitor bank modules.}
\label{spec_cond}
\centering
\begin{tabular}{lccc}
\toprule
\textbf{Parameters} & \textbf{Main 5~MJ} & \textbf{Main 2~MJ} & \textbf{Sub-capacitor}\textsuperscript{1} \\
\midrule
Energy (MJ) & 5 & 2 & 2 \\
Voltage (kV) & 50 & 50 & 20 \\
Capacitance (mF) & 4.0 & 1.6 & 10 \\
Switch type & 10 RAG\textsuperscript{2} & 4 RAG & 4 Ignitron \\
Number of HV cables & 480 & 192 & 4 \\
$I_{\text{max}}$ (MA) & 8 & 3.2 & 0.03\textsuperscript{3} \\
$R_{\text{res}}$ (m$\Omega$)\textsuperscript{4} & 0.6 & --- & --- \\
$L_{\text{res}}$ (nH)\textsuperscript{4} & 40 & --- & --- \\
\bottomrule
\end{tabular}

\vspace{5pt} 
\small
\begin{flushleft}
\textsuperscript{1} Sub-capacitor modules are used for generating the seed field. \\
\textsuperscript{2} RAG stands for a rotating air-gap switch. \\
\textsuperscript{3} Corresponding to 30~kA. \\
\textsuperscript{4} Residual resistance and inductance measured by setting a short-bar at the load position.
\end{flushleft}
\end{table} 
\subsubsection{Record Indoor Highest Magnetic Field 1,200~T}
\label{sec:record1200T}
\begin{figure}[htbp] 
\centering
 \centerline{\includegraphics[width=0.6\columnwidth]{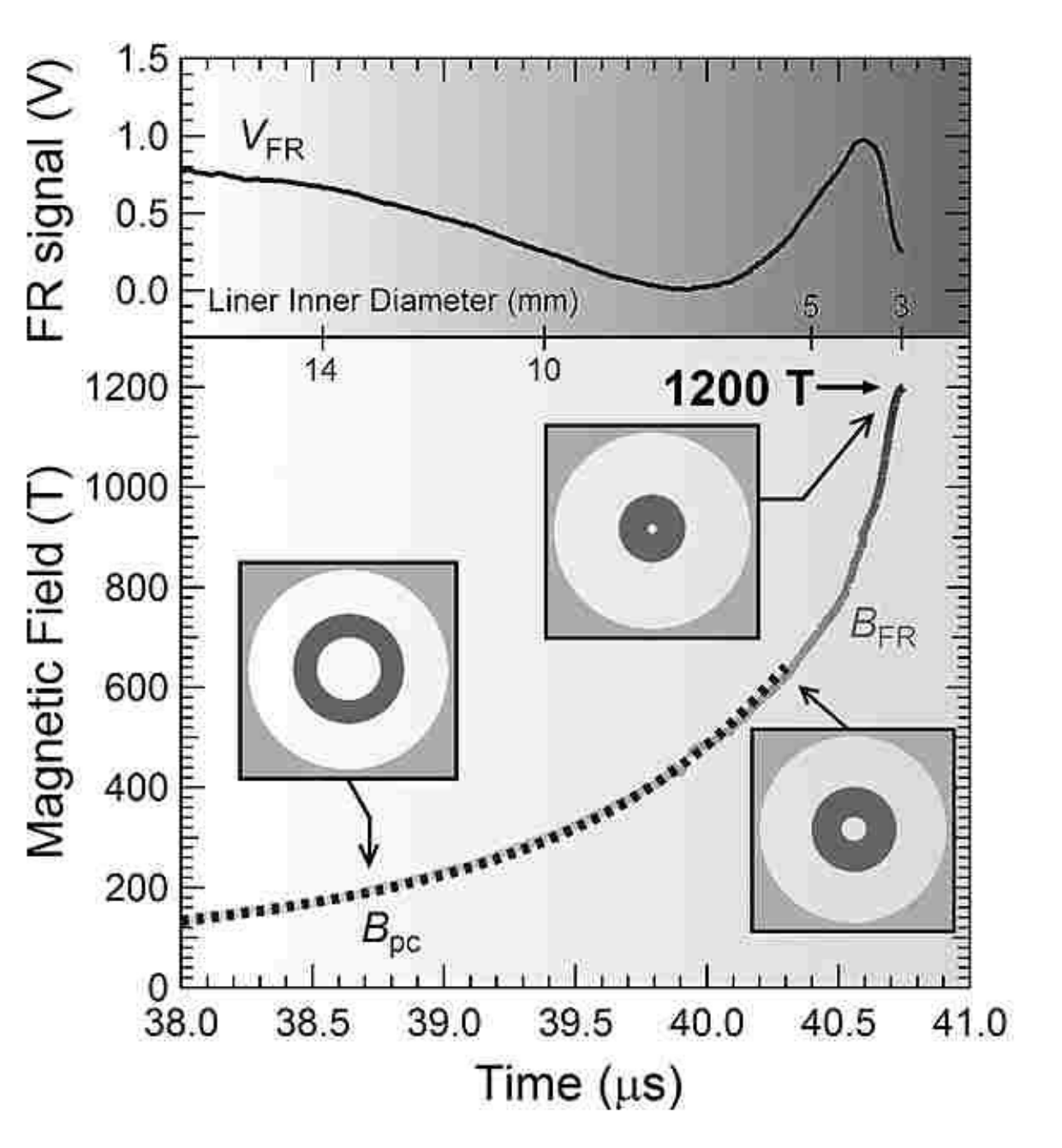} }
   \caption{Magnetic fields recorded up to 1,200~T. Upper panel: FR signal ($V_{\text{FR}}$) synchronized with the magnetic field intensity. The dotted line recorded from the pickup coil ($B_{\text{pc}}$) up to 600~T, and the green solid line represents magnetic fields ($B_{\text{FR}}$) up to 1,200~T converted from the FR angle ($V_{\text{FR}}$). The liner imploding images are illustrated by black-colored thick rings. The estimated liner inner diameter (which corresponds to the magnet bore) at each magnetic field intensity is guided on the axis of abscissa in the middle of the graph. 
[Reproduced from Ref.~\cite{Takeyama2018} (2018) with permission from IEEE.]}
    \label{1200T}
\end{figure}

A record-breaking indoor magnetic field of 1,200~T has been achieved by these capacitor bank modules discharging into the CL coil described in Section~\ref{sec:clc}. The coil parameters and experimental conditions used in the experiment are listed in Table~\ref{parameter1200T}. 
The recorded signals from this experiment are presented in Figure~\ref{1200T}. In the figure, the dotted line represents the signal from a pickup coil, which was functional only up to 600~T. Beyond this point, the magnetic field measurement reached its limit due to the dielectric breakdown of the ultrafine wire used in the pickup coil. This breakdown was triggered by the high induced voltage resulting from the extremely rapid increase of the magnetic field. Details are discussed in Section~\ref{sec:FaradayR}.


\begin{table}[htbp]
\caption{Experimental parameters employed for the generation of 1,200~T magnetic fields.}
\label{parameter1200T}
\centering
\begin{tabular}{lccc}
\toprule
\textbf{Component} & \textbf{Inner Diameter (mm)} & \textbf{Width (mm)} & \textbf{Thickness (mm)} \\
\midrule
Primary (steel) coil & 135 & 45 & 25 \\
CL plate & 130 & 45 & 2 \\
Copper liner & 119 & 50 & 1.5 \\
\midrule
\textbf{Operational Parameter} & \multicolumn{3}{c}{\textbf{Value}} \\
\midrule
Main bank charging voltage & \multicolumn{3}{c}{45~kV} \\
Main bank stored energy & \multicolumn{3}{c}{3.2~MJ} \\
Liner chamber vacuum & \multicolumn{3}{c}{0.06~Pa} \\
Initial seed field & \multicolumn{3}{c}{3.2~T} \\
\bottomrule
\end{tabular}
\end{table}
Simultaneously, the Faraday rotation (FR) angle was measured, with detection successfully maintained until the very end of the liner implosion. The solid green line represents the magnetic field converted from the FR angle. A characteristic slowdown is observed just before the ``turn-around'' phenomenon of the imploding liner.
In the inset, the concentric black circles at each data point illustrate the estimated cross-sectional images of the liner during implosion, based on the same simulation model reported in Ref.~\cite{NakamuraTakeyama2014}. The calculated inner diameter of the liner, which defines the magnet bore, is plotted on the center horizontal axis (abscissa). 
In Figure~\ref{1200T}, the liner imploding images (illustrated by black-colored thick rings) clearly show an increase in the cross-sectional thickness during the implosion process. This thickening of the liner as the implosion progresses is due to the fact that the volume of the liner before acceleration is substantially conserved throughout the process.
At the peak magnetic field of 1,200~T, the final inner diameter is estimated to be 3~mm. This bore size is sufficient to accommodate samples of the dimensions required for various optical measurements at room temperature. In conclusion, these results demonstrate the feasibility of performing optical measurements at room temperature in magnetic fields reaching 1,200~T.

\section{Magnetic Field Precision Measurement Probes}
\label{sec:B_fieldMeasur}

{How is the magnetic field applied to or utilized in science? 
Primarily, magnetic fields are used to investigate material properties by providing a physical environment that plays a key role in uncovering the essential nature of matter. Electronic states, which determine these properties, can be precisely probed using resonance techniques such as nuclear magnetic resonance (NMR), cyclotron resonance, and~electron spin resonance (ESR). The~intrinsic properties of materials are revealed by lifting the quantum degeneracy of electronic states through either cyclotron motion or the Zeeman effect, both induced by the magnetic field.
Furthermore, magnetic fields can fundamentally ``change'' material properties, as~exemplified by magnetic-field-induced insulator-to-metal transitions}. 
By inducing electronic phase transitions, magnetic fields cause drastic shifts in material behavior (see for details in Ref.~\cite{Berthier2002}). In~either case, the~magnetic field provides a crucial axis (abscissa) for the phase diagrams that depict the electronic states of matter. 
Therefore, it is of paramount importance in physics experiments to evaluate the strength of the magnetic field as precisely as possible.
In the ultra-high magnetic field regime spanning from 100~T to 1000~T, however, the~absolute determination of the field strength itself becomes exceptionally challenging. Historically, it has often been conventionally assumed that a 10\% uncertainty in the field value is acceptable~\cite{HMF-ST2003}. Nevertheless, to~achieve truly reliable and high-precision physical property measurements under these extreme environments, it is imperative to suppress this experimental uncertainty to less than 3\%.

\subsection{Pickup Coil Method}

Despite recent advancements in magnetic flux compression techniques, insufficient attention has been paid to the precision and reliability of the magnetic flux measurements themselves. As the peak field increases, the time derivative of the magnetic flux density ($dB/dt$), which is proportional to the induced voltage in the pickup coil, also grows substantially. This escalation demands rigorous verification to ensure that the measurements remain both accurate and reliable under such extreme conditions.

\begin{figure}[tbp] 
\centering
 \centerline{\includegraphics[width=0.5\columnwidth]{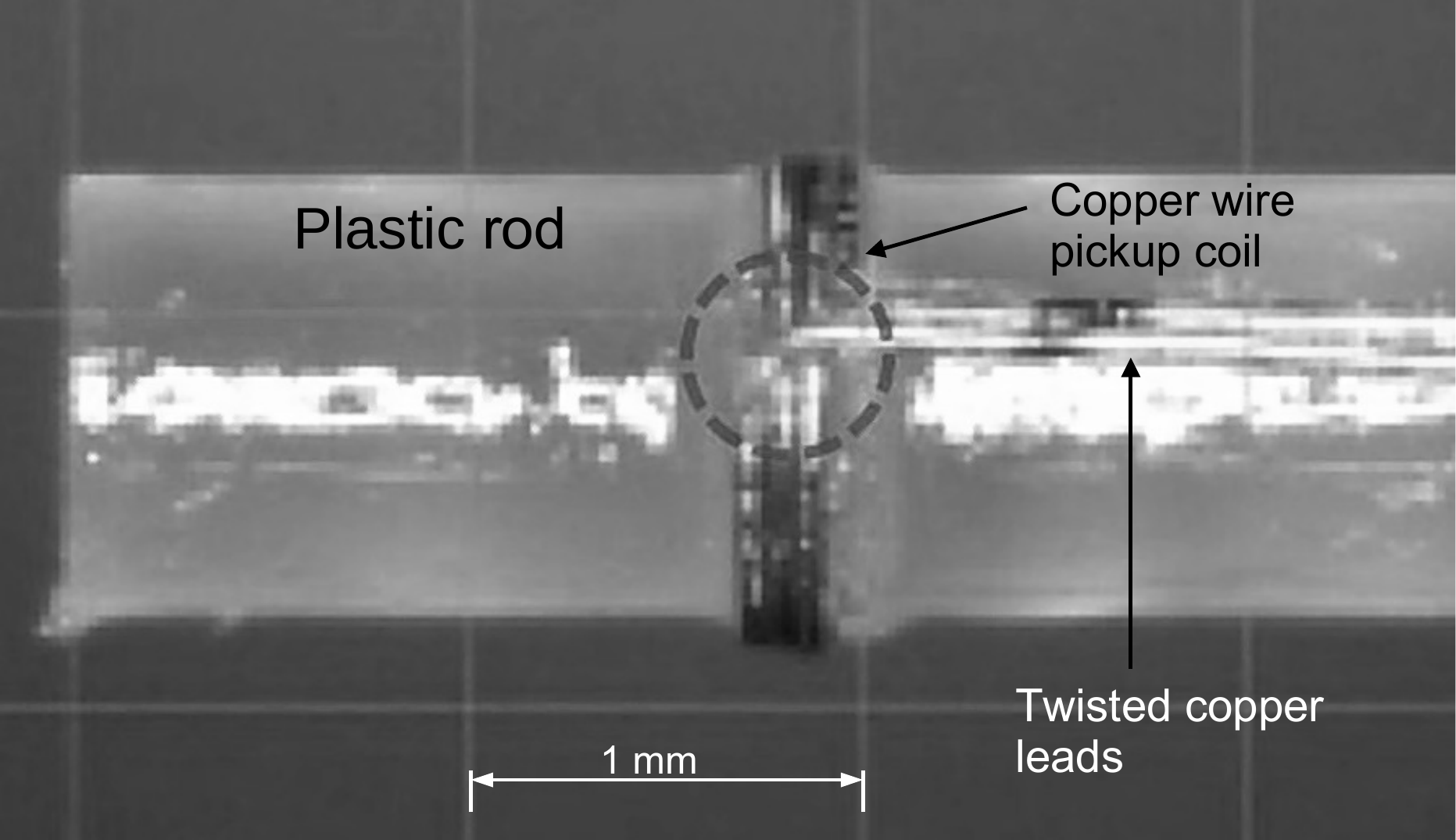} }
   \caption{A typical magnetic field pickup coil consisting of three turns wound around 
a 1-mm-diameter plastic rod (FRP or G10), used for EMFC 
experiments. The wire used is 60-$\upmu$m-thick polyamide-imide enameled copper wire, 
which provides high resistance to electrical insulation breakdown. The area 
enclosed by the dashed line indicates the most critical part, where the highest 
precision is required during the winding process, as it is a decisive factor 
in determining the performance of the pickup coil.
[Photograph taken by the author at ISSP, The University of Tokyo.]}
    \label{pickupcoil}
\end{figure}

Pulsed magnetic fields are generally measured using a pickup coil, typically constructed by winding thin copper wire (30--80~$\upmu$m in diameter) around a sample holder. In destructive pulsed-field experiments, these coils usually consist of only 1--3 turns on tubes with diameters of 1--3~mm (see Figure~\ref{pickupcoil}). Due to their diminutive scale, even minor structural deviations can significantly alter the effective area, making it impractical to determine precise dimensions through direct geometric measurement. Consequently, the calibration of these coils requires exceptional precision, which is typically performed by comparing the induced voltage of the miniature coil with that of a pre-calibrated reference coil. Both coils are subjected to a highly stable, high-frequency AC magnetic field (typically several kHz) generated by a multi-turn Helmholtz coil.

Despite their utility, measuring ultrastrong magnetic fields with pickup coils presents significant technical challenges due to the enormous induced voltages resulting from the massive $dB/dt$. Since the induced voltage can easily exceed 1,000~V, the operational range is strictly limited by the dielectric breakdown of the insulation on the ultra-thin copper wires. Such electrical failure frequently occurs at these high-voltage thresholds, making it difficult to rely solely on the pickup coil method as the field strength approaches the megagauss regime. In particular, the terminal lead-out point of the winding (indicated by the dashed line in Figure~\ref{pickupcoil}) requires the most meticulous attention during fabrication. If the wire is bent at a sharp right angle to achieve a precise geometry, the insulation coating weakens, leading to immediate dielectric breakdown under high induced voltages. Therefore, it is crucial to provide a certain curvature at this point; a bend radius ($R$) of 0.2--0.3~mm is highly recommended to ensure electrical integrity and overall performance.

%
\subsection{Faraday Rotation Method}
\label{sec:FaradayR}

A systematic survey of magnetic field measurement probes was conducted, leveraging the enhanced controllability and reproducibility of magnetic field generation afforded by the high-performance CL coil recently employed in EMFC experiments \cite{NakamuraFRPickup2013}. The use of this specific coil geometry allows for the pulse waveform of the magnetic field to be precisely tailored and consistently reproduced by adjusting the total capacitance or the charging voltage of the capacitor modules.
Another crucial feature is the observation of the ``turn-around phenomenon,'' which exhibits a peak-like structure resulting from the magnetic flux leakage from the imploding liner during the final stages of compression. The successful observation of this turn-around point serves as a definitive indicator for the integrity of the measurement throughout the entire compression process. Specifically, it confirms that the pickup coil remained functional without suffering from dielectric breakdown or other electrical failures even at the moment of peak field intensity. Consequently, the presence of a complete turn-around waveform validates that the measured values represent the actual maximum field achieved before the probe was eventually destroyed by the imploding liner.
\begin{figure}[tbp] 
\centering
 \centerline{\includegraphics[width=0.85\columnwidth]{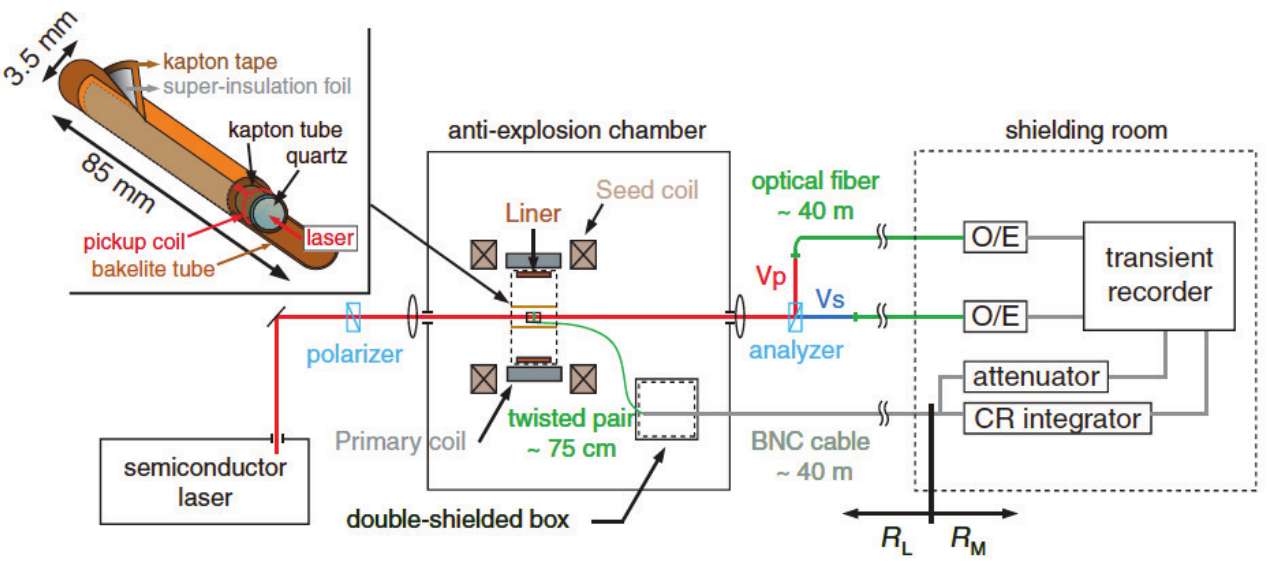} }
   \caption{Experimental setup for simultaneous magnetic field measurements using Faraday rotation and a pickup coil.
A Faraday rotator rod (quartz or crown glass) is inserted into a Kapton tube and mounted within a Bakelite sample holder, which is positioned at the center of the CL magnet coil. 
To protect against intense electromagnetic noise and stray light from the imploding liner, the sample holder is wrapped in super-insulation foil.
Linearly polarized light is transmitted through the Faraday rod and subsequently split into $s$- and
$p$-polarized components by a Wollaston prism (analyzer). The intensities of these components are converted into electrical signals, $V_s$ and $V_p$, via an optical-to-electrical (O/E) converter and recorded by a transient digital recorder. 
Simultaneously, the magnetic field pickup coil is wound around the Faraday rod. To maintain signal integrity, the coil leads---consisting of a 75-cm-long twisted pair of the same fine copper wire---are extended and soldered to a BNC terminal within a small shielded box, located away from the magnet coil. This shielded terminal serves as the interface to the BNC cables leading to the recording room.
[Reproduced from Ref.~\cite{NakamuraFRPickup2013} (2013) with permission from AIP Publishing.]}
    \label{faradaypickup}
\end{figure}

The experimental setup for this survey is illustrated in Figure~\ref{faradaypickup}.
The detection of the Faraday rotation becomes highly sensitive by splitting the transmitted light into two components ($s$- and $p$-polarized light); this differential method is particularly effective when the rotation angle is very small. The intensity of each component is converted into electrical signals, $V_s$ and $V_p$, via an O/E converter. These signals are then used to calculate the Faraday rotation angle, $\theta_{\text{F}}$, as follows:
\begin{equation}
\theta_{\text{F}}~[\text{deg.}] = \frac{180}{2\pi} \times \arccos \left( \frac{V_s - V_p}{V_s + V_p} \right).
\end{equation}
The magnetic field $B$ is obtained from the FR angle ($\theta_{\text{F}}$),
\begin{equation}
 B_{\text{FR}} = \frac{\theta_{\text{F}}}{v(\lambda) L}, 
 \label{BFR}
\end{equation}
where $v$ is the Verdet constant at a given wavelength of an incident light and $L$ is the length of the Faraday rod.

In EMFC experiments, data acquisition is performed in an electromagnetically shielded room located several meters away from the explosion-proof chamber. A pickup coil, featuring a 0.7--1~m long twisted copper wire, is connected to a 40-m-long BNC cable that extends to the shielded measurement and control room.
The magnetic field is calculated from the voltage recorded by a digital recorder, taking into account the signal transmission loss along the aforementioned line. Here, $B_p$ represents the field calculated considering only the DC component (resistive loss). Since the recorded signals contain high-frequency components reaching the MHz range, $B_p$ must be further corrected to $B_{p,c}$ to account for high-frequency transmission losses. A comprehensive comparative study regarding this correction can be found in Ref.~\cite{NakamuraFRPickup2013}.

\begin{figure}[tbp] 
\centering
 \centerline{\includegraphics[width=0.6\columnwidth]{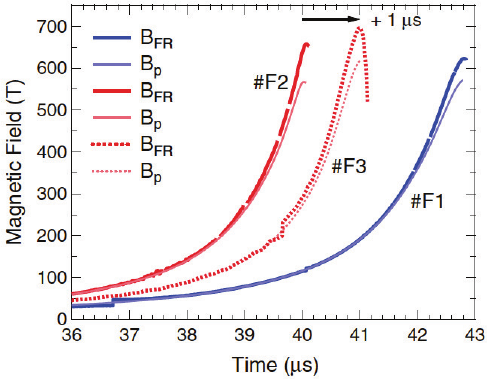} }
   \caption{
Magnetic field curves obtained from the pickup coil induced voltage (thin lines, $B_p$) and from the Faraday rotation ($\theta_{\text{F}}$) of fused quartz (thick lines, $B_{\text{FR}}$). Labels \#F1--\#F3 denote the data obtained from different experimental conditions (see text).
The curves for \#F3 are offset by 1~$\upmu$s along the time axis for clarity of comparison.
[Reproduced from Ref.~\cite{NakamuraFRPickup2013} (2013) with permission from AIP Publishing.]}
    \label{compFRandV}
\end{figure}

Figure~\ref{compFRandV} compares the magnetic fields obtained from the Faraday rotation of fused quartz rods with the fields measured under various experimental conditions. Two laser wavelengths, 404~nm (\#F1, \#F2) and 638~nm (\#F3), were used as light sources.
Additionally, the magnetic field growth rate was varied---slow for \#F1 and rapid for others---by selecting different capacitances in the capacitor modules.
In all experiments, the peak value of $B_{\text{FR}}$ is 50--100~T larger than that of $B_p$.
The fact that $B_{\text{FR}}/B_p$ ratios are similar for \#F2 ($\lambda = 404$~nm) and \#F3 ($\lambda = 638$~nm) suggests that the wavelength dependence of the Verdet constant is not the primary cause of this discrepancy. Instead, the major difference between $B_{\text{FR}}$ and $B_p$ is attributed to the high-frequency response of the electric circuit, which has been previously overlooked.

\begin{figure}[tbp] 
\centering
 \centerline{\includegraphics[width=0.5\columnwidth]{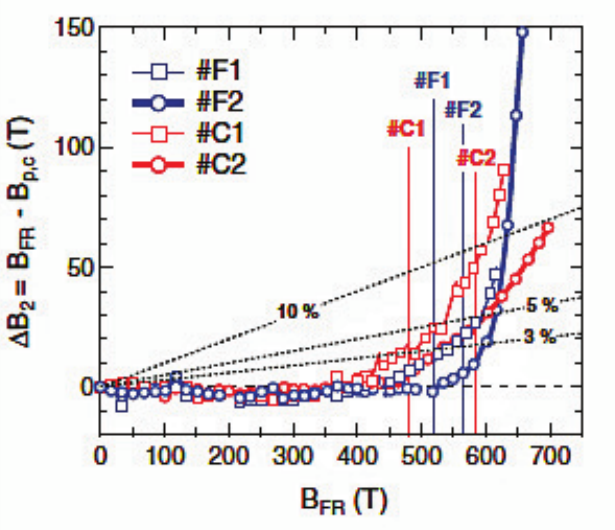} }
   \caption{
The dotted lines indicate the estimated error boundaries of 3\%, 5\%, and 10\%. The vertical thin lines indicate the positions of $B_{\text{FR}}$, where the induced voltage in the pickup coil reaches its maximum value. \#F1--\#C2 represent results under various experimental conditions, the details of which are described in Ref.~\cite{NakamuraFRPickup2013}.
[Reproduced from Ref.~\cite{NakamuraFRPickup2013} (2013) with permission from AIP Publishing.]}
    \label{FRvsBpc}
\end{figure}

According to their study, the magnetic field $B_p$ obtained from conventional calibrations using a resistance ratio is valid only up to 200~T; beyond this threshold, the high-frequency response of the signal transmission line must be accounted for. Crucially, as illustrated in Figure~\ref{FRvsBpc}, even the calibrated values $B_{p,c}$ exhibit insufficient accuracy above 500--600~T, which defines the practical limit for reliable measurements using this method. The authors concluded that magnetic fields in the megagauss regime can only be measured with sufficient precision using the FR of fused quartz (or crown glass). For these materials, Equation~(\ref{BFR}) remains valid up to at least 700~T, as supported by the experimental data presented in Figure~\ref{FRvsBpc} \cite{NakamuraFRPickup2013}.

\begin{figure}[tbp] 
\centering
\centering{\includegraphics[width=0.8\columnwidth]{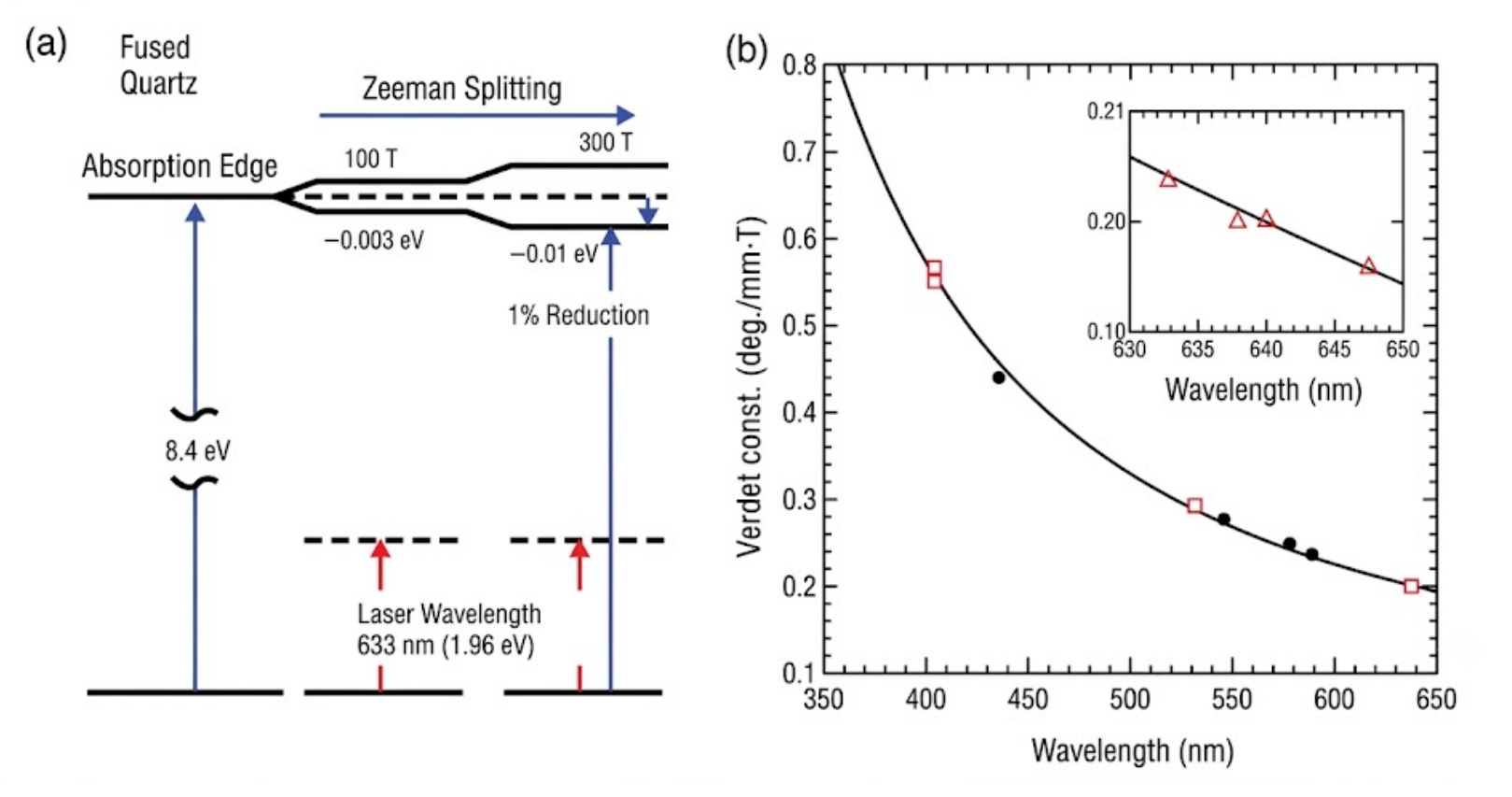} }
   \caption{
(a) Absorption edge of fused quartz (6.4~eV) and its shifts caused by the Zeeman effect at 600~T and 1,200~T. The laser wavelength of the incident light for Faraday rotation is set at 640~nm (photon energy = 1.94~eV).
(b) Verdet constant of fused quartz as a function of the incident light wavelength, $v(\lambda)$. The dependence around 640~nm is enlarged in the inset. The solid line represents the fitting curve obtained from the data ($\circ$) in Ref.~\cite{NakamuraFRPickup2013} and those data ($\bullet$) by Garn et al. \cite{Garn1968}. The dashed line represents the estimated Verdet constant curve accounting for the absorption-edge Zeeman shift of $-0.07$~eV at 1,200~T.
[Reproduced from the Supplementary Material in Ref.~\cite{Takeyama1200T} (2018) with permission from AIP Publishing.]}
    \label{faradayvalid}
\end{figure}

Then, to what extent can fused quartz be utilized as a reliable Faraday rod?
The Verdet constant in Equation~(\ref{BFR}) is not strictly a constant, but rather depends on the wavelength of the incident laser light. The wavelength ($\lambda$) dependence of the Verdet constant in optical glasses is known to follow the empirical formula \cite{Bach1995}:

\begin{equation}
\centering
 v(\lambda) = \frac{\pi}{\lambda} \left( a + \frac{b}{\lambda^2 - \lambda_0^2} \right),
\label{Verdetonlamda}
\end{equation}
where $a$ and $b$ are parameters obtained by fitting experimental data, and $\lambda_0$ is the wavelength corresponding to the optical absorption edge of the glass. This dependence is shown in Figure~\ref{faradayvalid}.
In extremely strong magnetic fields, the most significant factor potentially affecting the fused quartz Faraday rotator is the shift of the absorption edge due to the Zeeman shift, $\frac{1}{2} g \upmu_{\text{B}} B$, where $g$ is the $g$-factor and $\upmu_{\text{B}}$ the Bohr magneton.

The optical absorption edge of fused quartz is approximately 6.4~eV ($\lambda \sim 200$~nm). For instance, at a magnetic field of 1,200~T, the shift is estimated to be $-0.07$~eV, which accounts for only 1.1\% of the absorption edge energy. This is illustrated in Figure~\ref{faradayvalid}(a). A Zeeman shift of 0.07~eV causes a negligible change (represented by the dashed line in Figure~\ref{faradayvalid}(b)), and the resulting change in the Verdet constant $v$ (at $\lambda = 640$~nm) is estimated to be 0.1\%, which corresponds to an error of 1.2~T at a magnetic field of 1,200~T. 
It is important to note from Figure~\ref{FRvsBpc} that the 500--600~T range represents the definitive boundary where the measurement error exceeds 5\%, highlighting the operational limit of the current pickup coil system. Beyond this threshold, accurate measurements can only be achieved by utilizing the Faraday rotation of a quartz rod or crown glass, which maintains its reliability in higher magnetic fields up to at least 1,200~T.

\section{Cryogenics and Sample Environments}
\label{sec:cryo}
In solid-state physics experiments, it is frequently necessary to conduct measurements at cryogenic temperatures. Samples are typically inserted into a miniature cryostat designed to fit within the narrow bore of a destructive magnet. Under the condition of short-pulse magnetic fields with a microsecond-order ($\upmu$s) duration, metallic components must be strictly avoided. This is because the time derivative of the field can easily reach the order of $10^{10}$~T/s, inducing explosive eddy-current heating that can lead to the immediate burnout of the apparatus.

%
%

\subsection{Miniature Cryostat Sample Holder for the Horizontal Single-Turn Coil System}

The sample is placed within the optical path of the central tube, which is separated from the liquid helium flow by a thin Bakelite wall (indirect cooling). Consequently, the lowest achievable temperature in the sample space, which is kept under vacuum, is limited to approximately 5~K. 
By employing this all-Bakelite type sample cryostat, it has become possible to withstand 10--15 consecutive shots even under 150~T-class magnetic field generation. In optical measurements, even a minute crack in the sample cryostat causes liquid helium to leak, which obstructs the optical path and renders further measurements impossible.
\begin{figure}[tbp]
\centering
\includegraphics[width=0.8\columnwidth]{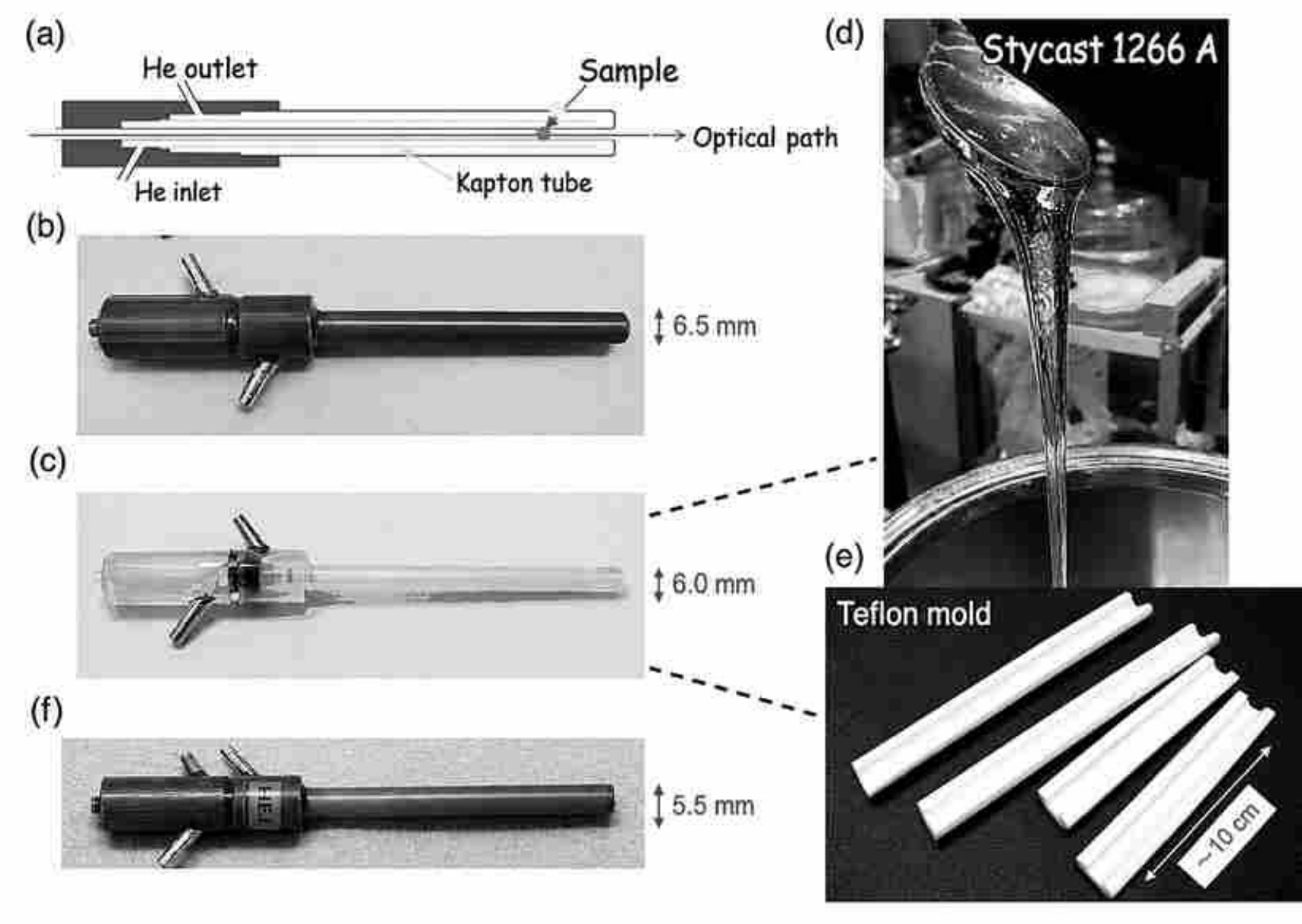}
\caption{
Liquid $^4$He flow-type miniature all-plastic cryostats for optical measurements in megagauss magnetic field generation (STC and EMFC).
(a) Cross-sectional view showing thin Kapton tubes dividing the space into coaxial sections for the $^4$He inlet and outlet. (b)--(f) Various cryostat versions optimized for specific experimental requirements: (b) All-Bakelite type (assembled using Crest 2170 cryogenic epoxy); (c) All-Stycast type; and (f) All-FRP (fiber-reinforced plastic) type, with all surfaces coated with Nitofix SK-229. (d), (e) Detailed fabrication process of the All-Stycast cryostat (c) using Teflon molds. These cryostats, originally designed and hand-crafted by the author, represent a breakthrough in cryogenic engineering, enabling stable measurements down to 5~K within the extremely confined and harsh environments of megagauss fields. [All figures and photographs were taken or created by the author; portions have been adapted from the author's previous publications.]}
\label{cryohyst}
\end{figure}

In EMFC experiments, a 1-mm reduction in the final inner diameter of an imploding liner results in an increase of the peak magnetic field by approximately 100~T. To overcome the thickness limitations of conventional vacuum-tight tubes, a Stycast-Teflon molding method was developed (Figure~\ref{cryohyst}(c--e)), enabling the reduction of the cryostat's outer diameter to 6~mm. The thermal conductivity of Stycast 1266 is superior to that of Bakelite, providing a distinct advantage for reaching lower temperatures. This all-Stycast-1266 cryostat has been specifically designed for optical measurements in EMFC experiments. It has enabled measurements in magnetic fields up to 600~T at 5~K, representing the highest magnetic field and lowest temperature at which magnetization data have ever been obtained \cite{MiyataPRL, MiyataJpsj}.

Unlike EMFC, where the all-Stycast cryostats are viable, magnetic field generation via STC presents a different challenge. In STC experiments, all-Stycast cryostats are easily damaged by the shockwaves emerging from single-turn coil explosions when fields exceeding 130~T are applied. FRP (specifically G10) thin tubes are more resilient against such shockwaves, but they are prone to liquid or gaseous $^4$He leaks. 
Recently, it was found that such leaks can be prevented by coating the tube surfaces with a cryogenic epoxy adhesive, ``Nitofix SK-229,'' which is also used for bonding all components. The diameter of the cylindrical part of this cryostat has been further reduced to 5.5~mm. This cryostat is designed to fit within a single-turn coil with an 8-mm inner bore, leading to the successful measurement of optical Faraday rotation and magnetization in magnetic fields up to 210~T at 5~K \cite{Otsuka2018}. The FRP cryostat shown in Figure~\ref{cryohyst}(f) remained intact and survived after 10 experimental shots in magnetic fields ranging from 200--210~T. This enhancement in the magnetic field limit---from 150--160~T to over 200--210~T---was achieved not only through the improvements to the sample cryostat but also via significant modifications to the explosion-proof chamber. Specifically, the front access sliding door was replaced with a polycarbonate version, and a shockwave-absorbing balloon was installed on the ceiling (see Figure~\ref{fig:panoramaHSTC}). This configuration successfully minimized shockwave back-reflection toward the sample holder, protecting the experimental setup from structural failure under extreme conditions.

\begin{figure}[tbp]
\centering 
 \centerline{\includegraphics[width=0.6\columnwidth]{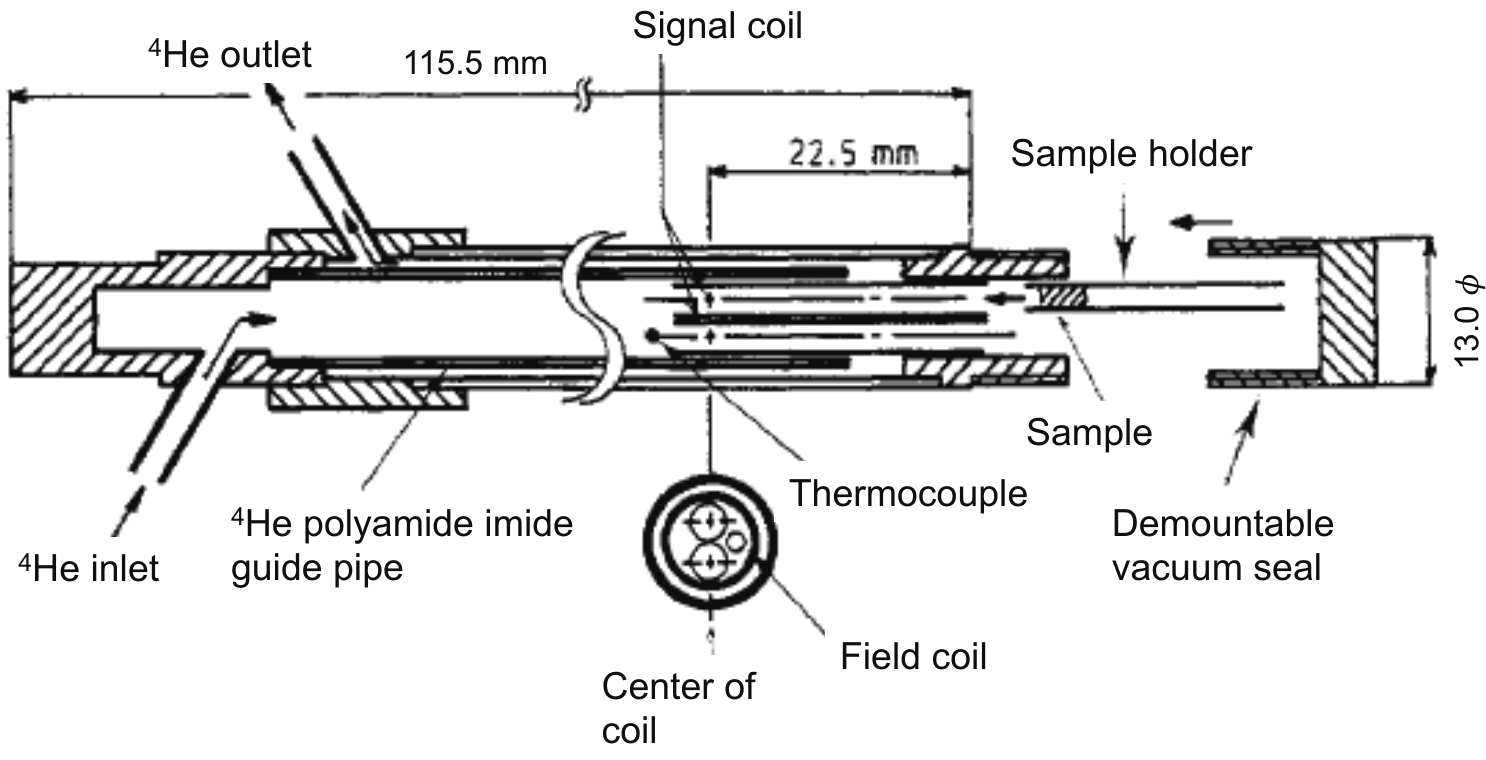} }
   \caption{
He cryostat for magnetization measurements in the horizontal STC system.
All components are made of Stycast 1266 and bonded together using the same material. The signal coil (a pair of magnetic pickup coils) is designed for magnetization measurements. ``Field coil'' denotes the pickup coil used to measure the magnetic fields. Details are described in Section~\ref{sec:MagMeasure}; see also Figure~\ref{Amayapickupcoil}(c). 
[Reproduced from Ref.~\cite{TakeyamaAmaya1988} with permission from IOP Publishing (1988).]}
    \label{amayacryo}
\end{figure}

The liquid $^4$He cryostat was designed for magnetization measurements in the horizontal STC system (Figures~\ref{fig:HSTC}, \ref{fig:panoramaHSTC}); its detailed cross-sectional view is illustrated in Figure~\ref{amayacryo} \cite{TakeyamaAmaya1988}. All cryostat components are fabricated from Stycast 1266, which also serves as the adhesive for the entire assembly. Constructing the entire body from the same material avoids distortions and cracks that could otherwise arise from mismatched thermal contraction rates.

A complete set of magnetization measurements requires two identical magnetic field applications. This is performed by moving the sample from one of the twin pickup coils to the other, which is easily accessible via a screw cap (indicated as the ``demountable vacuum seal'' in Figure~\ref{amayacryo}) at one end. A reliable vacuum seal is maintained by applying silicone grease to the screw threads.
The sample is directly immersed in liquid $^4$He, maintaining a temperature of 4.2~K during measurements. The outer diameter of the cryostat can be reduced to 8~mm, which is small enough to fit within a 12-mm-bore STC magnet. This setup enables magnetization measurements in magnetic fields up to 120~T \cite{Goto1994}.

\subsection{Cryostat for the Vertical Single-Turn Coil}
\label{sec:cryoV}
While the cryostats for the HSTC system are based on a $^4$He flow-type cooling method, which makes it technically challenging to achieve temperatures below 4.2~K, the VSTC system allows for direct vacuum pumping of the liquid $^4$He bath. This design facilitates efficient cooling down to 1.6--2~K, enabling high-precision measurements in the superfluid phase below the $\lambda$-point (2.17~K).
In this section, we describe the development and structural optimization of the VSTC cryostat specifically designed for these sub-$\lambda$-point temperature experiments.

\begin{figure}[tbp]
\centering
\includegraphics[width=0.7\columnwidth]{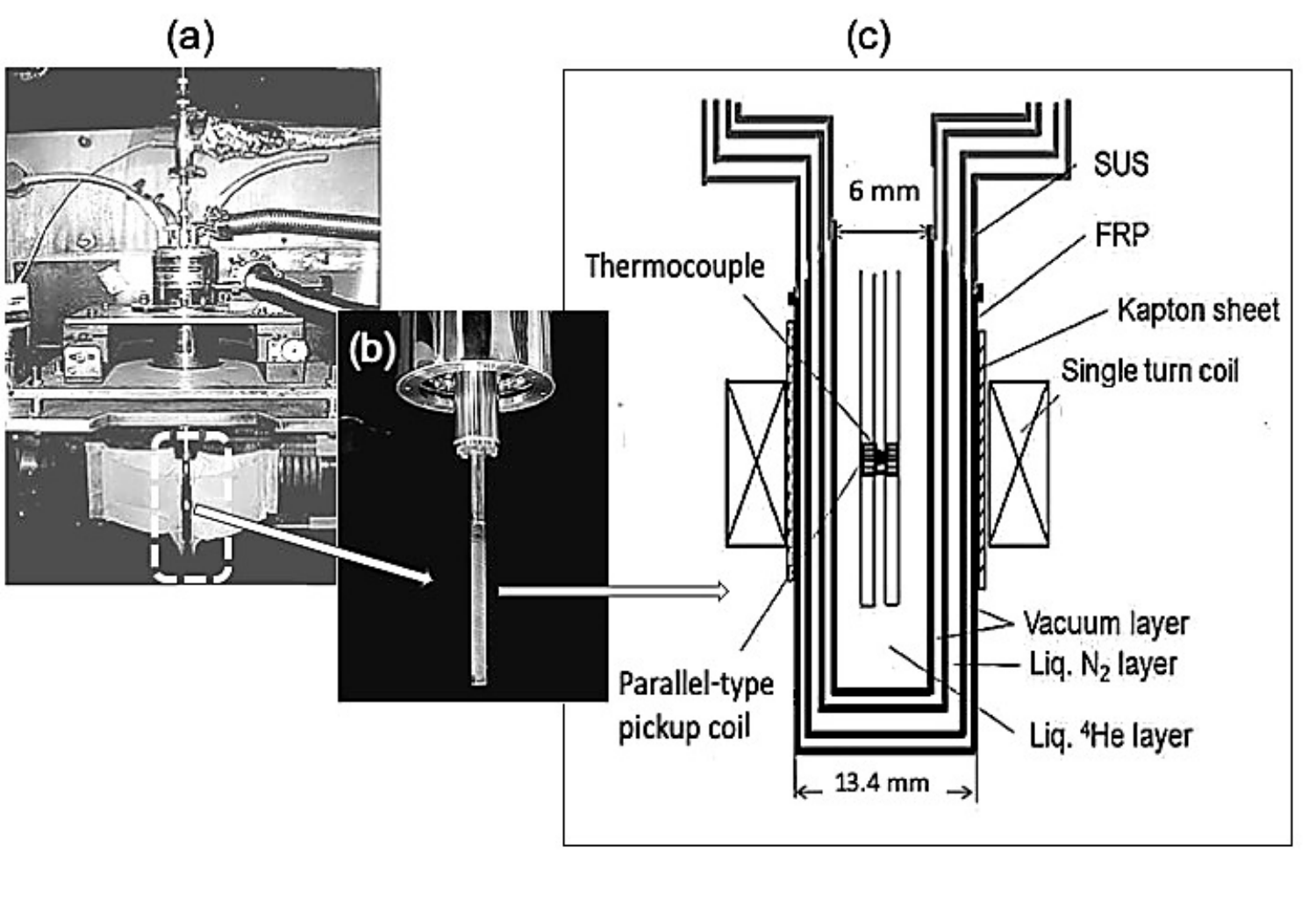}
\caption{
$^4$He cryostat used for magnetization measurements in the VSTC system.
(a) The cryostat positioned within the STC magnet coil. (b) Enlarged view of the cryostat tail section, composed of four glass-epoxy thin tubes. (c) Detailed cross-sectional illustration of the tail section. The ``parallel-type pickup coil'' is used to measure the magnetization of a sample. The magnetization measurement technique is described in Section~\ref{inductionm}.
[All figures and photographs were taken or created by the author. Portions of the images have been presented in the author's previous publications and presentations.]}
\label{vstcCryo}
\end{figure}

A high-performance cryostat was specifically developed and optimized for the vertical single-turn coil (VSTC) system~\cite{TakeyamaSakakura2012}. The dimensions of the cryostat are precisely adjusted to fit the 14-mm bore of the VSTC, which is capable of generating a maximum magnetic field of 100~T. As shown in Figure~\ref{vstcCryo}(a), the upper part of the cryostat is constructed from stainless steel and contains a 0.36-L liquid $^4$He reservoir. This is connected to a tail section made of thin FRP tubes. To ensure structural integrity and vacuum tightness, the entire surface of the FRP tail is pre-coated with a cryogenic epoxy adhesive (Nitofix SK-229). Furthermore, the tail section is joined to the upper stainless-steel section using the same adhesive, as illustrated in Figure~\ref{vstcCryo}(b).

Insulating Kapton sheets are placed inside the coil, leaving a 13.4-mm-diameter space for the cryostat tail. With the sample space set at 6~mm in diameter, four thin separating walls---comprising two vacuum insulation layers and one liquid nitrogen (LN$_{2}$) jacket for thermal shielding---must be incorporated within a radial gap of only 3.7~mm [calculated as (13.4--6.0) / 2]. The innermost and outermost tube walls are 0.7~mm thick, while the two intermediate walls are 0.5~mm thick. The spacing for the inner vacuum layer, the LN$_{2}$ reservoir layer, and the outer vacuum layer is designed to be 0.3~mm, 0.5~mm, and 0.5~mm, respectively (Figure~\ref{vstcCryo}(c)). Liquid $^4$He can be maintained for 2--3~h after the reservoir is fully filled. Furthermore, a temperature of 1.7~K can be maintained for approximately one hour by pumping the liquid $^4$He bath. For a visual overview of the actual experimental environment and its operation, please refer to Figure~\ref{fig:VSTC} and the corresponding caption.

\subsection{Sample and Cryostat Assembly in Electromagnetic Flux Compression}
\label{sec:sampleset}

\begin{figure}[tbp] 
\centering
 \centerline{\includegraphics[width=0.4\columnwidth]{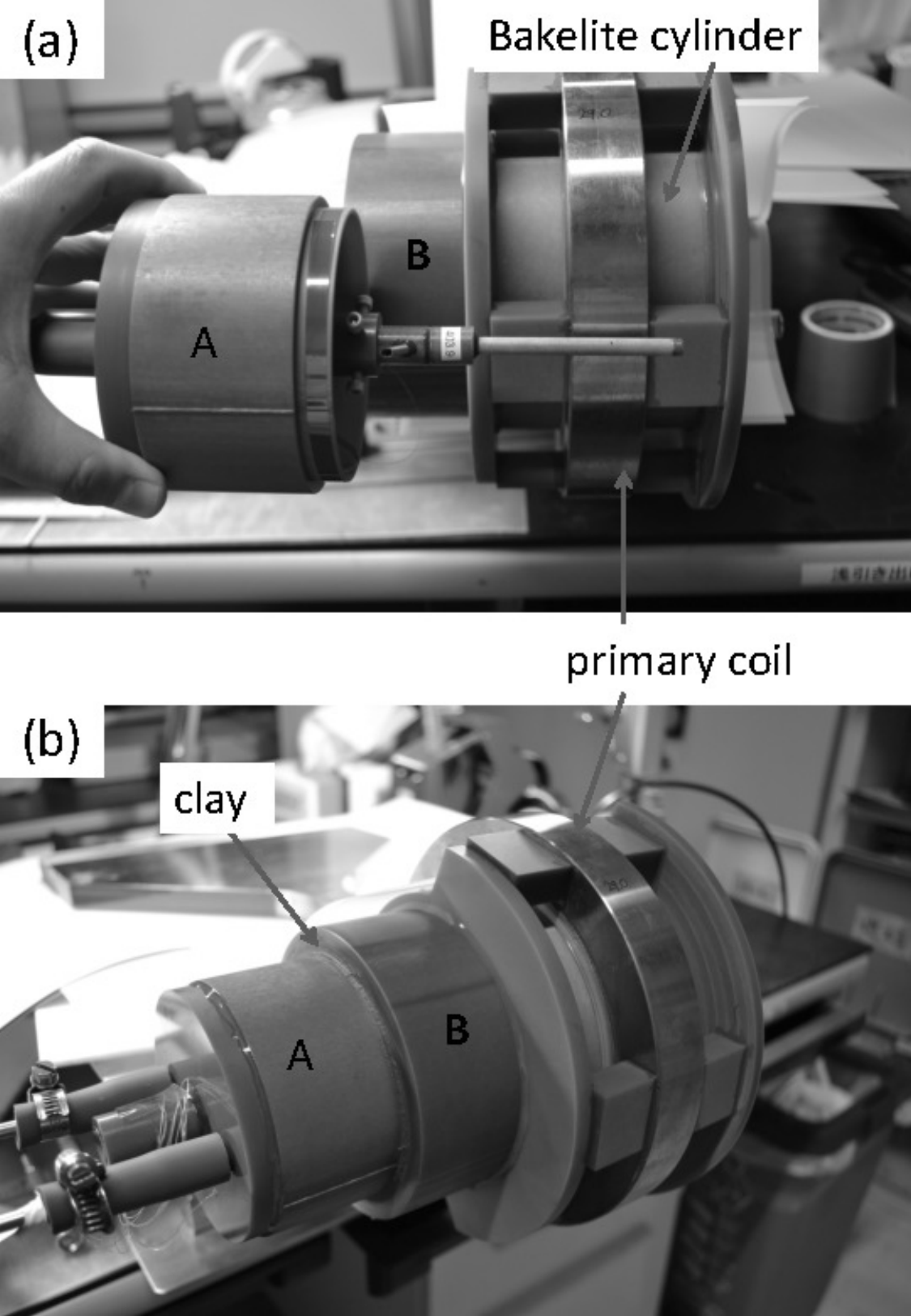} }
   \caption{
 Setup of the CL coil and cryostat with a sample holder for solid-state physics measurements at low temperatures.
(a) The $^4$He flow-type cryostat shown in Figure~\ref{cryohyst}(f) is mounted on flange A. (b) Flange A, along with the cryostat, is inserted into flange B and sealed for vacuum-tightness using clay. The photograph shows a scene of the off-site vacuum testing of the entire chamber assembly installed within the primary coil prior to the EMFC experiment.
 [Photographs were taken or created by the author. Portions of the images have been presented in the author's previous publications and presentations.]}
    \label{emfcsample}
\end{figure}

Solid-state physics experiments in most cases are carried out at cryogenic temperatures \cite{MiyataPRL, MiyataJpsj}. 
The primary CL coil with the liner housed in a Bakelite cylinder is clamped to the electrodes, followed by setting the seed magnetic coils (see Figures~\ref{fig:Princip} and \ref{collectorp}(a)).
A cryostat (type (c) or (f) in Figure~\ref{cryohyst}) is mounted on flange A, as seen in Figure~\ref{emfcsample}(a).
Flange A is then inserted into chamber B through a bore hole of the seed magnetic coil and sealed with clay for vacuum tightness (Figure~\ref{emfcsample}(b)).

The cryostat type (c) is employed for the lowest temperature measurements down to 5~K, whereas type (f) is used for measurements aiming at a higher magnetic field but compromised with a lowest temperature around 10--20~K. 
This is in contrast to the case of the STC system where 5~K is the lowest temperature achieved when type (f) is adopted, as is detailed in Section~\ref{sec:cryo}. 
This difference arises from the difference in the vacuum space around the cryostat. 
In the case of EMFC, the cryostat is directly exposed to thermal radiation from the inner wall of the metal liner, which is at room temperature.
This thermal load cannot be mitigated simply by extending the cooling time. To address this issue, a superinsulation sheet is wrapped around the cryostat, which also serves to shield the sample space from the intense arc light of the imploding liner (see Figures~\ref{faradaypickup} and \ref{fig:cr_emfc}).
The sheet is prepared by bonding the superinsulation onto a slightly larger rectangular Kapton film using a cryogenic adhesive. Specifically, the Kapton base is made 10--20~mm wider than the superinsulation in the circumferential direction, leaving a 3--4~mm gap between the edges of the insulation layer when wrapped. This ensures that the aluminum-evaporated layer does not form a closed loop, thereby preventing explosive eddy-current heating due to the rapid magnetic field changes.

\subsection{Sample Temperature During a Pulsed Magnetic Field}

The temperature of a sample mounted in the cryostat is monitored using a thermocouple placed adjacent to the sample, both in the STC and EMFC experiments. The standard thermocouple employed consists of Au/0.07\%Fe-chromel Teflon-coated wires with a diameter of 80~$\upmu$m. When space permits near the sample, a calibrated RuO$_2$ thin-film resistor is used to monitor temperatures below 10~K, down to the liquid $^4$He pumping range. Above 10~K, copper-constantan thin wires are also utilized.
In most cases, the temperature recorded just before the ignition of the pulsed magnetic field is defined as the measurement temperature, often cited as the initial temperature, $T_i$.
The temperature environment surrounding the sample is considered to remain constant during the extremely short pulse durations---typically 6--7~$\upmu$s for STC and 40--50~$\upmu$s for EMFC. During such short intervals, the sample is effectively held in an adiabatic state due to poor thermal contact; it is generally mounted on an insulating plastic or glass substrate and suspended in vacuum or a $^4$He gas atmosphere for optical measurements, or otherwise immersed in liquid $^4$He.
$T_i$ can only be regarded as the actual sample temperature during the pulse if the sample is an insulator, thereby avoiding eddy-current Joule heating induced by the massive $dB/dt$.

For conducting materials, the temperature rise due to Joule heating must be added to $T_i$. 
The total eddy-current Joule heating $Q(B,t)$ induced by the pulsed magnetic field is given by
\begin{equation}
Q(B,t) = \frac{\sigma(B) S_{\perp}}{8\pi} \left( \frac{dB(t)}{dt} \right)^2 V_s,
\end{equation}
where $\sigma(B)$ is the electrical magnetoconductivity, $V_s$ is the volume of the sample, and $S_{\perp}$ is the effective cross-section perpendicular to the magnetic field $B$. 
The actual sample temperature at time $t = \tau$ is then expressed as:
\begin{equation}
\label{eq:jouleheat}
T_s(B(\tau)) = T_i + \frac{S_{\perp}}{8\pi} \int^{\tau}_{0} \frac{\sigma(B)}{c_p(T)} \left( \frac{dB(t)}{dt} \right)^2 dt,
\end{equation}
where $c_p(T)$ is the specific heat of the sample. 
To minimize Joule heating when measuring conducting materials, the reduction of $S_{\perp}$ is essential.
In cases where the sample thickness is of submicron scale, heat transfer to the substrate must also be accounted for. For instance, in the case of 15-nm V$_{1-x}$W$_x$O$_2$ thin films deposited on TiO$_2$ substrates, the temperature increase derived solely from Equation~(\ref{eq:jouleheat}) leads to an overestimation.
In practice, the substrate acts as an effective heat sink; therefore, in ultra-thin films, the temperature rise during a pulsed magnetic field can be significantly suppressed even for conducting materials.
Specifically, in the case of V$_{1-x}$W$_x$O$_2$ thin films with $T_i$ = 14~K, the temperature rise due to the eddy-current Joule heating is limited to approximately 4~K even at a magnetic field of 500~T (see Supplementary Information of Ref.~\cite{Matsuda2020} for a detailed discussion).

%

\section{Magnetization Measurements}
\label{sec:MagMeasure}

\subsection{Induction Pickup Coil Method}
\label{inductionm}

\begin{figure}[tbp] 
\centering
 \centerline{\includegraphics[width=0.5\columnwidth]{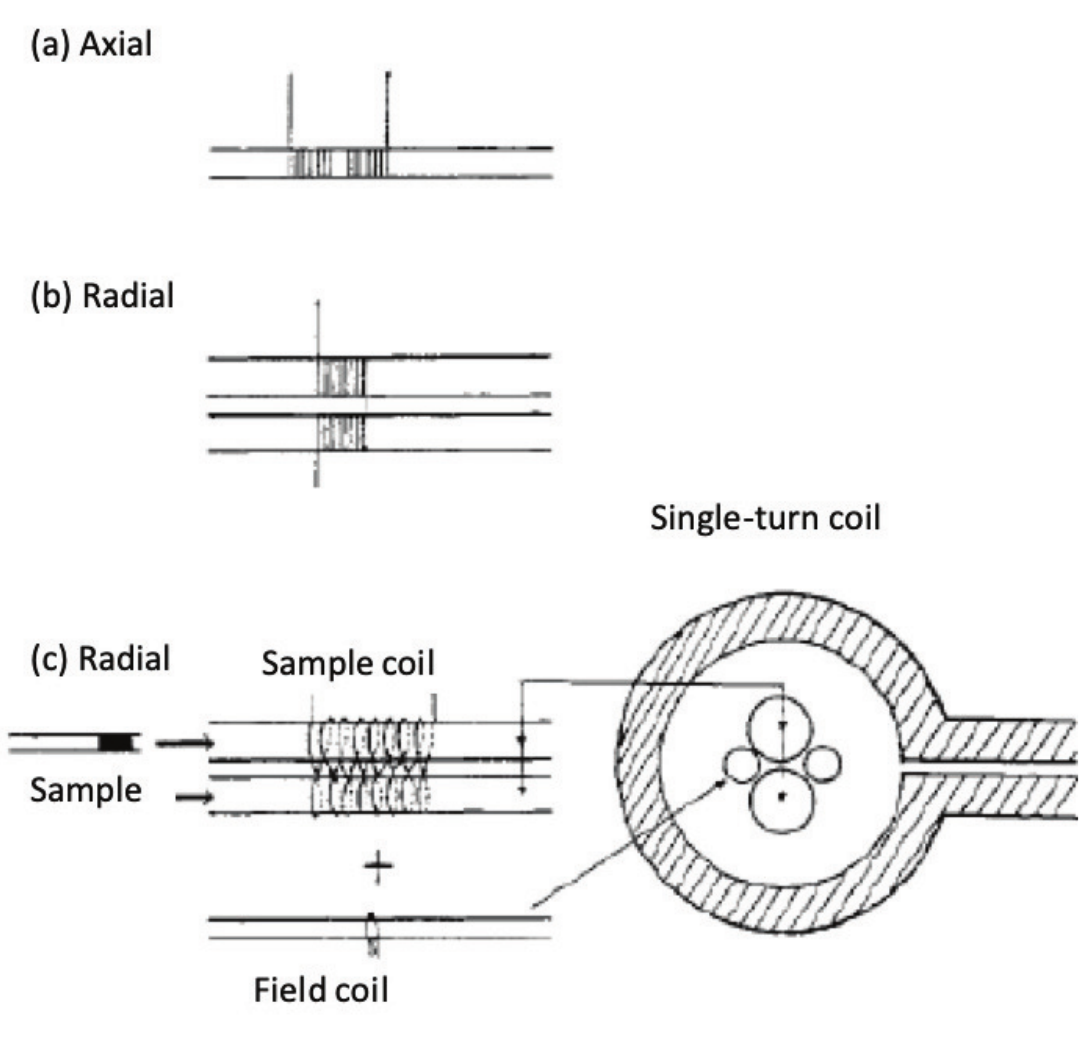} }
   \caption{
Magnetic pickup coil configurations: (a) axial-, (b) radial-, and (c) radial-type with a pair of counter-wound coils. The field pickup coil is wound around a 1-mm-diameter plastic rod and positioned adjacent to the sample magnetization pickup coil. The field coil is placed on the opposite side of the feed-gap of a single-turn coil (on the left-hand side in panel (c)). 
[Reproduced from Ref. \cite{TakeyamaAmaya1988} with permission from IOP Publishing  (1988).]}
    \label{Amayapickupcoil}
\end{figure}

Magnetization measurements were first attempted in ultrastrong magnetic fields produced by an STC system \cite{TakeyamaAmaya1988, AmayaTakeyama1989}. These measurements involve several difficulties arising from the destructive nature of the coil: extremely high induced voltages due to the large $dB/dt$ values, poor magnetic field homogeneity, and limited working space. The spatial homogeneity of the magnetic field in an exploding coil was investigated, revealing that the field homogeneity along the radial direction is significantly better than that along the axial direction.
The voltage induced across the pickup coil reaches approximately 860~V/turn for a 100-kJ capacitor discharge into an 18-mm-diameter single-turn coil. A 20-turn pickup coil spans a length of 2.5~mm, and the field homogeneity exceeds 99\% when it is positioned at the center of the single-turn coil, as illustrated in Figure~\ref{Amayapickupcoil}. The pickup coils are wound with copper wire of 100~$\upmu$m overall diameter, which includes a 10-$\upmu$m-thick layer of insulating ``Formvar.''
Several types of magnetic pickup coils were tested, as shown in Figure~\ref{Amayapickupcoil}. It was found that the axial type (a) is overly sensitive to misalignment along the axial direction. The radial type (b) is more susceptible to insulation breakdown caused by the induced high voltage. In contrast, type (c) is wound with a pair of counter windings to cancel the induced voltage at each turn. Stable performance and the most reliable magnetization ($M$) data were obtained using the type (c) pickup coil.

A pair of coils is connected in series with opposite polarity, which in principle cancels the voltage induced by an external pulsed magnetic field. In the first shot, a sample is inserted into coil No. 1 while No. 2 is left empty, yielding the signal $V_{\text{first}}$. In the second shot, the sample is moved to coil No. 2, yielding $V_{\text{second}}$. These signals are expressed as:
\begin{align}
V_{\text{first}}  &= (A_1 - A_2) \frac{dB}{dt} + A_1 \frac{dM}{dt}, \\
V_{\text{second}} &= (A_1 - A_2) \frac{dB}{dt} - A_2 \frac{dM}{dt},
\end{align}
where $A_1$ and $A_2$ are the effective areas of pickup coils No. 1 and No. 2, respectively. 

\begin{figure}[tbp] 
\centering
 \centerline{\includegraphics[width=0.7\columnwidth]{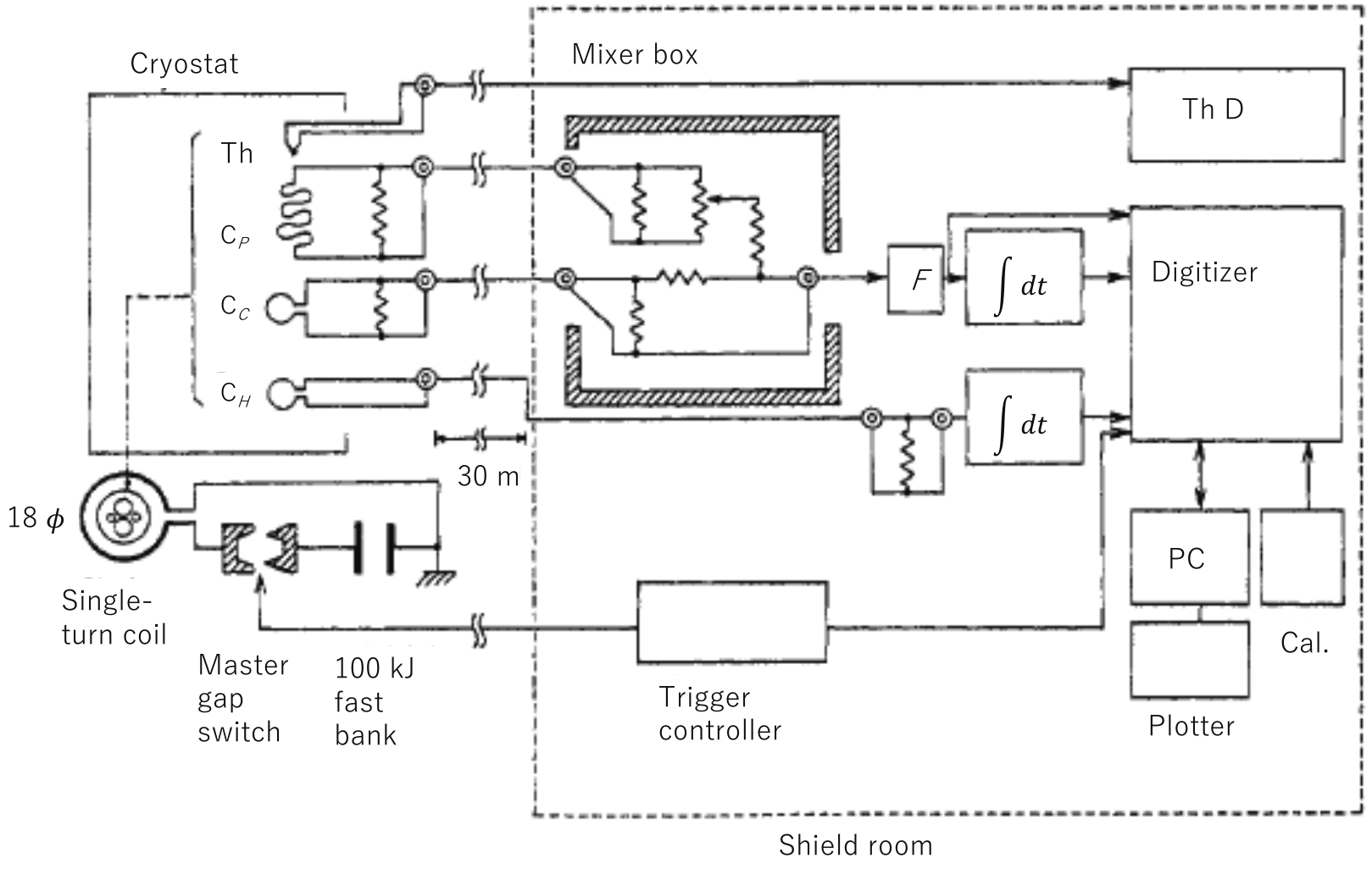} }
   \caption{
Block diagram of the measurement circuit system. Th: thermocouple; C$_{\text{p}}$
: pickup coil; C$_{\text{e}}$
: compensation coil; C$_{\text{H}}$
: field pickup coil. 
 [Reproduced from Ref. \cite{TakeyamaAmaya1988} with permission from IOP Publishing  (1988).]}
    \label{compecircuit}
\end{figure}

Ideally, $A_1$ should be identical to $A_2$ to cancel the $dB/dt$ term. However, in practice, a residual background term $(A_1 - A_2) dB/dt$ always remains. Due to the exceptionally high $dB/dt$ inherent in STC experiments, this background is typically orders of magnitude larger than the magnetization signal itself. To mitigate this, the hardware-level compensation rate $r = (A_1 - A_2) / A_1$ is first reduced to the order of $10^{-2}$ through meticulous manual adjustment, and further improved to $10^{-4}$ using an auxiliary compensation coil C$_{\text{e}}$ and a mixer box (Figure~\ref{compecircuit}).

Despite such precise hardware balancing, a dual-shot subtraction is essential to extract the finalized $M$ signal. By calculating the difference between the two shots:
\begin{equation}
V_{\text{sub}} = V_{\text{first}} - V_{\text{second}} = (A_1 + A_2) \frac{dM}{dt},
\label{eq_subtraction}
\end{equation}
the background term $(A_1 - A_2) dB/dt$ is precisely canceled, provided that the magnetic field environment remains perfectly reproducible between the shots. At the time, this combination of hardware-level balancing and dual-shot numerical subtraction was considered the most reliable method for extracting small magnetization signals from the overwhelming $dB/dt$ background.

\begin{figure}[htb] 
\centering
 \centerline{\includegraphics[width=0.5\columnwidth]{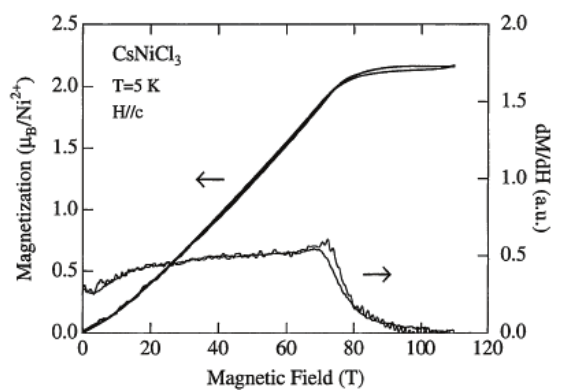} }
   \caption{
 Time derivative of the magnetization $dM/dt$
(signal directly from the pickup coil) of the triangular-lattice antiferromagnet CsNiCl$_3$
and its time-integrated magnetization $M=\int (dM/dt)dt$
plotted against magnetic fields. The magnetic fields were applied up to 110~T. The measurement was performed at a temperature of 5~K. 
[Reproduced with permission from Ref. \cite{Katori1995} \copyright (1995) The Physical Society of Japan. ]}  
    \label{fig:CsNiCl_3}
\end{figure}

The fast sweep rate of the magnetic flux produced by the STC technique is advantageous for the sensitive detection of magnetic phase transitions. This technical innovation served as a milestone in the advancement of high-field magnetism. Magnetization measurements were performed in magnetic fields up to 80--120~T generated by the STC system.
The magnetization processes of Co-based intermetallic compounds (such as YCo$_2$, YCo$_3$, LuCo$_2$, and Y(Co$_{1-x}$Al$_x$)) unveiled metamagnetic phase transitions, including full saturation of the magnetic moment in ultrastrong magnetic fields \cite{Goto1989, Goto1992, Goto1994, Goto2001}. Figure~\ref{fig:CsNiCl_3} presents one of the successful magnetization datasets obtained under magnetic fields exceeding 110~T. 
The magnetization process up to full saturation was investigated in the quasi-one-dimensional triangular lattice antiferromagnet CsNiCl$_3$, which was well interpreted using the Heisenberg antiferromagnetic chain $S=1$ interaction \cite{Katori1995}.

Measurements were further extended to magnetic fields of up to 120~T applied to ferrimagnetic R$_2$Fe$_{14}$B, where field-induced transitions from ferrimagnetic to ferromagnetic phases were observed \cite{Kato1995}. Magnetization in ultrastrong magnetic fields is a highly efficient tool for evidencing the suppression of spin fluctuations underlying competing antiferromagnetic exchange interactions; for example, in random mixtures of two antiferromagnets such as Fe$_x$Mn$_{1-x}$TiO$_3$ \cite{Katori1992}.

\begin{figure}[tbp] 
\centering
 \centerline{\includegraphics[width=0.6\columnwidth]{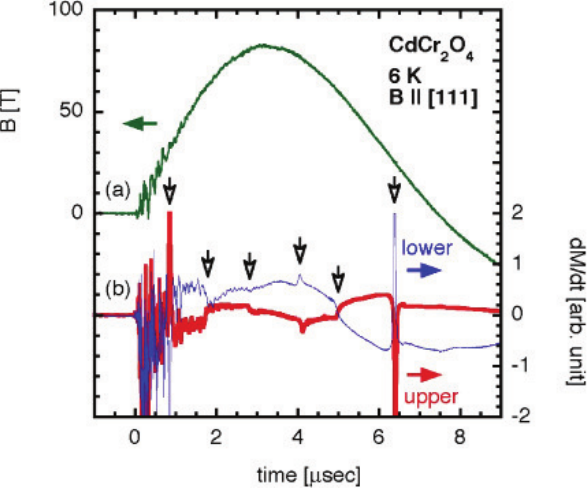} }
   \caption{
(a) Time profile of pulsed magnetic fields up to 80~T produced by the STC system; (b) Induced voltage signal from a magnetization pickup coil. The labels ``upper'' and ``lower'' indicate the sample positions as explained in the text. Downward arrows indicate the positions of the magnetic phase transitions observed in the CdCr$_2$O$_4$ sample measured at 6~K. Significant discharge noise with ringing oscillations persists almost until the field peak. 
[Reproduced from Ref.~\cite{Mitamura2007} (2007) with permission from The Physical Society of Japan. ]}  
    \label{fig:mitamuraCdCrO}
\end{figure}

Despite these advancements, a common challenge persists across these studies.
Due to substantial starting noise, primarily arising from the air-gap switches of the STC system, the magnetization $M = \int (dM/dt) dt$ is, in most cases, only reliably obtained from the raw data during the descending part of the magnetic field pulse. This disturbance caused by the discharge ringing noise is clearly visible in Figure~\ref{fig:mitamuraCdCrO} \cite{Mitamura2007}, where the gap-switch starting noise induces a sizable disturbance in the magnetization pickup signals during the rising phase of the magnetic field pulse.

\begin{figure}[htbp] 
\centering
 \centerline{\includegraphics[width=0.6\columnwidth]{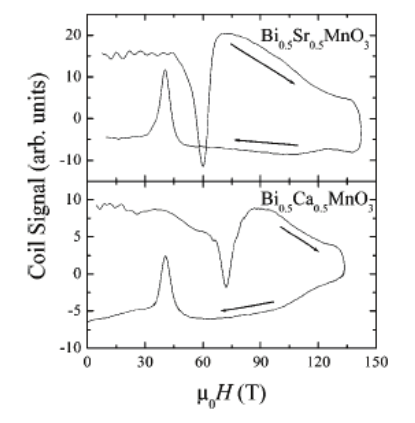} }
   \caption{
 Signals from the magnetization pickup coil during the rising and descending phases of a pulsed magnetic field applied to Bi$_{1/2}$Sr$_{1/2}$MnO$_{3}$ (upper panel) and Bi$_{1/2}$Ca$_{1/2}$MnO$_{3}$ (lower panel) perovskites. The magnetic field was applied up to 130--140~T. The starting electromagnetic noise is significantly lower than that shown in Figure~\ref{fig:mitamuraCdCrO}. 
[ Reproduced from Ref.~\cite{Kirste2003} (2003) with permission from APS.]}
      \label{BiSrMnO}
\end{figure}

The quality of the high-field magnetization data was improved by one order of magnitude compared to that obtained in experiments performed so far with the system shown in Figure~\ref{fig:HSTC} (at ISSP, U-Tokyo), by utilizing the vertical-type STC system (see Figure~\ref{fig:vstcPortugall} in Section~\ref{STC}) installed at Humboldt University, Germany \cite{Portugall1999}.
Field-induced magnetic phase transitions were investigated in various magnetic materials, including Van Vleck paramagnetic rare-earth zircons such as PrVO$_4$ and TmPO$_4$ \cite{Kazei2001, Kirste2001}, and bismuth-based manganite perovskites, Bi$_{1/2}$(Sr,Ca)$_{1/2}$MnO$_{3}$ (the pickup coil signal, $dM/dt$, is shown in Figure~\ref{BiSrMnO}) \cite{Kirste2003}. 
The STC megagauss generator, which incorporates strip-line collector plates and rail-gap switches as mentioned in Section~\ref{STC}, is advantageous for magnetization measurements due to its reduced starting discharge ringing noise. As a result, both the rising and descending phases of the magnetic field pulse provide valid and meaningful $dM/dt$ signals (see Figure~\ref{BiSrMnO}).

All of the methods developed above were restricted to magnetic materials that exhibit abrupt changes in magnetization, such as metamagnetic transitions, which can be detected as peaks in the time derivative of magnetization, $dM/dt$. The signal-to-noise ratio and the level of background signals remain insufficient to provide detailed absolute values of magnetization, $M$, in the ultrastrong magnetic field region; they are only useful for determining magnetic phase diagrams through distinct transitions.

There are three major hurdles associated with measurements in the exploding environment of an STC system. The first is the electromagnetic noise generated by the high-voltage and high-current gap-switch discharge. The second is the poor spatial homogeneity of the magnetic field within the small single-turn coil. The third is the deformation of the coil during the pulse duration due to its explosive nature, which induces time-dependent field inhomogeneity. This third point is particularly distressing for attaining the high compensation ratios that are indispensable for accurate and precise magnetization measurements using pickup coils.
\begin{figure}[htbp] 
\centering
 \centerline{\includegraphics[width=0.6\columnwidth]{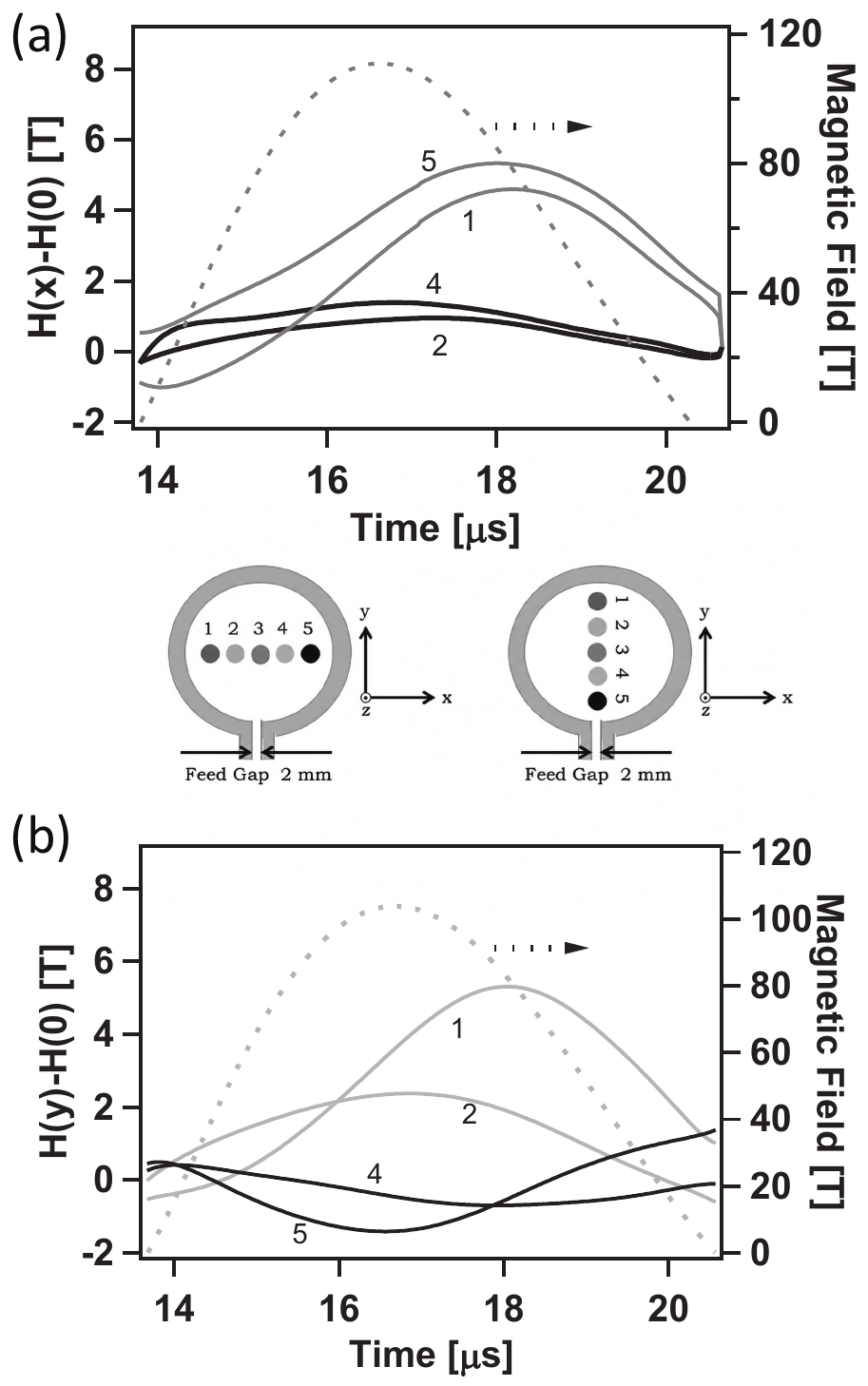} }
   \caption{
Time evolution of the magnetic field deviation from the value at the coil center for (a) the $x$-direction and (b) the $y$-direction, measured at positions 1--5 as shown in the inset. The inset illustrates the positions of the pickup coils inside a 14-mm-diameter single-turn coil. 
The pulsed magnetic field (absolute value, not the deviation) at the center of the coil (position No. 3) is indicated by a dashed line. (a) The curves for No. 1 and No. 5, as well as those for No. 2 and No. 4, can be regarded as almost identical within the errors of pickup coil calibration and possible minor deviations from exact symmetrical positions. (b) The time-dependent deviation curves are complex due to the presence of the feed-gap in the $y$-direction; the magnetic field intensity tends to decrease as it approaches the feed-gap.
[Panels (a) and (b) are adapted from Ref. \cite{TakeyamaSakakura2012} with permission from The Physical Society of Japan. The central schematic is an original addition by the author.]}
    \label{dynamichomo}
\end{figure}
Figure~\ref{dynamichomo} shows the time evolution of the magnetic field distribution within a single-turn coil during a magnetic field pulse up to 100~T. The magnetic fields were measured by placing calibrated pickup coils at five positions, both perpendicular ($x$) and parallel ($y$) to the direction of the current feed-gap, in a 14-mm-inner-diameter single-turn coil \cite{TakeyamaSakakura2012}. Figure~\ref{dynamichomo} exhibits the complex behavior of this time-and-space-dependent magnetic field distribution. While auxiliary compensation coils have been conventionally employed for magnetization measurements in non-destructive pulsed magnetic field experiments, field compensation in STC experiments is significantly more challenging. This difficulty is further compounded when using a resistance compensation network based on the block diagram illustrated in Figure~\ref{compecircuit}.

\begin{figure}[htb] 
\centering
\includegraphics[width=0.45\columnwidth]{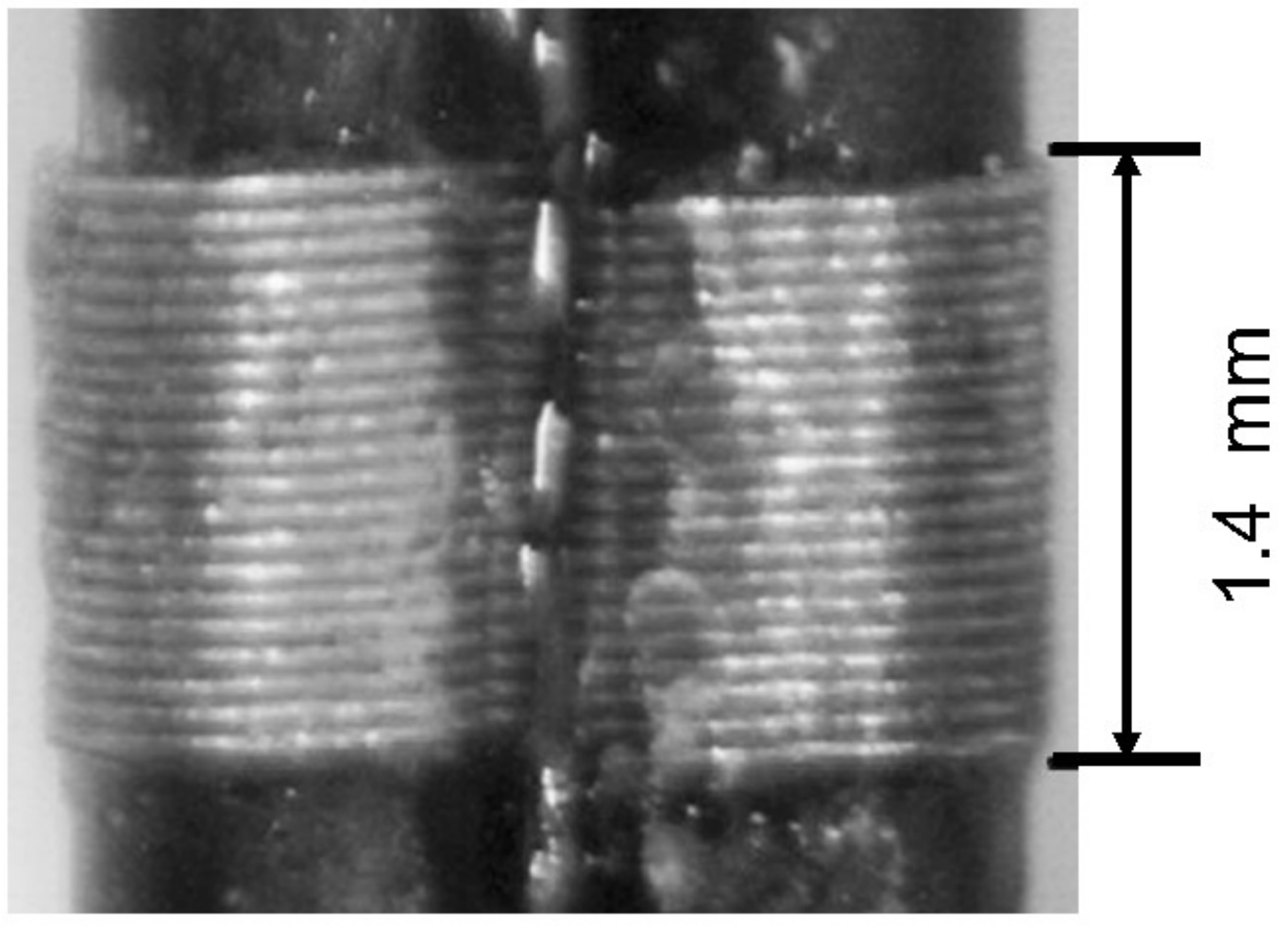}
\caption{Photograph of the self-compensated magnetization pickup coil. The~radial-type pickup coils, as~illustrated in Figure~\ref{Amayapickupcoil}b, feature counter-windings of 20 turns each. A~thin copper wire (60~$\upmu$m diameter) is wound around a Kapton tube with an outer diameter of 1.12~mm. The~pair of pickup coils is glued and solidified using ``Stycast 1266.'' The total coil length is 1.4~mm. [This photograph is selected and adapted from Figure~1 in Ref.~\cite{TakeyamaSakakura2012}. Reproduced with permission from the Physical Society of Japan].}
\label{fig:comp_coil}
\end{figure}

High-precision compensation was achieved solely through a pair of identical, counter-wound pickup coils, intentionally omitting an auxiliary compensation coil. The exclusion of the auxiliary coil is critical, as its presence within the magnet introduces an additional source of noise and electromagnetic interference, thereby degrading the signal-to-noise ratio. Furthermore, as revealed by the data in Figure~\ref{dynamichomo}, the auxiliary coil is susceptible to the dynamic, time-dependent inhomogeneity of the pulsed magnetic field, which often deteriorates the final compensation quality rather than improving it. 
To achieve an exceptional degree of intrinsic compensation, the coils were meticulously wound under a microscope \cite{TakeyamaSakakura2012}. These parallel-type pickup coils (structured similarly to type (b) in Figure~\ref{Amayapickupcoil}) consist of 20--23 turns around a Kapton tube with an outer diameter of 1.12~mm and a wall thickness of 0.06~mm. A photograph of the coil fabricated in this manner is shown in Figure~\ref{fig:comp_coil}. In the field of pulsed magnets, it has been a long-standing convention for decades to rely on auxiliary compensation coils for Ms. However, the decision to eliminate these auxiliary components represents a paradigm shift.

With precise positioning within the magnet, the induced voltage---initially 2~kV for a 20-turn coil---was reduced to less than 1~V. This remarkable performance yielded a compensation ratio (defined in Equation~\eqref{eq_subtraction}) of less than $5 \times 10^{-4}$ without the need for any external electronic or auxiliary compensation, proving that high-precision mechanical winding can supersede traditional electronic compensation methods. 
Under zero-field conditions, this intrinsic compensation ratio is comparable to that achieved with the external auxiliary coils shown in Figure~\ref{compecircuit}. However, during actual field generation, the spatial inhomogeneity of the magnetic field evolves dynamically over time. In systems utilizing external auxiliary coils, this time-dependent inhomogeneity causes the effective compensation ratio to deteriorate by more than an order of magnitude, typically exceeding $10^{-3}$, whereas the current method maintains its high precision.

The selection of the winding wire material is equally crucial. The wire must be sufficiently flexible for precise winding while possessing appropriate electrical resistance and a high-performance insulation coating. Generally, high-resistance wires tend to be brittle, whereas low-resistance wires are susceptible to burnout caused by eddy current heating, especially given the shallow skin depth at high frequencies. It was experimentally confirmed that parallel, counter-wound coils with up to 30 turns can withstand magnetic fields up to 100~T by employing a polyamide-imide-coated copper wire with a diameter of 60~$\upmu$m (AIW, SWCC Showa Holdings Co., Ltd., Tokyo, Japan). This wire guarantees electrical insulation up to 6~kV for DC voltages, although this threshold decreases for high-frequency AC signals in the MHz range.

An innovative technique for ultrastrong-field magnetization measurements has been established through the precise and manual assembly of  a ``parallel self-compensated induction pickup coil'' (hereafter referred to as ``S-C pickup coil''), effectively eliminating the requirement for an auxiliary compensation coil~\cite{TakeyamaSakakura2012}. This technique was initially deployed to capture metamagnetic phase transitions in the bismuth-based manganite Bi$_{1/2}$Ca$_{1/2}$MnO$_{3}$ perovskite, as~well as novel magnetic phase transitions in the geometrically frustrated chromium spinel CdCr$_2$O$_4$. A~key advantage is that the induced $dM/dt$ signal remains virtually background-free during both the rising and descending phases of the megagauss pulsed field (for details, see Figures~7 and 8 in  Ref.~\cite{TakeyamaSakakura2012}). 
Consequently, the~resulting magnetization curves $M(B)$ exhibit a high-fidelity quality that is directly comparable to data obtained using conventional, non-destructive long-pulse magnets with millisecond-range durations (e.g., Figure~2 in Ref.~\cite{HUeda2006}).

The S-C pickup coil technique has driven significant breakthroughs in high-field magnetism research, particularly in probing systems where extreme conditions are essential to lifting quantum degeneracy. A~prominent class of such materials includes ``geometrically frustrated magnets,'' where the spatial arrangement of magnetic spins---such as two-dimensional triangular/Kagome networks or three-dimensional pyrochlore lattices---leads to a macroscopically degenerate quantum spin ground state. In~these systems, long-range magnetic ordering is strongly suppressed, making them highly sensitive to subtle perturbations like spin--lattice coupling, as~well as quantum or thermal fluctuations. Unveiling the enigmatic magnetic phases emerging from this frustration strictly demands a combination of multiple extreme environments, namely ultra-high magnetic fields and cryogenic temperatures.

As a prototypical three-dimensional frustrated magnet belonging to the chromium spinel family $A$Cr$_2$O$_4$ ($A$=Hg, Cd, Zn, Mg), a~cubic chromium spinel, CdCr$_2$O$_4$ features a pyrochlore lattice of corner-sharing tetrahedra where Cr$^{3+}$ spins form a macroscopically degenerate ground state. For~such complex systems, measuring the magnetization all the way up to the full saturation moment is indispensable to understand the true essence of geometrical spin frustration. Historically, early magnetization measurements using non-destructive long-pulsed fields up to 40~T had revealed a metamagnetic transition into a broad 1/2 magnetization plateau~\cite{HUeda2005}.
%

\begin{figure}[htb]
\centering
\includegraphics[width=0.6\columnwidth]{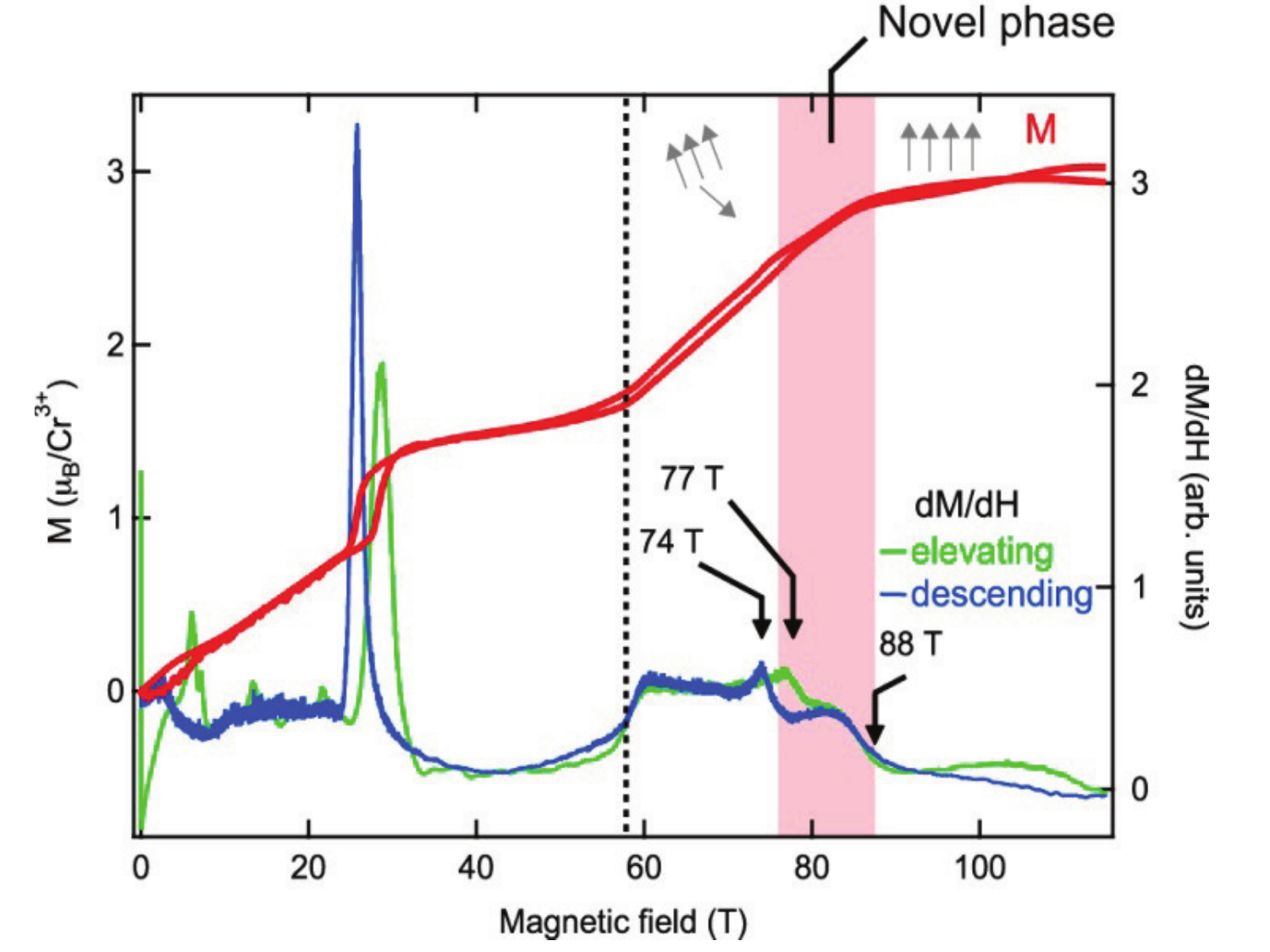}
\caption{Time derivative of magnetization ($dM/dt$) and magnetization ($M$) of CdCr$_2$O$_4$ measured in the VSTC system under magnetic fields up to 120~T. The~measurement temperature of 4.2~K was maintained using the cryostat shown in Figure~\ref{vstcCryo}. The~$dM/dt$ signal detected by the pickup coil (Figure~\ref{fig:comp_coil}) is almost background-free in both the rising and descending phases of the pulsed field. Details regarding the physics of the discovered magnetic phases are described in a later section. [Reproduced from Ref.~\cite{Miyata2013} (2013) with permission from APS].}
\label{fig:cdcr2o4M100T}
\end{figure}

%
%
%
However, fully exploring the underlying physics required much higher magnetic fields to overcome the robust exchange interactions. By~implementing the S-C pickup coil technique within the STC megagauss generator, the~complete magnetic phase diagram of CdCr$_2$O$_4$---including the elusive spin-nematic and the full span of the half-plateau phases---was unambiguously mapped out in fields up to 120~T, as~shown in Figure~\ref{fig:cdcr2o4M100T}~\cite{Miyata2013}.
Another milestone achieved with this technique is the investigation of strontium copper borate, SrCu$_2$(BO$_3$)$_2$. This compound is celebrated as a prototypical two-dimensional orthogonal dimer system that realizes the theoretical ``Shastry--Sutherland model'' \cite{SS_model}. Its complex spin structures and field-induced plateau states had long been a subject of intense debate from both experimental and theoretical perspectives. Magnetization measurements performed up to 118~T using the S-C pickup coil definitively revealed, for~the first time, the~long-predicted 1/2 magnetization plateau phase spanning from 84~T to 108~T~\cite{Matsuda2013}. This landmark discovery, complemented by large-scale numerical simulations, provided a consistent and comprehensive understanding of the Shastry--Sutherland quantum antiferromagnet (details are provided in the Supplemental Material of Ref.~\cite{Matsuda2013}).

The utility of this measurement framework extends further to molecular crystals, where strong magnetic energies can directly compete with structural cohesive forces. This is beautifully demonstrated in the investigation of solid oxygen under ultra-high magnetic fields~\cite{Nomura2014}. Magnetization measurements performed across a broad temperature range unambiguously revealed a distinct field-induced phase transition accompanied by a giant, striking hysteresis between 70~T and 180~T at temperatures below 44~K. This phenomenon signifies a structural metamorphosis from the conventional low-temperature $\beta$ or $\alpha$ phase into a novel cubic state, termed the $\theta$ phase. This dramatic field-induced molecular rearrangement is driven by the cooperative reorientation of the O$_2$--O$_2$ dimer system, highlighting the intricate interplay between magnetic exchange, lattice condensation, and~van der Waals interactions in solid oxygen~\cite{Nomura2014, Nomura2015}.

In addition, the~S-C pickup coil technique was subsequently transferred and implemented in the STC system at LNCMI in Toulouse. This has enabled high-precision magnetization measurements at the facility, with~notable results including the following. Local Ising anisotropy is a key characteristic of the spin-ice state in the well-known classical spin-ice pyrochlore Ho$_2$Ti$_2$O$_7$. Magnetization measurements in magnetic fields up to 130~T were performed to probe this robust local Ising anisotropy. Although~magnetic fields of 130~T were still insufficient to reach full saturation, the~data quality was high enough to evaluate the details of the strong crystal electric field created by the surrounding oxygen ions~\cite{Opherden2019}.

\subsection{Faraday Rotation Technique for Magnetization Measurements}

The optical Faraday rotation method can be applied not only to detect magnetic field intensity ($B$), as described in Sec.~\ref{sec:FaradayR}, but also to measure the magnetization of magnetic materials. Since Faraday rotation measurements are based on optical detection, they are less susceptible to electromagnetic noise compared to the induction-based magnetization pickup coil technique. The linear polarization of an incident light beam rotates by an angle $\theta$ after transmission through a material lacking time-reversal symmetry. When magnetization $M(B)$ is induced in a material by an external magnetic field $B$, the time-reversal symmetry is broken, resulting in a finite Faraday rotation angle
in the transmitted light. The induced Faraday rotation angle $\theta(B)$ is related to $M(B)$
as follows:

\begin{equation}
\theta(B) =\theta_0(B)+A_1M(B)+A_3M(B)^3+A_5M(B)^5+ \dots\dots \\
\label{eqFaradayangle}
\end{equation}
Here, $\theta_0(B)$ is a background contribution irrelevant to $M(B)$ or a contribution from a substrate.
$A_1, A_3, A_5, \dots$ are coefficients of each order of $M(B)$.
The higher-order terms in Eq.~(\ref{eqFaradayangle}) can only be neglected with an appropriate choice of a wavelength of the incident light on the material, and the relation can be simplified to
\begin{equation}
\theta_M(B) \simeq AM(B).
\label{FRpropM}
\end{equation}
The Faraday rotation angle can be considered proportional to the magnetization $M(B)$ after the background contribution $\theta_0(B)$ is subtracted, as will be demonstrated in the following section.

To investigate the magnetic phases of the geometrically frustrated chromium spinel oxide CdCr$_2$O$_4$, the Faraday rotation measurement was pioneered by implementing Eq.~(\ref{FRpropM}) (in place of Eq.~(\ref{eqFaradayangle})) for the first time \cite{Kojima2008}. 
This approach enabled the full magnetization process to be revealed with high accuracy in magnetic fields up to 140~T generated by the single-turn coil system. 
Although this magneto-optical approach appears after the description of the S-C pickup coil technique in Section~\ref{inductionm} for structural consistency, this optical measurement historically preceded the application of the pickup coil method. 
Notably, the~high-precision Faraday rotation profile obtained up to 140~T served as an indispensable baseline to cross-check and validate the legitimacy and accuracy of the subsequent magnetization curves measured via the S-C pickup coil method, as~comprehensively discussed in Ref.~\cite{TakeyamaSakakura2012}.
For this specific compound, tracking the magnetization profile all the way up to its full saturation moment is of paramount importance to elucidate the true essence of its geometrical spin frustration. In~the following, we discuss the detailed features of the magnetization process in CdCr$_2$O$_4$ revealed by these high-precision Faraday rotation~measurements.

To maintain the proportionality expressed in Eq.~(\ref{FRpropM}), the laser wavelength for Faraday rotation experiments must be carefully selected by tuning the incident light. Faraday rotation originates from the off-diagonal terms of the dielectric tensor, which are closely related to the optical conductivity and absorption characteristics of the material. High-field Faraday rotation data for the antiferromagnet EuTe were previously examined and compared with conventional magnetization data in the context of intra-atomic $d$--$d$ absorption spectra \cite{Hori1994}. That study reported a sizable discrepancy between the two measurements depending on whether the incident light was resonant or non-resonant with the intra-atomic $d$--$d$ optical transitions. It was demonstrated that while the Faraday rotation angle measured with resonant light shows excellent agreement with the magnetization curve, it deviates significantly when non-resonant light is used. This underscores that precise selection of the incident wavelength is crucial for ensuring that the Faraday rotation accurately reflects the magnetization.
%
\begin{figure}[tbp]
\centering
\includegraphics[width=0.5\columnwidth]{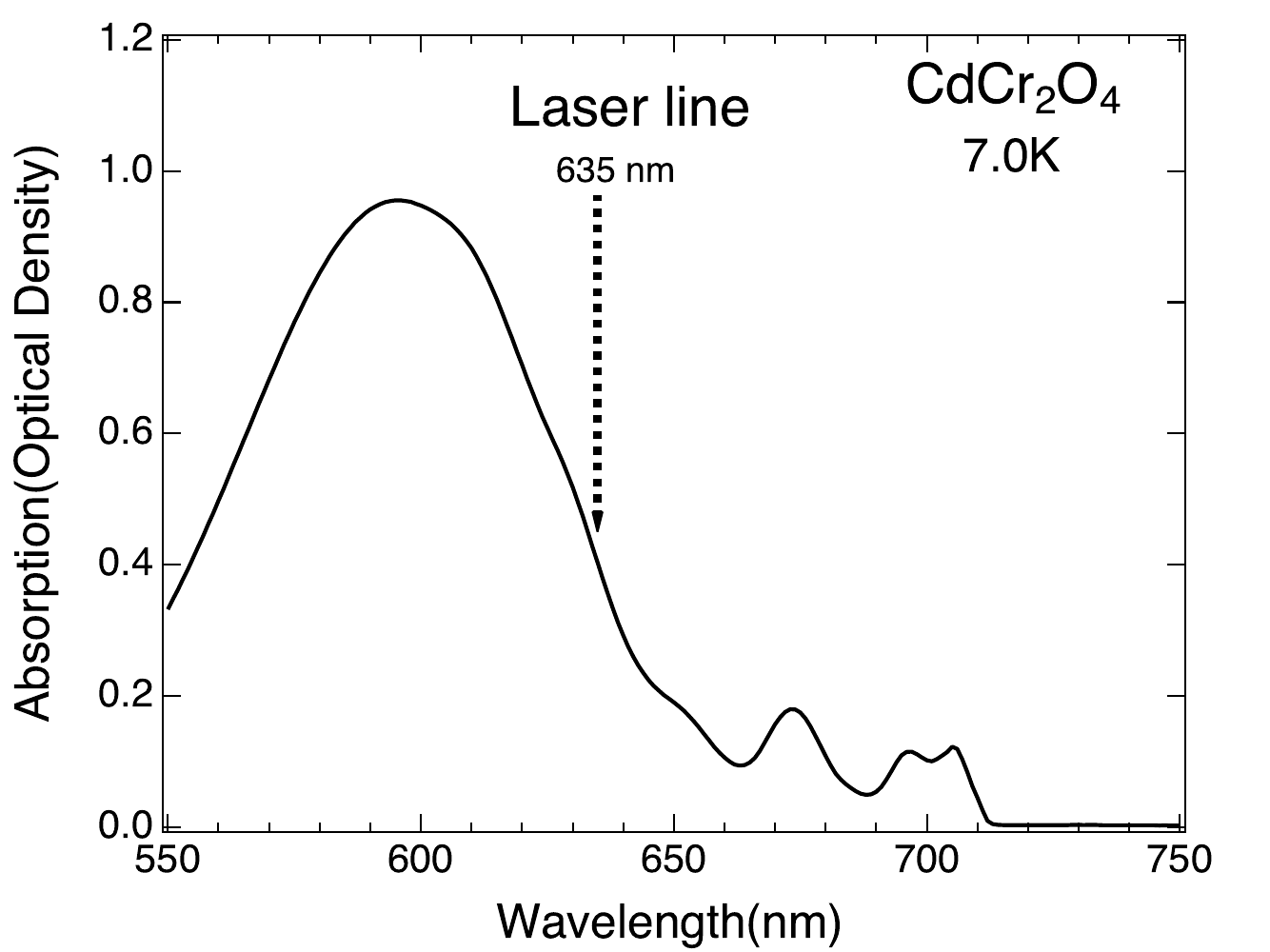}
\caption{
Absorption spectra of $d$--$d$ intra-atomic optical transitions of Cr$^{3+}$ ions in a frustrated spinel lattice (CdCr$_2$O$_4$), measured using an optical spectrometer at 7~K. The broad absorption peak around 600~nm arises from the optical transitions ${^4}\mathrm{A}_2 \to {^4}\mathrm{T}_1$ and ${^4}\mathrm{A}_2 \to {^4}\mathrm{T}_2$. The double absorption peaks centered at 700~nm are attributed to exciton--magnon--phonon transitions. The dotted arrow indicates the wavelength of the laser used for the Faraday rotation magnetization measurements.}
\label{fig:cdcr2o4spectra}
\end{figure}

The absorption spectrum of CdCr$_2$O$_4$ is shown in Figure~\ref{fig:cdcr2o4spectra}, exhibiting Cr$^{3+}$ $d$--$d$ intra-atomic absorption bands. The~prominent peak observed around 600~nm is attributed to the $d$--$d$ intra-atomic $t_{2g} \to e_g$ optical transition, which reflects the spin configurations influenced by the crystal electric field of the spinel lattice~\cite{Szymczak1980}. In~Faraday rotation measurements, the~light passing through a sample must have sufficient transmission intensity to maintain a high signal-to-noise (S/N) ratio. Consequently, the~sample thickness is typically adjusted to achieve an optical density (OD) (The optical density is defined as $\text{OD} = \alpha d$, where $\alpha$ is the absorption coefficient and $d$ is the sample thickness. This value determines the contrast of the spectral features.) of approximately 1.0 (e.g., a~thickness of approximately 40~$\upmu$m for CdCr$_2$O$_4$).

Furthermore, the laser wavelength is ideally tuned to the tail of the main absorption peak, specifically where the OD is approximately 0.5. This selection is crucial for optimizing the S/N ratio of the transmitted light. If the OD is too high, the transmitted intensity becomes insufficient for reliable detection. Conversely, if the OD is too low, the light does not interact sufficiently with the $d$--$d$ transitions, resulting in a diminished Faraday rotation signal. Tuning the wavelength to a region with an OD of 0.5 provides an ideal balance, ensuring both adequate transmission and a significant magneto-optical response. As indicated by the dotted arrow in Figure~\ref{fig:cdcr2o4spectra}, the 635~nm emission from a semiconductor laser is positioned ideally for this purpose.

\begin{figure}[tbp]
\centering
\includegraphics[width=0.75\columnwidth]{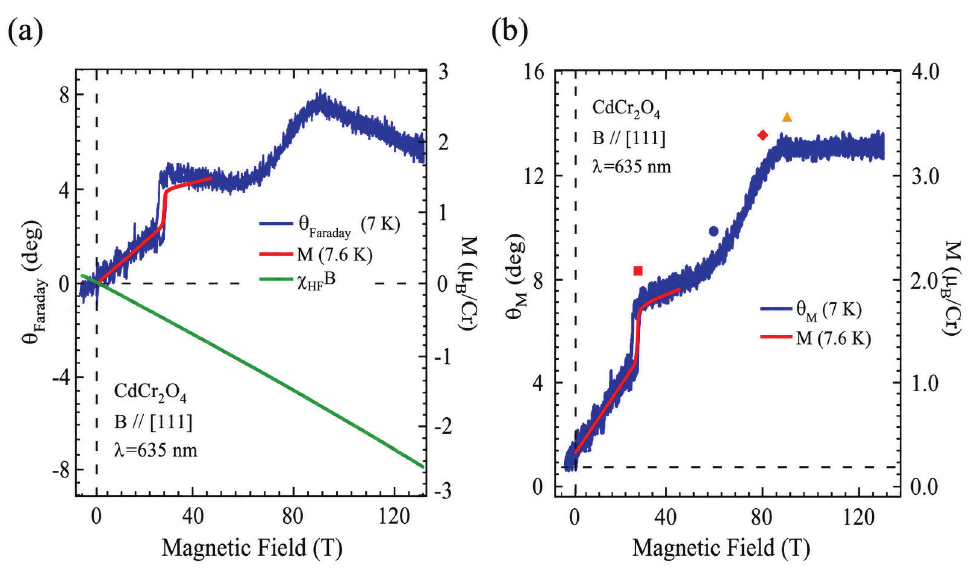}
\caption{\label{fig:cdcr2o4FR_M}
(a) Raw data of the Faraday rotation angle as a function of magnetic field at 7~K (blue line). Magnetization measured using the induction coil method in a non-destructive magnet up to 50~T at 7.6~K is shown by the red line, which almost overlaps with the former signal. The background contribution ($\chi_{HF} B$) is indicated by the straight green line. (b) The Faraday rotation angle $\theta_M(B)$ obtained after subtracting (effectively adding) the background term $\chi_{HF} B$ shown in (a). 
[Panel (b) is modified from Ref.~\cite{Kojima2008} for better clarity, with permission from APS.]}
\end{figure}

The experimental setup is identical to that used for the magnetic field measurements shown in Figure~\ref{faradaypickup}. 
The time evolution of the Faraday rotation angle, $\theta(t)$, is plotted against the measured magnetic field, $B(t)$, and the resulting raw data $\theta(B)$ is displayed in Figure~\ref{fig:cdcr2o4FR_M}(a). 
The signal contains a background component $\theta_0(B)$, which includes contributions from the quartz substrate (approximately 0.2~mm thick), the adhesive between the sample and the substrate, and the diamagnetic response of the sample material. 

All these background components are expected to exhibit a linear dependence on the magnetic field, represented by the solid gray line denoted as $\chi_{HF}B$ in Figure~\ref{fig:cdcr2o4FR_M}. After subtracting $\theta_0(B) = -\chi_{HF}B$, the Faraday rotation angle satisfies the relation in Eq.~(\ref{FRpropM}), yielding $\theta_M(B)$, which is proportional to $M(B)$ as shown in Figure~\ref{fig:cdcr2o4FR_M}(b). 
$\theta_M(B)$ is calibrated using the magnetization data from the non-destructive pulsed magnet (gray dashed line) and converted to $M(B)$ (right ordinate) in Figure~\ref{fig:cdcr2o4FR_M}. 
The results show excellent agreement with the data in Figure~\ref{fig:cdcr2o4M100T} obtained by the S-C pickup coil method. 
The obtained $\theta_M(B)$ fits well with the magnetization data independently measured using a non-destructive pulsed magnet at approximately the same temperature \cite{HUeda2005}.
%
\begin{figure}[htbp]
\centering
\includegraphics[width=0.45\columnwidth]{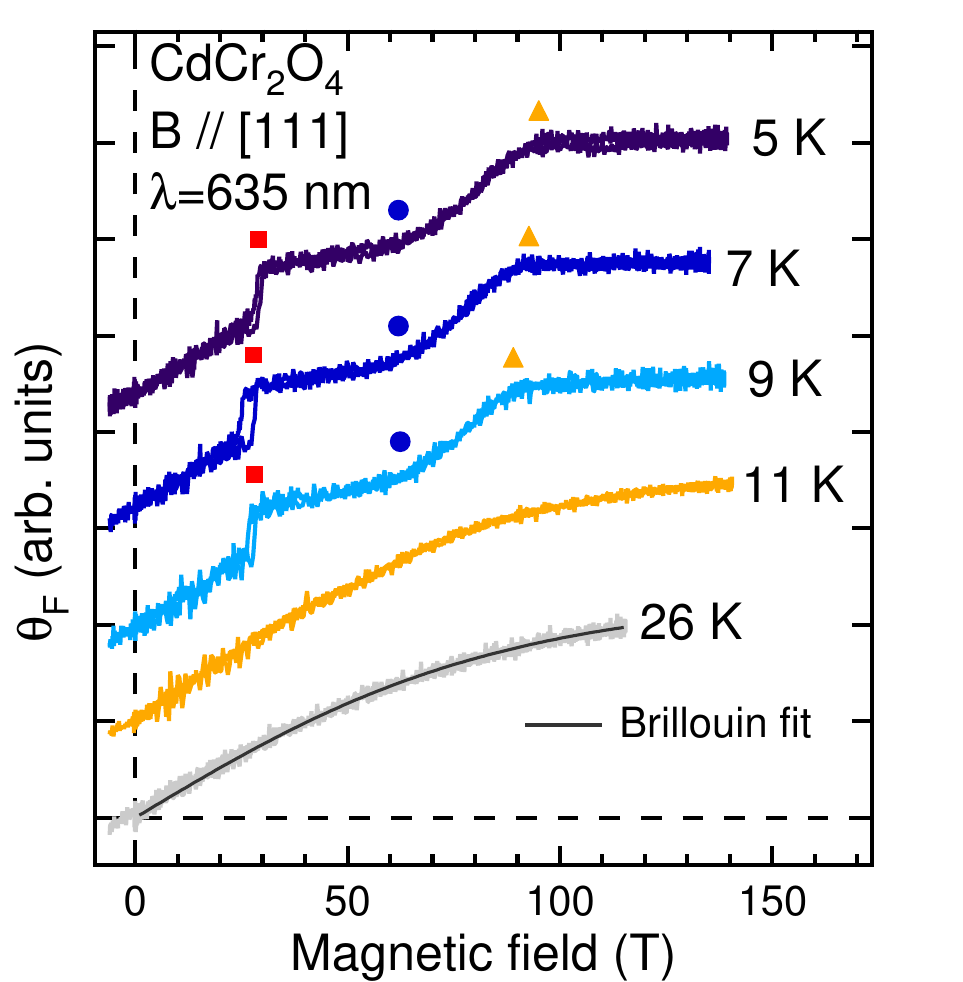}
\caption{\label{cdcr2o4FR_temp}
Magnetization curves $M(B)$ (derived from the Faraday rotation angle $\theta_M$) measured at various temperatures. The 1/2 magnetization plateau and the full-saturation state are clearly observed at temperatures below 10~K. At 26~K, the magnetic phase enters a paramagnetic state, where the magnetization curve is well described by the Brillouin function. The magnetic phase diagram of the frustrated spinel oxide CdCr$_2$O$_4$ is constructed based on these data. 
[Reproduced from Ref.~\cite{Kojima2008} (2008) with permission from APS.]}
\end{figure}

In this manner, Faraday rotation measurements in ultrastrong magnetic fields were conducted at various temperatures, as displayed in Figure~\ref{cdcr2o4FR_temp}, to map out the temperature--magnetic field phase diagram. This approach facilitated a quantitative comparison with existing theories \cite{Penc2004, Motome2006}.

\begin{figure}[htbp]
\centering
\includegraphics[width=0.6\columnwidth]{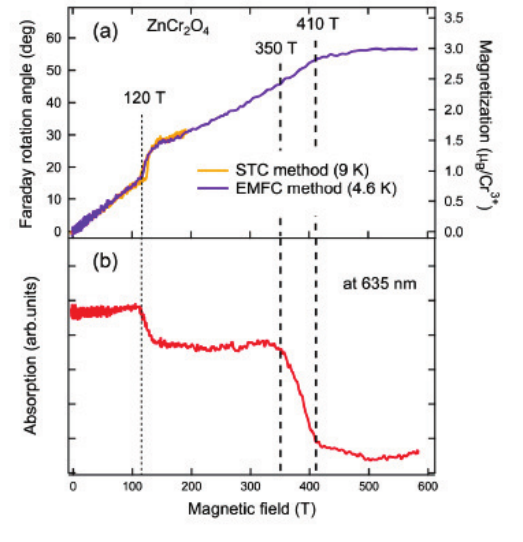}
\caption{\label{zncroMandAbs}
(a) Magnetization obtained by the Faraday rotation method in magnetic fields of up to 600~T generated by the EMFC method, measured at a temperature of 4.6~K. The Faraday rotation angle obtained from STC experiments in magnetic fields up to 190~T at 9~K is also plotted for comparison \cite{MiyataJpsj2011}. (b) Optical absorption intensity at a fixed wavelength of 635~nm in magnetic fields, measured at 4.6~K. The broken and dashed lines at 120, 350, and 410~T distinguish pronounced transitions observed in the absorption intensity. 
[Reproduced from Ref.~\cite{MiyataPRL} (2011) with permission from APS.]}
\end{figure}

Faraday rotation measurements were extended to magnetic fields of approximately 600~T generated by the EMFC method. ZnCr$_2$O$_4$ belongs to the same family of $A$Cr$_2$O$_4$ spinels as CdCr$_2$O$_4$ discussed above, possessing similar magnetic properties associated with a highly frustrated pyrochlore lattice. Due to its smaller spin--lattice coupling and stronger antiferromagnetic interactions between neighboring Cr$^{3+}$ ions compared to those in CdCr$_2$O$_4$, substantially higher magnetic fields are required to achieve magnetic saturation. 
To achieve the lowest possible temperatures, a liquid $^4$He flow-type cryostat, as shown in Figure~\ref{cryohyst}(c), was employed and positioned with high precision within the copper-lined coil of the EMFC system. The Faraday rotation angle $\theta_{M}(B)$, which is proportional to the magnetization $M(B)$, was measured in magnetic fields up to 600~T with high resolution and an excellent signal-to-noise ratio, as shown in Figure~\ref{zncroMandAbs}(a). These data represent the first precise solid-state physics measurements ever conducted under the combined extremes of a 600~T magnetic field and a low temperature of 4.6~K. Successive magnetic phase transitions, including a clear 1/2 magnetization plateau, were unambiguously observed, culminating in the fully polarized phase (forced ferromagnetic state) at magnetic fields between 410~T and 600~T.
\section{Streak Magneto-Optical Absorption Spectroscopy}
\label{sec:StreakSpec}
\begin{figure}[htbp]
\centering
\includegraphics[width=0.7\columnwidth]{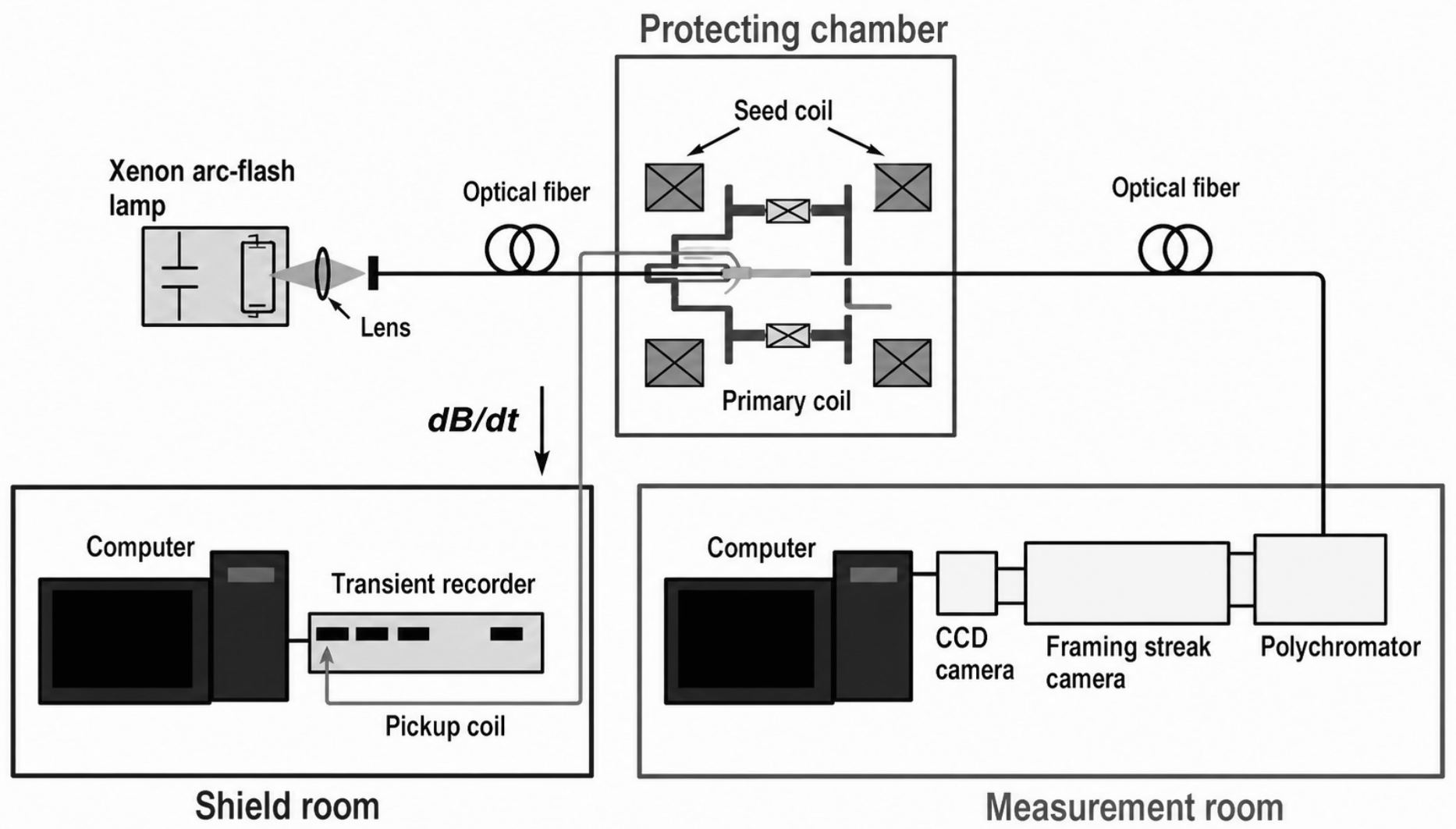}
\caption{Experimental setup for streak spectroscopy employed in magneto-optical absorption measurements under EMFC experiment. The double shield box comprises an aluminum inner box and an iron outer box to house a digital transient recorder and a computer, providing protection against intense electromagnetic interference. This box is positioned adjacent to the explosion protection chamber. In contrast, the data acquisition shield room, which accommodates various precision instruments, is a larger shielded facility located several meters away from the explosion chamber. 
Incident light from a Xenon short-arc flash lamp is guided to the sample via a long, large-core optical fiber (600--1000~$\upmu$m in diameter). The light transmitted through the sample is directed into a polychromator coupled with a streak camera and a CCD detector. 
The streak scanning is precisely synchronized with the duration of the pulsed magnetic field.
\label{streakspecEMFC}
[Reproduced from Ref.~\cite{MiyataJpsj} (2012) with permission from The Physical Society of Japan.]}
\end{figure}
The experimental setup for streak magneto-absorption spectroscopy, employed for experiments using the EMFC or STC methods, is illustrated in Figure~\ref{streakspecEMFC}. Streak spectroscopy using an image converter camera is the most effective technique for capturing magneto-optical spectra in ultrastrong magnetic fields with one-pulse and very short durations. 
High-speed sweeping of the optical image is essential to synchronize with the short-pulsed magnetic fields, which occur on a microsecond timescale. Detailed descriptions of streak spectroscopy applied to measurements under ultrastrong magnetic fields are available in the literature \cite{HMF-ST2003, Miurastreak1988}.

\subsection{\textit{d--d} Intra-atomic and Exciton--Magnon--Phonon Magneto-absorption Spectroscopy}

Magneto-optical absorption streak spectroscopy was performed in magnetic fields up to 600~T. The absorption spectral peaks of the intra-atomic $d$--$d$ transitions in Cr$^{3+}$ ions, as well as the exciton--magnon--phonon transitions, were monitored to track changes in the crystal structure and spin configurations under extreme magnetic fields. Phase transitions associated with lattice distortions and spin states manifested as pronounced changes in absorption intensity, as displayed in Figure~\ref{zncroMandAbs}(b). Focusing on the exciton--magnon--phonon spectral peak---typically observed on the lower-energy (longer-wavelength) side of the main $d$--$d$ transition peak (e.g., the peak around 700~nm for CdCr$_2$O$_4$ in Figure~\ref{fig:cdcr2o4spectra})---provides a highly effective method for tracking magnetic phase transitions, as exemplified below.

\begin{figure}[htbp]
\centering
\includegraphics[width=0.4\columnwidth]{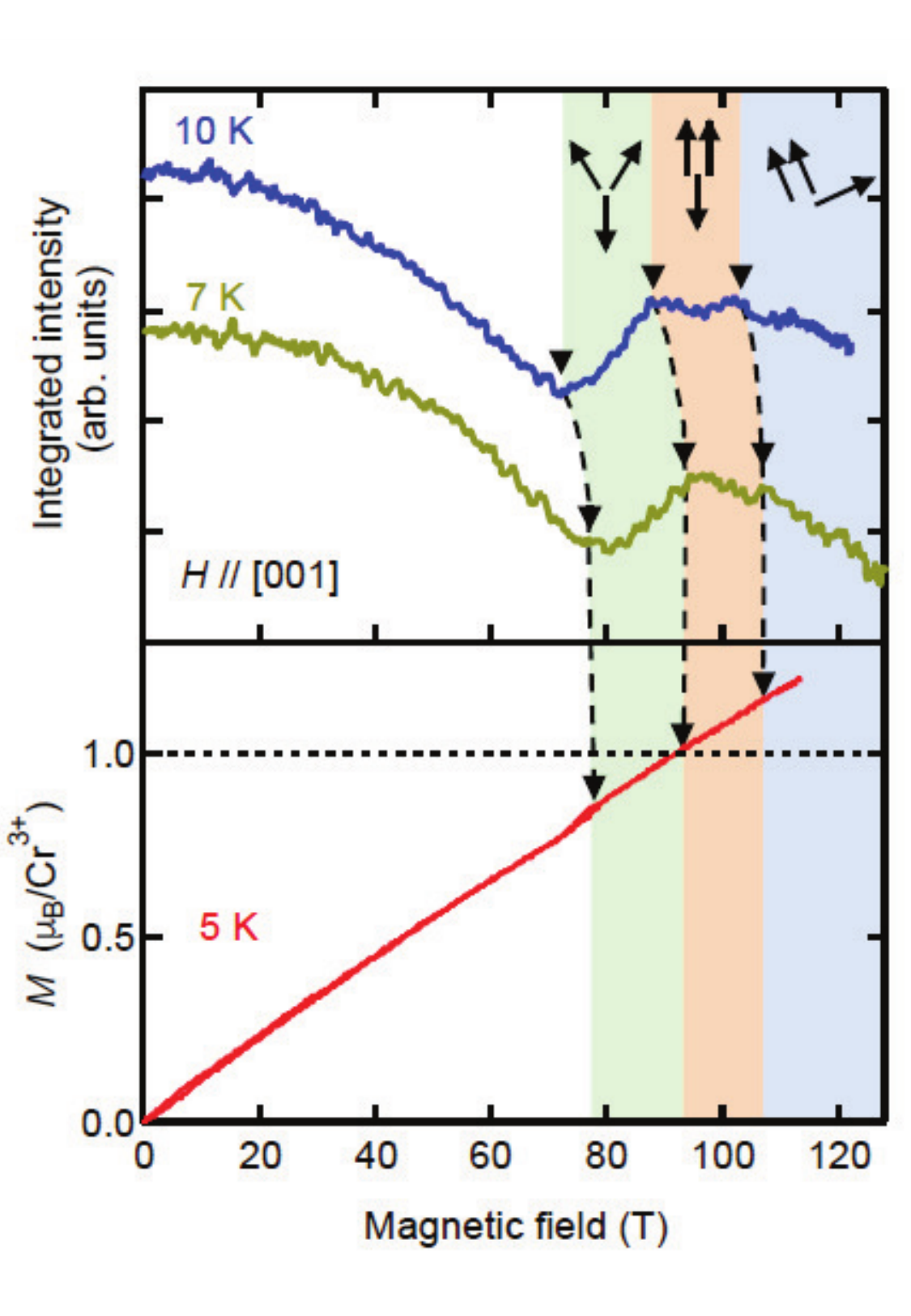}
\caption{\label{fig:cucro2}
(Upper panel) Integrated intensity of the magneto-absorption spectral peak attributed to the exciton--magnon--phonon optical transition in CuCrO$_2$ measured at temperatures of 7 and 10~K. (Lower panel) Magnetization curve at 5~K deduced from Faraday rotation angles ($H \parallel [001]$; magnetic field applied parallel to the $[001]$ crystal axis). Dashed arrows are a guide to the eyes for magnetic phase boundaries. The spin structures of each magnetic phase are illustrated by solid arrows. 
[Reproduced from Ref.~\cite{Miyata2017} (2017) with permission from APS.]}
\end{figure}

The triangular-lattice antiferromagnet CuCrO$_2$ was investigated using both Faraday rotation and magneto-absorption spectroscopy to reveal the diverse magnetic phases expected in ultrastrong magnetic fields \cite{Miyata2017}. CuCrO$_2$ is categorized as a multiferroic delafossite oxide, in which ferroelectricity emerges from unique spin configurations stemming from geometrical spin frustration. The exciton--magnon--phonon optical transition, observed at wavelengths between 660~nm and 689~nm, reflects magnon creation and annihilation processes in accordance with the underlying spin structures. Accordingly, the integrated magneto-absorption intensity of the exciton--magnon--phonon spectral peak is highly sensitive to changes in the spin configuration. 
Figure~\ref{fig:cucro2} compares the magneto-absorption intensity of the exciton--magnon--phonon transition (upper panel) with the magnetization derived from Faraday rotation (lower panel) for CuCrO$_2$. While the latter exhibits only a tiny deviation associated with the phase transition at 75~T, the changes at 90~T and 105~T are much more pronounced and obvious in the magneto-absorption intensity. This demonstrates that magneto-absorption can effectively distinguish phase boundaries to which the magnetization is relatively insensitive.

\subsection{Magneto-optical Absorption of Carbon Nanotubes}
Magnetic flux quanta threading through a material break the time-reversal symmetry of electronic states and lift the degeneracies of charge, spin, and orbital degrees of freedom that determine the functional properties of the material. In nanoscale systems, these effects are only fully elucidated through the application of ultrastrong magnetic fields. Since their discovery in the early 1990s \cite{Iijima}, single-walled carbon nanotubes (SWCNTs) have been the subject of intensive research from both fundamental and applied perspectives, owing to the unique one-dimensional tubular structures of their carbon networks. 

A graphene sheet rolled into a nanometer-sized cylinder (with a diameter typically between 1~nm and several nm) forms an SWCNT with lengths on the order of micrometers. A single magnetic flux quantum ($\phi_0 = ch/e$, where $e$ is the elementary charge and $h$ is Planck's constant) threading through a 1~nm diameter nanotube corresponds to a magnetic field of 5200~T. This flux induces one full period of the bandgap oscillation via the Aharonov--Bohm (AB) effect. This phenomenon in SWCNTs was first predicted theoretically by Ajiki and Ando and has since been widely referred to as the Ajiki--Ando splitting \cite{AjikiAndo}.

It is generally recognized that the optical properties of SWCNTs are dominated by band-edge excitons with extremely large binding energies induced by their one-dimensional nature \cite{Ando1997}. Owing to the degeneracy of the $K$ and $K'$ valleys in the Brillouin zone, combined with electron spin degrees of freedom, the band-edge excitons in a single-walled carbon nanotube comprise 16 unique excitonic states. Among these, only the spin-singlet zero-momentum excitons are optically allowed (bright excitons), while the remaining 15 states are optically forbidden (dark excitons) \cite{Ando1997, Ando2004}. 

A magnetic field penetrating the cross-section of a nanotube lifts the valley degeneracy at the $K$ and $K'$ points and induces mixing of the excitonic states. Consequently, the bright exciton and its dark counterpart are known to convert into two bright excitons at the $K$ and $K'$ points upon the application of an ultrastrong magnetic field. According to the $k \cdot p$ theory by Ando \cite{Ando2006}, the absorption spectral peaks of these $K$ and $K'$ excitons evolve differently with increasing magnetic field, depending on their respective energy levels; this leads to the distinct splitting of the excitonic absorption peak into two components.

In an attempt to unveil the effects of the Aharonov--Bohm flux, absorption and photoluminescence (PL) spectroscopy have been performed by several groups on various types of SWCNTs in magnetic fields up to 45--78~T \cite{Zaric2004, Zaric2006, Mortimer2007, Shaver2008, Takeyama2011}. These nanotubes are characterized by their ``chirality.''\footnote{Single-walled carbon nanotubes are specified by the chiral vector, which defines the direction and magnitude along which a monolayer graphene sheet is rolled into a cylindrical structure.} 
Photoluminescence involves relaxation processes from the excited state to the final emissive state and is highly susceptible to extrinsic factors such as impurities and localized states. Furthermore, photoluminescence (PL) spectra are significantly influenced by thermal distributions and the choice of excitation photon energies. Consequently, magneto-PL spectroscopy may not always provide a direct probe of intrinsic and coherent excited states, such as excitons, in their pristine form.

\begin{figure}[htbp]
\centering
\includegraphics[width=0.4\columnwidth]{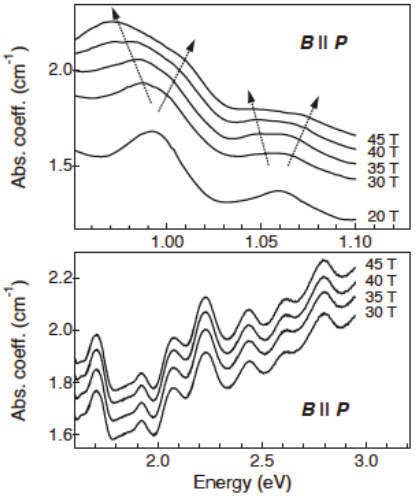}
\caption{\label{CNT45T}
First observation of the splitting of excitonic absorption spectral peaks under steady magnetic fields up to 45~T, generated by the hybrid magnet at the National High Magnetic Field Laboratory in Tallahassee. The upper and lower panels display the first-subband ($E_{11}$) and second-subband ($E_{22}$) transitions, respectively. The notation $B \parallel P$ in the figure indicates that the linear polarization of the incident light ($P$) is parallel to both the nanotube axis and the magnetic field ($B$). With increasing magnetic fields, the $E_{11}$ spectral peak splitting evolves into peak broadening. 
[Reproduced from Ref.~\cite{Zaric2004} with permission from AAAS.]}
\end{figure}
The first observation of exciton splitting in single-walled carbon nanotubes (SWCNTs) suspended in sodium dodecyl sulfate (SDS)/D$_2$O was achieved via magneto-absorption measurements in magnetic fields up to 45~T, as presented in Figure~\ref{CNT45T} \cite{Zaric2004}. In such specimens, the nanotubes are randomly oriented and comprise a mixture of various chiralities. This randomness significantly reduces the effective magnetic flux component parallel to the tube axis ($B_\parallel$). 
For nanotubes suspended in liquid media (such as sodium cholate or aqueous surfactant solutions), magnetic-field-induced reorientation \cite{Takeyama2004Orient, Shaver2009Orient} introduces additional ambiguity into the spectral analysis. Furthermore, the coexistence of multiple nanotube species obscures individual peak structures due to the overlapping of their respective absorption features. Consequently, the spectral splitting often manifests as simple peak broadening as the magnetic field increases, making it difficult to determine exact peak positions. 
Additionally, strong background signals from the host material deteriorate the spectral quality by reducing the signal-to-background ratio (i.e., the optical contrast of the exciton peaks), further complicating the resolution of individual features (see Figure~\ref{CNT45T}). Similar limitations were encountered in pulsed magnetic field measurements up to 78~T \cite{Takeyama2011}. Overall, the data obtained to date in fields below 78~T remain insufficient for a rigorous quantitative understanding of the Aharonov--Bohm effect in SWCNTs.

\begin{figure}[htbp]
\centering
\includegraphics[width=0.35\columnwidth]{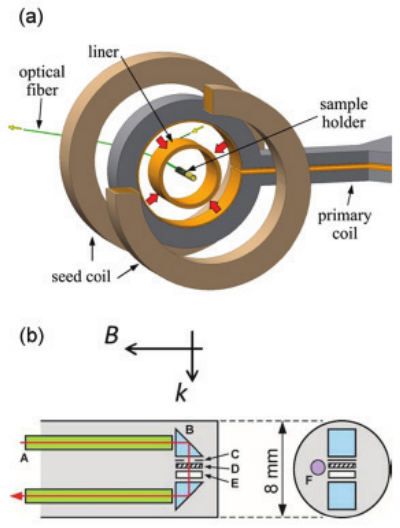}
\caption{\label{CNTandEMFC}
(a) Sample holder with an optical fiber positioned at the center of the primary coil, and the imploding liner compressing the seed magnetic flux (approximately 3~T) generated by the Helmholtz-type seed-field coils. (b) Schematic view of the sample holder and optics in the Voigt geometry. A: optical fibers; B: prisms; C: optical mask; D: linear polarizer; E: sample specimen; F: magnetic field pickup coil. 
[Reproduced from Ref.~\cite{Nakamura2015} (2015) with permission from APS.]}
\end{figure}

\begin{figure}[tbp]
\centering
\includegraphics[width=0.75\columnwidth]{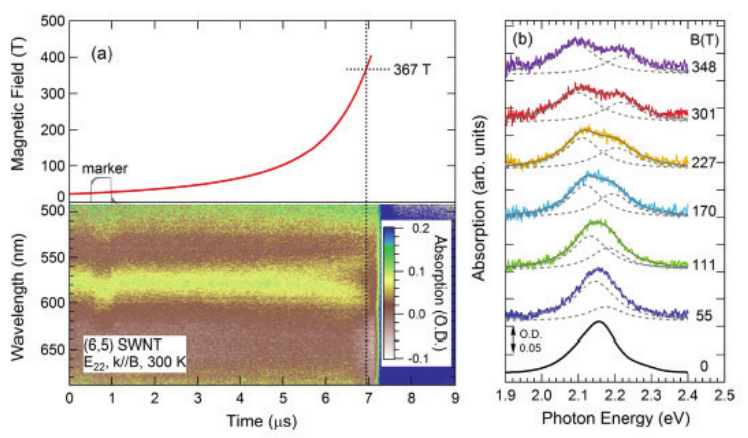}
\caption{
Spectral evolution of the band-edge exciton absorption peak in magnetic fields up to 367~T ($B_\parallel = 321$~T) generated by the EMFC method. (a) Upper panel: Time profile of the magnetic field. Lower panel: Streak image of the absorption spectra as a function of time, precisely synchronized with the magnetic field waveform. Time markers, captured simultaneously by the streak camera and the digital recorder, were employed to ensure rigorous temporal alignment. (b) Absorption spectra of the band-edge exciton at various magnetic field strengths. The peak splitting becomes clearly discernible above 200~T. Notably, an excellent signal-to-noise ratio is maintained despite the low optical density (OD $\approx$ 0.1). The dashed curves represent the results of Lorentzian deconvolution, identifying the two singlet band-edge excitons at the $K$ and $K'$ valleys. [Reproduced from Ref.~\cite{Nakamura2015} (2015) with permission from APS.]}
\label{CNTStreak}
\end{figure}

For definitive magneto-absorption measurements, the following factors are absolutely vital: the selection of a single chirality (type) of nanotubes, the use of a host material with minimal background absorption, and the synthesis of highly aligned nanotubes to maximize the effective parallel magnetic field component ($B_\parallel$). Highly isolated carbon nanotubes of a specific chirality have recently become available through advanced nanotechnology, such as single-surfactant multicolumn gel chromatography \cite{LiuKataura2011}. 
To maximize $B_\parallel$ relative to the applied field $B$, the single-walled carbon nanotubes were dispersed in a sodium deoxycholate solution and subsequently mixed with an aqueous solution of polyvinyl alcohol (PVA) to form a thin film. The polyvinyl alcohol (PVA) film containing the nanotubes was stretched to five times its original length at a constant speed using a stepping-motor-driven stretching apparatus at temperatures between 80 and 100~$^\circ$C. Consequently, the average angle of the nanotubes with respect to the magnetic field direction was reduced to $\theta_{\text{ave}} = 29^\circ$, achieving a high effective field of $B_\parallel = B \cos 29^\circ = 0.87 \times B$.
The nanotube-embedded film was mounted in a sample holder equipped with optical fibers in the Voigt geometry, as illustrated in Figure~\ref{CNTandEMFC}, and subjected to magneto-absorption measurements using the EMFC method with magnetic fields reaching $B \approx 400$~T \cite{Nakamura2015}. The results are summarized in Figure~\ref{CNTStreak}. The adoption of the Voigt geometry and the precise alignment of the linear polarizer relative to the nanotube axis are crucial for selectively probing the $E_{11}$ transitions; further technical details and the physical rationale are provided in Ref.~\cite{Nakamura2015}.

A magneto-absorption streak image of (6,5) single-chirality nanotubes was captured in magnetic fields up to 367~T. This field strength corresponds to a parallel flux component of $B_\parallel = 321$~T, which translates to a magnetic flux of $\phi = 0.035 \phi_0$ for (6,5) nanotubes (with a diameter of $d = 0.757$~nm). The spectral evolution shown in Figure~\ref{CNTStreak}(b) provides a rigorous verification of the theoretical prediction by Ando~\cite{Ando2006}: a new peak emerges on the higher-energy side of the primary peak and eventually develops into two clearly resolved peaks at the highest magnetic fields. This behavior occurs when the bright exciton level initially lies below the dark level; subsequently, both levels split and evolve into two bright exciton states under the influence of the magnetic field~\cite{Ando2006}.

It was discovered in this study that the conventional two-level model, previously used to interpret magneto-optical spectra, remains valid only up to approximately 150~T. Above this threshold, the Ajiki--Ando linear splitting term must be explicitly incorporated. Under the limited resolution of measurements below 150~T, the observed spectral splitting was primarily attributed to quantum mixing that facilitates the brightening of dark excitons. However, the Aharonov--Bohm effect in SWCNTs (the Ajiki--Ando splitting) was unambiguously detected for the first time here, made possible by the well-resolved peak evolution observed in ultrastrong magnetic fields exceeding 200~T, as demonstrated in Figure~\ref{CNTStreak}(b).

\section{Infrared and Near-Infrared  Cyclotron Resonance Laser Spectroscopy in Ultrastrong Magnetic Fields}
\label{sec:InfraredCR}

A free electron subjected to an external magnetic field undergoes cyclotron motion, where the cyclotron energy can be tuned to resonance with incident light in the millimeter or sub-millimeter wavelength range. 
Cyclotron resonance (CR) has been extensively used to determine the band structures of bulk and low-dimensional semiconductors. Cyclotron resonance measurements require stronger magnetic fields and higher-energy incident radiation (such as far-infrared, infrared, or even near-infrared light) when applied to materials with large effective masses or low carrier mobilities. To date, cyclotron resonance measurements in ultrastrong magnetic fields have been successfully applied to semiconductors such as silicon carbide (SiC), aluminum arsenide (AlAs) \cite{Takeyama1993, KonoPhysica1993, KonoPRB1993}, and diamond \cite{KonoPRB_D1993}, all of which are promising materials for electronic devices operating at high temperatures.

\subsection{Diluted Ferromagnetic Semiconductors}
In most cases, magnetic semiconductors exhibit low carrier mobilities; consequently, a very high magnetic field is required to observe well-resolved CR, satisfying the condition $\upmu B \gg 1$. This necessitates the condition $\omega_{c}\tau \gg1$, where $\omega_{c}$ is the cyclotron frequency and $\tau$ is the relaxation time. CR enables the direct determination of free-carrier effective masses and scattering times, providing detailed information on carrier itinerancy---especially regarding its role in carrier-induced ferromagnetism.
Diluted ferromagnetic semiconductors, such as InMnAs and InMnSb, are known for their carrier-induced ferromagnetism. In recent years, the Curie temperature ($T_C$) of these materials has been increased above room temperature, thanks to advanced thin-film growth technologies such as molecular beam epitaxy (MBE) and metal-organic vapor phase epitaxy (MOVPE). 
\begin{figure}[htbp]
\centering
\includegraphics[width=0.45\columnwidth]{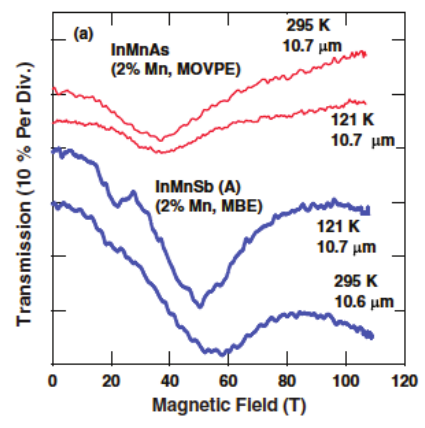}
\caption{
Cyclotron resonance spectra for InMnAs and InMnSb films measured in ultrastrong magnetic fields. The cyclotron resonance of InMnSb at 295~K was measured at a wavelength of 10.6~$\upmu$m, while the other three resonances were measured at 10.7~$\upmu$m using a CO$_2$ gas laser. [Reproduced from Ref.~\cite{Khodaparast2013} (2013) with permission from APS.]}
\label{fig:inmnas}
\end{figure}

Ultrastrong magnetic field CR measurements were conducted on $p$-type InMnAs and InMnSb films with high carrier concentrations and low mobilities to precisely determine their hole-band parameters. Such information is crucial for understanding the mechanisms underlying high-$T_C$ ferromagnetism~\cite{Khodaparast2013}. Magnetic fields exceeding 100~T were generated using the horizontal-type single-turn coil (STC) system described in Figures~\ref{fig:HSTC} and \ref{fig:panoramaHSTC} (Section~\ref{STC}).
A CO$_2$ gas laser with wavelengths of 10.6~$\upmu$m and 10.7~$\upmu$m was employed as the incident light source. A mechanical chopper gated the incident laser light in synchronization with the 7~$\upmu$s magnetic field pulse, thereby minimizing unnecessary radiative heating of the sample.
The light transmitted through the sample was collected by a liquid-nitrogen-cooled HgCdTe photovoltaic detector mounted inside a ``double shield box.'' The electrical signal from the detector was converted into an optical signal by an E/O converter and transmitted via optical fiber to a digital recorder located in a shielded measurement room, safely isolated from the exploding single-turn coil. Figure~\ref{fig:inmnas} shows the cyclotron resonance transmission spectra for In$_{0.98}$Mn$_{0.02}$As and In$_{0.98}$Mn$_{0.02}$Sb in magnetic fields exceeding 100~T. The CR signals are sufficiently resolved despite the hole carrier concentration being as high as 10$^{18}$~cm$^{-3}$ and the mobility as low as 100~cm$^2$/(V~s).

\subsection{Graphene Monolayers}
\label{sec.GMonolayer}
When the cyclotron motion of a free electron is quantized in a magnetic field, it forms discrete Landau levels with energy separations of $\hbar\omega_c$, where $\omega_c = eB/m^*c$ ($\hbar$ is the reduced Planck constant, $m^*$ is the effective mass, and $c$ is the speed of light). While Landau levels in a parabolic potential form a linear fan-chart, those in a non-parabolic band exhibit sub-linear behavior. In the latter case, the effective mass $m^*$ increases with the magnetic field.

\begin{figure}[htbp]
\centering
\includegraphics[width=0.4\columnwidth]{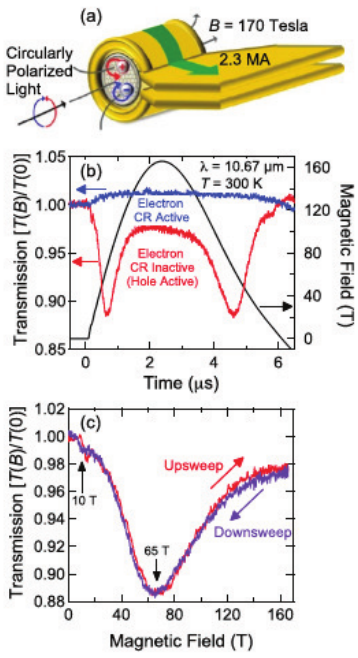}
\caption{
(a) Experimental setup for magneto-transmission measurements using the single-turn coil magnet at Los Alamos National Laboratory. Right- or left-circularly polarized infrared light ($\lambda = 10.67~\upmu$m) from a CO$_2$ laser is used in the Faraday geometry. A magnetic field of 170~T was generated by injecting a 2.3~MA pulse current into the coil. Measurements were conducted at room temperature. 
(b) Time evolution of the 170~T magnetic field pulse and the magneto-transmission signal of a graphene film. The major absorption appears only in the ``hole-inactive'' polarization, indicating that the sample is $p$-type. 
(c) The hole-active transmission signal from (b) plotted against the magnetic field for both upsweep and downsweep. Two cyclotron resonance features are observed at 10~T (small dip) and 65~T (main dip), corresponding to the $n=0 \to n=-1$ and $n=-1 \to n=-2$ inter-Landau level transitions, respectively. 
[Reproduced from Ref.~\cite{Booshehri2012} (2012) with permission from APS.]}
\label{fig:Kono_CRgraphene}
\end{figure}
In the extreme limit of non-parabolicity, the energy band dispersion becomes linear in $k$-space, analogous to the dispersion of light (the so-called Dirac cone). This situation is realized in graphene, an ultimate two-dimensional monatomic layer, where the energy dispersion is given by $E = v_F \hbar |k|$ ($v_F$ is the Fermi velocity, $v_F \approx c/1000$) starting from the Dirac point (the charge neutrality point). In this case, the cyclotron mass is defined as $m^* = E/v_F^2$. As a result of this ultimate non-parabolicity, the Landau levels are expressed as $E_n = \text{sgn}(n) v_F \sqrt{2e\hbar B|n|}$, where $n = 0, \pm 1, \pm 2, \dots$ is the Landau level index. These discrete levels are non-equidistant and obey a $\sqrt{B}$ dependence. This $\sqrt{B}$ dependence implies that the quantum limit can be reached even in relatively weak magnetic fields. However, high-field cyclotron resonance spectroscopy remains essential for large-area epitaxial graphene films---often used in industrial applications---to overcome low carrier mobility, which otherwise blurs the Landau quantization.

Cyclotron resonance spectroscopy was conducted on graphene films in magnetic fields up to 170~T, generated by the single-turn coil system at the National High Magnetic Field Laboratory in Los Alamos \cite{Booshehri2012}. The experimental setup and the resulting magneto-transmission spectra under the pulsed field are presented in Figure~\ref{fig:Kono_CRgraphene}. 
Chemical vapor deposition (CVD)-grown graphene films were used as specimens. The data in Figure~\ref{fig:Kono_CRgraphene}(c) indicate that the as-grown samples were highly $p$-doped, with a Fermi energy as high as $-0.295$~eV (corresponding to a carrier concentration of $p \approx 7 \times 10^{12}$~cm$^{-2}$). 
A drastic change in the cyclotron resonance spectra was observed in the same sample after an annealing process in vacuum. The observed shift toward lower magnetic fields and the narrowing of the resonance peaks were attributed to a substantial reduction in carrier density due to the annealing. This finding was made possible by the wide range of resonance fields accessible, which is a hallmark of ultrastrong magnetic field cyclotron resonance spectroscopy.

The magneto-optical properties of graphene are detailed in a comprehensive review by Orlita and Potemski \cite{Orlita2010}. 
A peculiarity of graphene is the cyclotron transition involving the $n=0$ Landau level. The quantum limit is easily achieved with magnetic fields of moderate strength, where the Fermi energy becomes pinned at the $n=0$ level. For samples with carrier concentrations as high as $p \approx 7 \times 10^{12}$~cm$^{-2}$, the Fermi level is pinned to the $n=0$ level in magnetic fields below 100~T. Consequently, both $n=0 \to 1$ (electron-active) and $n=-1 \to 0$ (hole-active) transitions become allowed in either $n$- or $p$-doped graphene. 
Furthermore, the oscillator strength of the cyclotron transition is proportional to $\sqrt{B}$, which is another significant advantage of ultrastrong-field cyclotron resonance spectroscopy. The increasing integrated intensity of the resonance spectra enhances the signal detection capability, even within the significant electromagnetic noise environment characteristic of exploding pulsed magnets.

\begin{figure}[tbp]
\centering
\includegraphics[width=0.6\columnwidth]{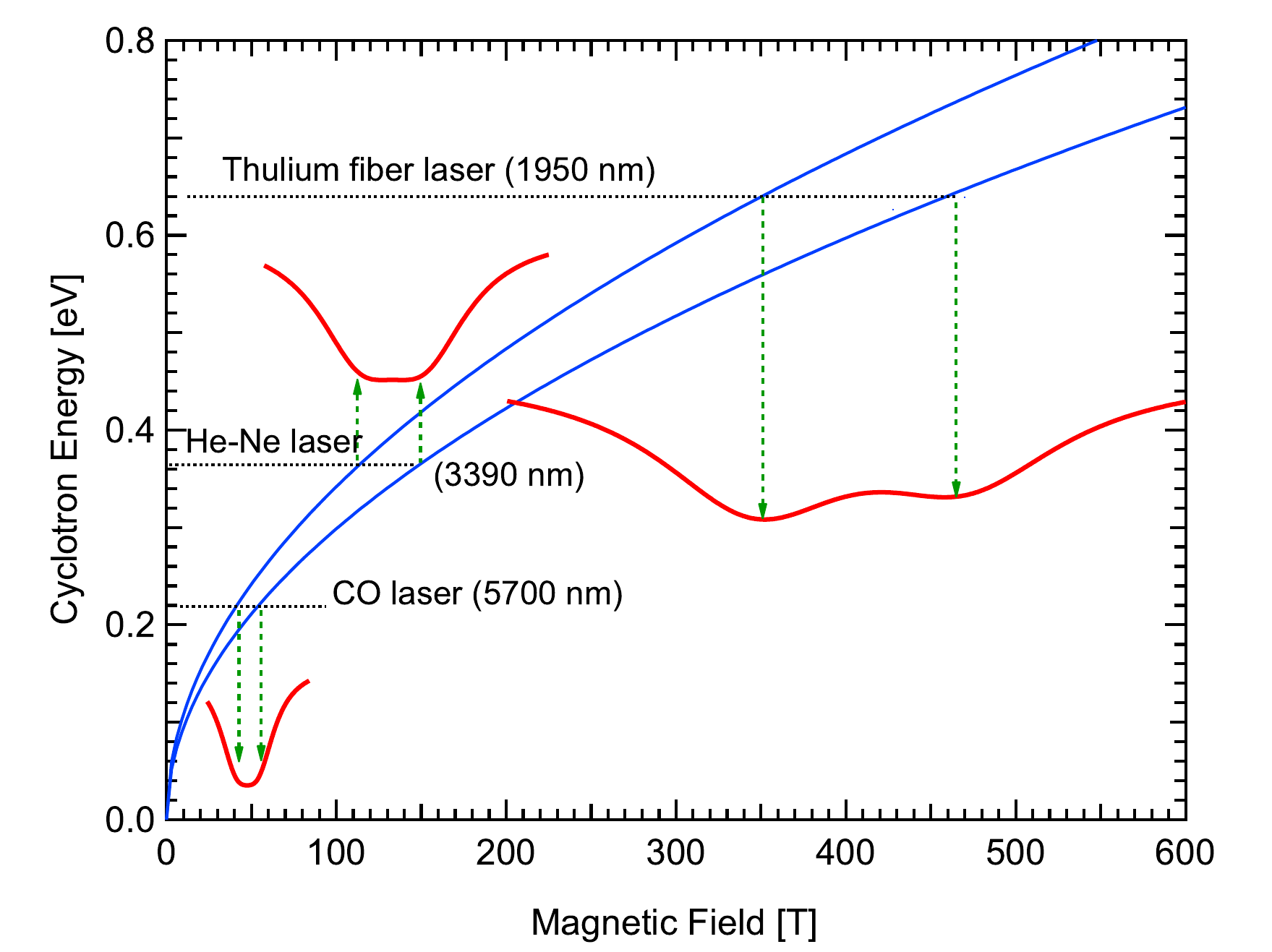}
\caption{\label{fig:CRspectraDemo}
Simulated cyclotron resonance spectra demonstrating the advantage of the $\sqrt{B}$ dependence in graphene for spectral resolution. Two independent resonance modes with similar energies are modeled using Lorentzian line shapes with a predefined width. In contrast to conventional semiconductors where Landau levels scale linearly with $B$, the $\sqrt{B}$ dependence in graphene leads to a more rapid expansion of the resonance field separation as the photon energy increases. Consequently, while the two transitions overlap to form a single broad dip at lower fields, they are clearly resolved into two distinct structures in magnetic fields exceeding 200~T.
The laser lines with different wavelengths (CO laser; 5.7~$\upmu$m, He-Ne laser; 3.39~$\upmu$m, Thulium fiber laser; 1.95~$\upmu$m) are indicated for each resonance field. 
[Reproduced from the Supplemental Material of Ref.~\cite{Nakamura2020} (2012) with permission from APS.]
}
\end{figure}
The $\sqrt{B}$ dependence of the cyclotron resonance in graphene necessitates very high photon energies to match the resonance conditions in ultrastrong magnetic fields. At higher energies far from the Dirac point, the energy dispersion is expected to deviate from linear behavior due to higher-order $k \cdot p$ corrections, specifically involving the overlap integral between nearest-neighbor $\pi$ orbitals and hopping integrals between next-nearest-neighbor orbitals. 

Deacon \textit{et al.} reported electron--hole asymmetry in exfoliated graphene, observing a 5\% difference between the electron and hole Fermi velocities at approximately 0.1~eV from the Dirac point. Based on this, they claimed clear evidence for the breaking of particle--antiparticle symmetry in the graphene system \cite{Deacon2007}. Plochocka \textit{et al.} also found subtle deviations from the ideal $\sqrt{B}$ dependence through far- and near-infrared cyclotron resonance measurements on multilayer epitaxial graphene \cite{Plochocka2008}. However, the spectral resolution was insufficient to confirm whether this deviation originated from electron--hole symmetry breaking.

\begin{figure}[htbp]
\centering
\includegraphics[width=0.7\columnwidth]{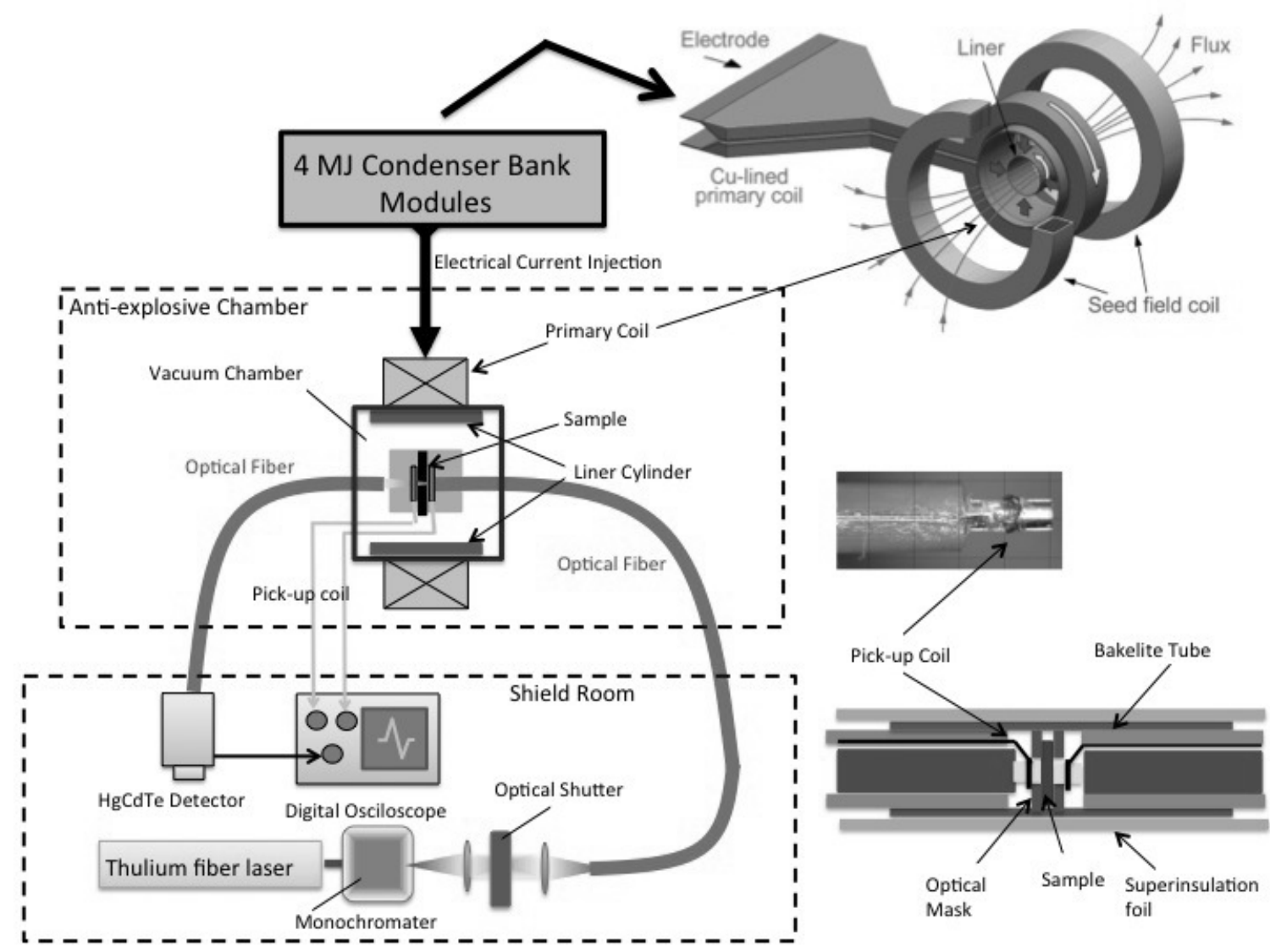}
\caption{\label{fig:cr_emfc}
Experimental setup for cyclotron resonance measurements using the EMFC technique. The measurement shield room is located approximately 15~m away from the explosion-proof chamber. A thulium fiber laser (center wavelength $\lambda \approx 1950$~nm) was used as the incident light source. Since the laser has a broad spectral width of 50~nm, a compact monochromator was employed to select a specific wavelength with a 5~nm bandwidth. An optical shutter is positioned at the laser output to prevent sample heating caused by continuous infrared radiation. (Upper right inset) The primary coil and the imploding liner, which compresses the seed magnetic field generated by a pair of seed-field coils; see Sections~\ref{sec:emfc_principle} and \ref{sec:clc} for operational details. (Right inset) Detailed views of the sample assembly. The sample is sandwiched between two optical fibers for light delivery and collection. The optical mask is fabricated from a black paper sheet. The sample holder is made of Bakelite (phenolic resin), and the outer tube is wrapped with super-insulation foil to protect the sample from the intense light arc generated by the imploding liner. The magnetic field pickup coil is wound around the optical fiber adjacent to the sample.
[All figures and photographs were taken or created by the author. Portions of the images have been presented in the author's previous publications and presentations.]}
\end{figure}
There are two primary approaches to cyclotron resonance spectroscopy. In steady magnetic fields, the photon energy of the incident light is typically scanned at a fixed magnetic field. In contrast, in pulsed magnetic fields, the magnetic field is swept while maintaining a fixed photon energy. Cyclotron resonance in pulsed ultrastrong magnetic fields generally functions as a high-resolution spectroscopy. In particular, when the Landau levels obey the characteristic $\sqrt{B}$ dependence, as in graphene, the cyclotron resonance spectra gain significantly improved resolution in ultrastrong magnetic fields, as demonstrated in Figure~\ref{fig:CRspectraDemo}. An integrated broad spectrum at lower fields is resolved into two distinct resonant dips in magnetic fields exceeding 200~T, effectively magnifying the spectral features and providing higher experimental resolution.

\begin{figure}[htbp]
\centering
\includegraphics[width=0.55\columnwidth]{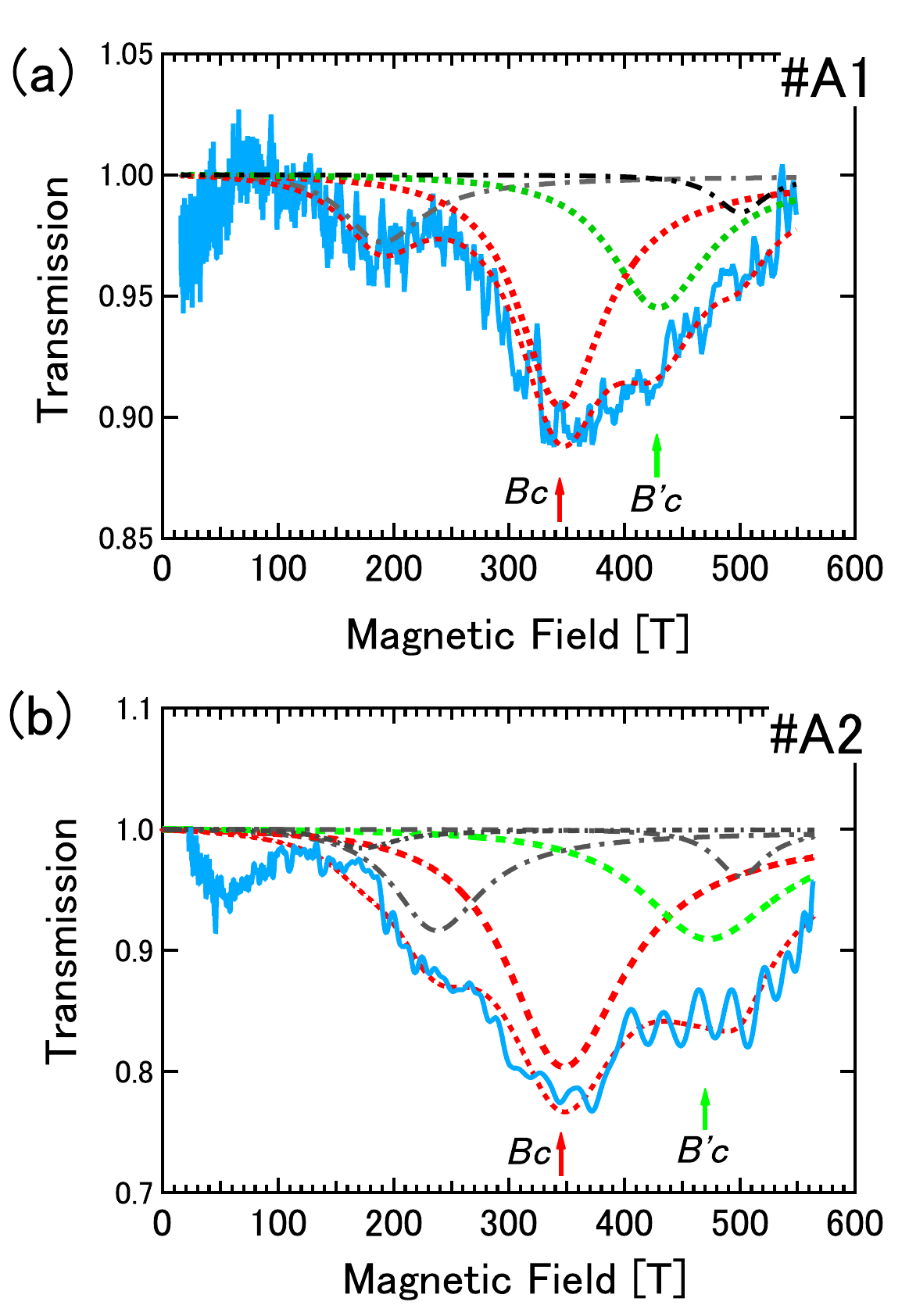}
\caption{\label{grapheneCR}
Cyclotron resonance transmission spectra for samples (a) $\sharp$A1 and (b) $\sharp$A2, measured at room temperature using EMFC generator. The dash--dot lines indicate contributions from bilayer graphene present in the sample. The red and green dashed lines represent the deconvoluted peaks for the two splitting cyclotron resonances, with peak positions at $B_{\rm c}$ and $B^{\prime}_{\rm c}$, respectively. The solid red line (or the sum of dashed lines) indicates the total fit to the experimental spectra. 
[Reproduced from Ref.~\cite{Nakamura2020} (2020) with permission from APS.]}
\end{figure}

Cyclotron resonance measurements were performed on epitaxial graphene in ultrastrong magnetic fields to achieve the high-resolution spectroscopy described above \cite{Nakamura2020}. Pulsed magnetic fields of up to 560~T were generated by EMFC technique. In magnetic fields exceeding 300~T, the resonance photon energy reaches as high as 0.64~eV (wavelength: 1.95~$\upmu$m), which falls within the near-infrared region. The experimental setup is illustrated in Figure~\ref{fig:cr_emfc}. 
The results for two samples, cut from different positions on a 10~mm $\times$ 10~mm wafer, are shown in Figure~\ref{grapheneCR}(a) and (b), respectively. With a nominal electron concentration of $n_e = 3 \times 10^{11}$~cm$^{-2}$, the carriers are fully degenerate; accordingly, the Fermi level is already pinned to the $n = 0$ Landau level in magnetic fields above 10~T. The two primary absorption peaks observed at $B_{\rm c}$ and $B^{\prime}_{\rm c}$ were identified as the cyclotron resonance transitions for $n = 0 \to 1$ (electron-active) and $n = -1 \to 0$ (hole-active), respectively. 

\begin{figure}[htbp]
\centering
\includegraphics[width=0.65\columnwidth]{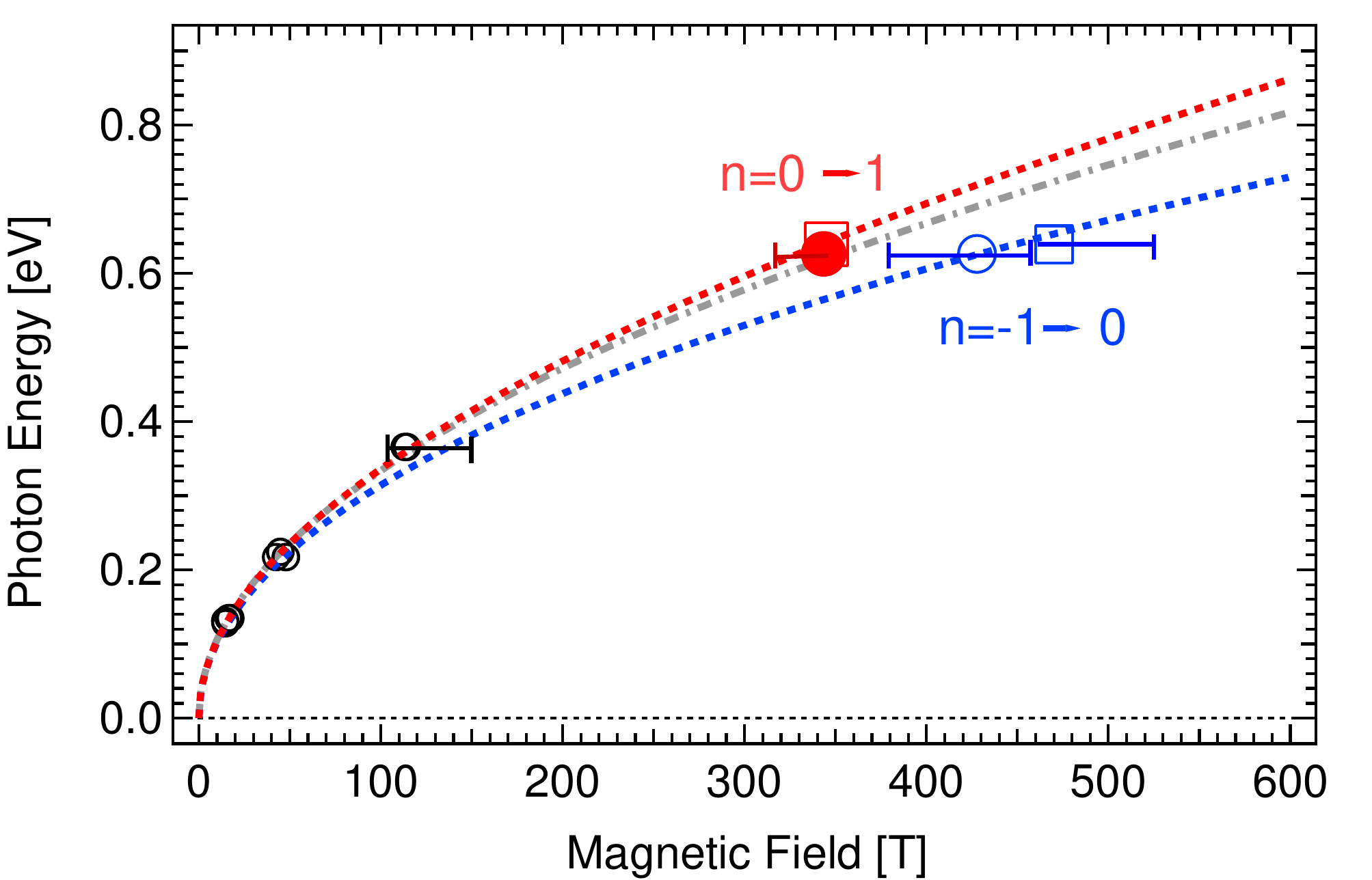}
\caption{
Landau fan-chart for the cyclotron resonance transitions $n = -1 \to 0$ and $n = 0 \to 1$. The peak positions $B_{\rm c}$ and $B^{\prime}_{\rm c}$ for samples $\sharp$A1 and $\sharp$A2 from Figure~\ref{grapheneCR} are plotted as large symbols. Small symbols at magnetic fields below 200~T represent data obtained using the single-turn coil generator. The dotted lines are fits to the resonance peaks: red for $n = 0 \to +1$ and blue for $n = -1 \to 0$. The dash--dot line represents a fit only to the resonance peaks below 170~T (small symbols), treating each resonant dip as a single broad absorption peak. 
[Reproduced from Ref.~\cite{Nakamura2020} (2020) with permission from APS.]}
\label{fig:graphCRline}
\end{figure}

According to Ando and Suzuura \cite{AndoSuzuura2017}, the cyclotron resonance splitting induced by electron--hole asymmetry can be linearly approximated with respect to the magnetic field $B$. The resonance energies for the $n = 0 \to +1$ and $-1 \to 0$ transitions are modified by a linear term as follows:
\begin{align}
\label{eqlinearCR1}
E_{0 \to +1} &= v_{\rm F} \sqrt{2e \hbar B} + cB, \\
\label{eqlinearCR2}
E_{-1 \to 0} &= v_{\rm F} \sqrt{2e \hbar B} - cB.
\end{align}
The second term $\pm cB$ in Eqs.~(\ref{eqlinearCR1}) and (\ref{eqlinearCR2}) represents the electron--hole asymmetry. The values $v_{\rm F} = 0.90 \times 10^6$~m/s and $c = 1.1 \times 10^{-4}$~eV/T were obtained as the best-fit parameters to the peak positions shown in Figure~\ref{fig:graphCRline}. 
The best-fit curves for the $n = 0 \to +1$ and $n = -1 \to 0$ transitions are plotted as dotted lines. 
An extrapolation of a simple $\sqrt{B}$ dependence fitted to magnetic fields below 120~T ($v_{\rm F} = 0.92 \times 10^6$~m/s) deviates slightly from $B_{\rm c}$ and stays far from $B^{\prime}_{\rm c}$, as indicated by the dash--dot line in Figure~\ref{fig:graphCRline}.
The ultra-high resolution achieved in the cyclotron resonance spectral range of 200--550~T revealed a clear spectral splitting near 400~T. These two split resonances are a direct consequence of the electron--hole asymmetry in graphene, manifesting as a difference between the electron ($n = 0 \to +1$) and hole ($n = -1 \to 0$) quantum-limit cyclotron resonances. This finding implies that the spectral structure of the cyclotron resonance below 200~T is also composed of two such resonances; however, they are smeared out by broadening and remain unresolved. Notably, the dip minimum of each resonance at magnetic fields below 200~T is shifted to the lower-field side of the dotted curves in Figure~\ref{fig:graphCRline}. This suggests that the resonance observed below 200~T consists of two overlapping peaks, where the lower-field component is dominant, causing the combined, unresolved peak to appear at a lower field.

\section{Magnetoconductivity Measurements in Ultrastrong Pulsed Magnetic Fields}
\label{sec:RF}

Magnetoconductivity (or magnetoresistance) measurements reveal essential physics regarding the electron transport properties of various materials. The observation of Shubnikov--de Haas quantum oscillations in magnetoconductivity is crucial for determining the Fermi surface and electron scattering mechanisms in semiconductors, semimetals, and the normal states of superconductors. Furthermore, magnetic-field-induced metal--insulator (or semiconductor) transitions are directly evidenced by magnetoconductivity measurements, providing insights into the underlying novel physics. The magnetic field--temperature phase diagram of the superconducting state, determined through magnetoconductivity, provides key information on Cooper pair-breaking mechanisms.

The noisy and extreme conditions of pulsed magnetic fields impose serious difficulties on standard four-contact magnetoconductivity measurements. Consequently, contactless techniques using radio frequencies (ranging from MHz to GHz) have been actively developed and applied to measurements under strong pulsed magnetic fields. In an attempt to overcome these experimental challenges, a transmission-type radio-frequency (hereafter abbreviated as RF) (20--30~MHz) contactless measurement system was developed by Sakakibara \textit{et al.} \cite{Sakakibara1989_1, Sakakibara1989_2}. In this setup, a sample thinner than its skin depth is sandwiched between a set of two small coils acting as an emitter and a receiver. Radio-frequency RF transmission signals through the sample were analytically modeled and calibrated with data from four-probe DC conductivity measurements. This technique proved highly successful when applied to the high-$T_{\rm c}$ superconductor YBa$_2$Cu$_3$O$_{7-\delta}$ under non-destructive long-pulse magnetic fields.

Since around 2000, there has been a trend among scientists working with pulsed magnets to adopt ready-made compact devices for measurement technologies. The tunnel diode oscillator (TDO), conventionally used in metal detectors, was adopted as an RF power source for miniature RF tank circuits. In this setup, the shift in resonant frequency induced by conductivity changes serves as a sensitive sensor for the magnetoconductivity of a sample. A change in electrical conductivity alters the inductance of a sample placed within the resonant $LC$ tank circuit, which in turn induces a frequency shift $\Delta f_0$ from the resonant frequency $f_0 = 1 / (2\pi\sqrt{LC})$. This measurement technique has been applied by several groups to observe quantum oscillations and phase transitions in organic semiconductors \cite{Drigo}, organic superconductors \cite{Coffey, Bayindir, Ohmich, Kamatsu, Mielke}, and iron-based superconductors \cite{Audouard} in pulsed magnetic fields up to approximately 40--50~T.

To address certain drawbacks of the TDO in pulsed magnet environments, proximity detector oscillators (PDOs) based on commercially available integrated circuits (ICs) have been adopted as a more convenient alternative \cite{Altarewneh2009, Ghannadzadeh}. These oscillators perform better in higher pulsed magnetic fields, where a wider dynamic range of resonant frequency variation is required. The effectiveness of PDO-based conductivity measurements is exemplified by the observation of discernible quantum oscillations in under-doped YBa$_2$Cu$_3$O$_{6+\delta}$ in magnetic fields between 40~T and 80~T \cite{Sebastian}. These high-frequency RF conductivity measurement techniques have also been applied to ultrastrong magnetic fields generated by single-turn coil (STC) systems \cite{Sakakibara1989_2, Mielke}. However, the data quality in such environments is substantially degraded by electromagnetic discharge noise and the extremely fast rise time of the pulsed field (on the order of $5 \sim 7 \times 10^7$~T/s).

Radio-frequency (RF) conductivity measurements, using a transmission-type setup similar to that employed by Sakakibara \textit{et al.} \cite{Sakakibara1989_1}, have been applied to significantly higher magnetic fields generated by chemical explosive-driven flux compression in Russian magnetocumulative generators. The upper critical fields ($H_{c2}$) of high-$T_{\rm c}$ superconducting materials, such as polycrystalline YBa$_2$Cu$_3$O$_7$ and Bi$_2$Sr$_2$CaCu$_2$O$_y$, as well as YBa$_2$Cu$_3$O$_x$ films, were determined in magnetic fields between 100~T and 200~T \cite{Golovashkin}. Later, in a quest to observe the semiconductor-to-metal phase transition in iron monosilicide (FeSi), magnetoconductivity measurements were conducted in magnetic fields up to 450~T \cite{Kudasov1998}. These measurements evidenced a gradual increase in the RF reflection signal above 400~T, which was ascribed to a transition to a highly conducting state rather than an abrupt semiconductor-to-metal transition. In contrast, at a temperature of 4.2~K, measurements using a compensated pickup coil revealed an abrupt increase in the magnetic moment above 355~T. This was attributed to a first-order phase transition associated with high electronic conduction \cite{Kudasov1999}.

In pursuit of transport measurements for solid-state physics in the ultrastrong magnetic fields generated by Russian chemical explosive-driven flux compression, an international joint research project known as the ``Dirac Series'' was organized. Experiments were conducted between 1996 and 1997 at Los Alamos National Laboratory, USA. A magnetoconductivity measurement system was instrumented using precise photolithographic patterning and GHz-range high-frequency conductivity sensors in an attempt to overcome the significant difficulties associated with the imploding process of magnetic flux compression~\cite{Kane1997}.

In the case of the semimetal bismuth, where a magnetic-field-induced semimetal-to-semiconductor transition is expected, the transmission signal from the sample was severely disturbed by massive electrical noise at each stage of the cascade fusion as the magnetic field increased. Nevertheless, they eventually obtained reliable magnetoconductivity data in magnetic fields up to 300~T. Furthermore, high-frequency alternating current (AC) conductivity measurements were applied to high-$T_{\rm c}$ cuprate superconductor YBa$_2$Cu$_3$O$_{\delta}$ films in magnetic fields up to 800~T, revealing an upper critical field of $H_{c2} = 240$~T and the onset of a dissipative state at $H_{c1} = 150$~T~\cite{Dzurak1998b}. It was concluded that the high-frequency AC conductivity measurement system, when applied to semiconductor materials, can yield reliable data up to 450~T before the fusion of the third cascade in the 1000~T-class MC-1 generator, where three liners (cascades) are used for flux compression.

Systematic high-frequency radio-frequency (RF) conductivity measurements were conducted in magnetic fields up to 600~T, generated by EMFC, to investigate the magnetic field--temperature ($B$--$T$) phase diagram of the high-$T_{\rm c}$ superconductor YBa$_2$Cu$_3$O$_{7-\delta}$~\cite{Sekitani2007}. The EMFC technique, as described in Sec.~\ref{sec:emfc_principle}, offers several advantages over chemical explosive-driven flux compression. First, only a single imploding liner is required to achieve the peak field, which minimizes the complex noise associated with multi-stage cascades. Second, and most importantly for systematic studies, the field generation process is more reproducible and efficient, allowing for multiple experimental runs by simply replacing the coil components while maintaining consistent measurement conditions.

\begin{figure}[htbp]
\centering
\includegraphics[width=0.5\columnwidth]{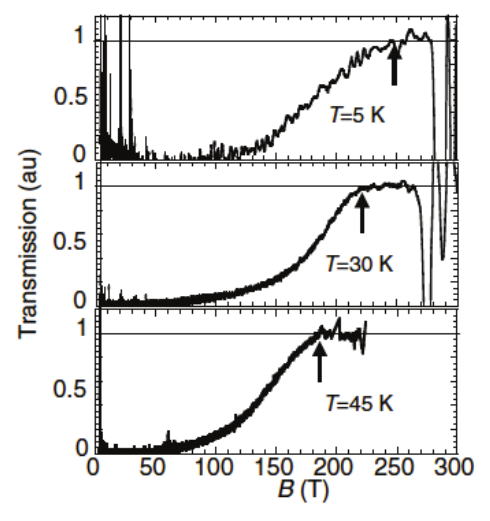}
\caption{Normalized RF transmission spectra at 5~K, 30~K, and 45~K in magnetic fields up to 300~T. At 5~K, the cryostat sample holder was destroyed by the impact of the imploding liner, causing the signal to be truncated at 290~T, even though the magnetic field measured by the pickup coil was recorded up to 600~T. The solid lines representing the superconducting (0) and normal state (1) levels were determined in the absence of a magnetic field below and above $T_{\rm c}$, respectively. [Reproduced from Ref.~\cite{Sekitani2007} with permission from IOP Publishing (2007).]}
\label{SekitaniHiTc}
\end{figure}

The use of a thin-film microcoil, fabricated with high precision via photolithography, along with a 60~MHz RF source, was key to obtaining reliable measurements in magnetic fields up to 280~T. At higher fields, the cryostat and signal transmission leads were destroyed by the imploding liner, although the pickup coil continued to record the magnetic field up to 600~T. (Prior to the development of the ``copper-lined'' EMFC coil described in Sec.~\ref{sec:clc}, limited reproducibility and the early destruction of the sample probe before the peak field restricted reliable physical measurements to values significantly lower than the maximum recorded magnetic field.)
\subsection{RF Self-induction Resonant Coil Method}
\label{rf_sirc}
There is a difference of approximately three to four orders of magnitude in pulse duration between non-destructive long-pulse magnets and destructive short-pulse magnets. Consequently, the difference in the rate of magnetic flux change, $dB/dt$, amounts to $10^4$--$10^5$~T/s. Alternating current (AC) conductivity measurements applied to destructive short-pulse magnetic field experiments require a much wider dynamic range, higher sensitivity, superior performance stability, and a robust capacity to discriminate signals from high-frequency electromagnetic noise.

To overcome these challenges, a novel approach to high-frequency contactless measurements was proposed by Altarawneh \cite{Altarawneh2012}, utilizing an inductance--capacitance ($LC$) band-stop filter circuit. The core concept involves employing the radio-frequency (RF) tank sensor coil as a ``self-resonant circuit'' that functions as a band-stop frequency filter. Rather than monitoring a shift in the resonant frequency $\Delta f_0$, this method measures the change in the resonance amplitude of the sensor RF-tank circuit. This amplitude variation directly reflects the changes in sample conductivity induced by external parameters, such as temperature or magnetic field.
This self-resonant coil (SRC) technique was revised for application in the destructive short-pulse magnetic field environment of the STC system \cite{Nakamura2018mst}.

\begin{figure}[htbp]
\centering
\includegraphics[width=0.6\columnwidth]{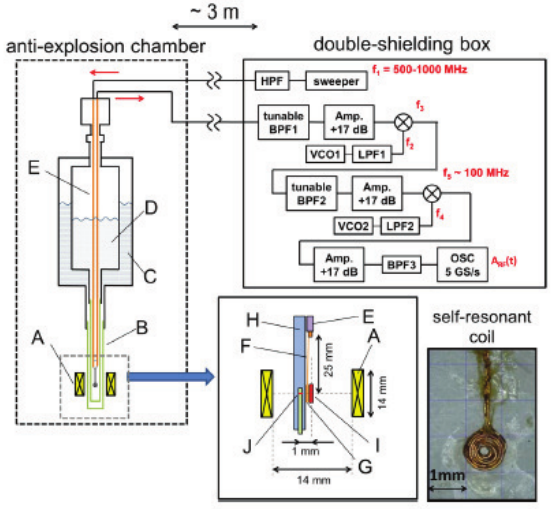}
\caption{
Arrangement of the experimental setup. (A) Single-turn coil (14~mm inner diameter); (B) Tail of the cryostat containing liquid $^4$He; (C) Liquid nitrogen dewar; (D) Liquid $^4$He dewar ($\sim$0.4~L); (E) Semi-rigid coaxial cable within the measurement probe rod; (F) Pair of twisted copper wires ($\sim$25~cm); (G) Self-resonant coil; (H) Sample stage; (I) Sample; (J) Magnetic field pickup coil. The spiral self-resonant coil (SRC) is shown in the inset photograph. The input frequency ($f_1 \approx 800$~MHz) supplied by an analog signal generator is tuned close to the resonant frequency of the SRC with the sample ($f_0$). The RF signal returned from the probe undergoes frequency down-conversion via a double-stage superheterodyne circuit using a voltage-controlled oscillator (VCO) and frequency mixers ($\otimes$). Low-pass filters (LPF1 and LPF2) and tunable narrow band-pass filters (BPF1 and BPF2) are employed for signal conditioning. The signal is finally down-converted to approximately 100~MHz and recorded by a high-resolution digital oscilloscope. 
[Reproduced from Ref. \cite{Nakamura2018mst} with permission from IOP Publishing (2018).]}
\label{srccircuit}
\end{figure}

\begin{figure}[hb]
\centering
\includegraphics[width=0.9\columnwidth]{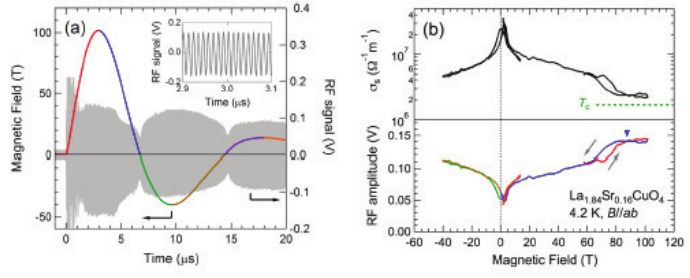}
\caption{\label{src100T}
Electrical transport characteristics of La$_{1.84}$Sr$_{0.16}$CuO$_4$ under ultrastrong magnetic fields up to 102~T. (a) Recorded RF down-converted waveform and the magnetic field pulse generated by the single-turn coil generator. The inset shows an expanded view of the detected RF wave between 2.9~$\upmu$s and 3.1~$\upmu$s. (b) (Bottom) Amplitude traces of the RF wave; the colored curves correspond to the respective $B(t)$ pulses in (a). (Top) Conductivity $\sigma_{\rm{s}}$ derived from the RF amplitude (bottom) plotted as a function of the magnetic field. The triangle symbol indicates the upper critical field $H_{c2}$, and the horizontal dotted line labeled $T_{\rm c}$ represents the conductivity $\sigma_{\rm{s}}$ at $T_{\rm c}$ in the absence of a magnetic field (normal-state conductivity). 
[Reproduced from Ref. \cite{Nakamura2018mst} with permission from IOP Publishing (2018).]}
\end{figure}

A flat spiral coil was adopted instead of the conventional ``hollow-type coil'' to improve the coupling between the sensor coil and the sample while minimizing inductance; this enabled the measurement frequency to extend above 800~MHz. 
The experimental setup of the SRC technique and its electronic circuitry are illustrated in Figure~\ref{srccircuit}. The input frequency ($f_1 \approx 800$~MHz) supplied by an analog signal generator is tuned close to the resonant frequency ($f_0$) of the SRC with the sample mounted. 
The RF signal returned from the probe undergoes frequency down-conversion via a double-stage superheterodyne circuit with a VCO. The SRC technique was applied to determine the upper critical field, $H_{c2}$, of the cuprate high-$T_{\rm c}$ superconductor La$_{1.84}$Sr$_{0.16}$CuO$_4$ in magnetic fields up to 100~T, as summarized in Figure~\ref{src100T}. 
A distinct hysteresis was observed in the magnetoconductivity between 62--82~T in Figure~\ref{src100T}(b). Such a phenomenon is rarely observed in magnetic fields close to the $H_{c2}$ of high-$T_{\rm c}$ superconductors. This provides definite evidence for a first-order phase transition from the superconducting to the normal state.

The upper critical field $H_{c2}$ was determined at several temperatures, and the resulting $B$--$T$ phase diagram was analyzed in detail using the Werthamer--Helfand--Hohenberg (WHH) theory, which incorporates the Pauli paramagnetic effect (characterized by the Maki parameter, $\alpha$) and the spin--orbit interaction ($\lambda_{\rm{SO}}$) \cite{WHH}. This analysis yielded a relatively large Maki parameter of $\alpha = 4.4$ and a minimal spin--orbit interaction term of $\lambda_{\rm{SO}} < 0.1$. These findings well explain the occurrence of the first-order phase transition, representing a rare case among high-$T_{\rm c}$ superconducting materials \cite{Nakamura2019sr}.

\subsection{Magnetoconductivity Measurements in Electromagnetic Flux Compression}
\label{sec:rf_sirc_emfc}

The single-turn coil (STC) method is highly advantageous for generating ultrastrong magnetic fields, as described in Section~\ref{STC}. However, it entails certain drawbacks, particularly when applied to magnetoconductivity measurements. First, the pulsed magnetic field exhibits an extremely short rise time ($dB/dt \sim 10^8$~T/s), which induces substantial eddy current Joule heating within and around the sample. Second, and most critically, massive discharge spike and ringing noise originating from the gap switches persist until the field reaches nearly half of its peak value, significantly distorting the output signal (see, e.g., Figure~\ref{fig:mitamuraCdCrO}). 

As shown in Figure~\ref{src100T}(a), intense initial spike noise is superimposed on the signal from 0 to 2~$\upmu$s. Consequently, data from the initial rising phase up to 60~T---exceeding half of the peak field in this case---are too noisy to be reliable and are typically excluded. This issue similarly degrades magnetization data obtained via the induction pickup coil method during the rising phase of STC experiments (see Section~\ref{inductionm}). Therefore, magnetization and transport data from STC experiments are frequently analyzed and plotted only during the descending phase of the magnetic field pulse.

In the case of EMFC, on the other hand, the magnetic field grows gradually over 35--45~$\upmu$s to reach 200~T, before rapidly increasing to peak fields (600--1200~T) within 2--3~$\upmu$s, followed by total destruction (see, e.g., Figure~\ref{compFRandV}). Accordingly, the critical data in the ultrastrong field region often remain unaffected by the initial discharge noise, as the spike noise typically decays within 1--3~$\upmu$s. This characteristic is a significant advantage for conductivity measurements in ultrastrong magnetic fields.

\begin{figure}[tbp]
\centering
\includegraphics[width=0.9\columnwidth]{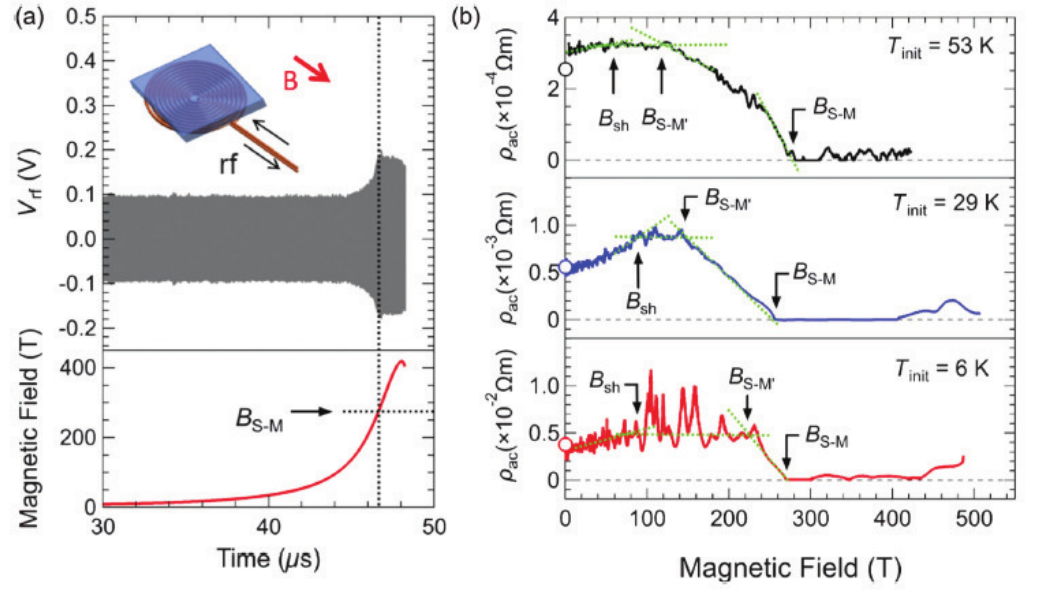}
\caption{
Results of self-resonant coil (SRC) RF-conductivity measurements performed at 700~MHz on a thin FeSi sample in magnetic fields up to 500~T generated by EMFC method. (a) Time evolution of the RF resonance signal $V_{rf}$ (upper panel) and the pulsed magnetic field (lower panel) at 53~K. The inset illustrates the self-resonant spiral probe coil loaded with the FeSi sample. (b) Magnetoresistance as a function of magnetic field, converted from the $V_{rf}$ signals in (a) at 53~K, 29~K, and 6~K. The arrows labeled $B_{sh}$, $B_{S-M'}$, and $B_{S-M}$ indicate the critical magnetic fields where the transport mechanisms undergo transitions. The semiconductor-to-metal phase transition is fully completed at $B_{S-M}$. 
[Reproduced from Ref. \cite{Nakamura2018mst} with permission from IOP Publishing (2018).]}
\label{fesidata}
\end{figure}

The SRC method was applied to magnetoconductivity measurements of iron monosilicide (FeSi) to investigate the semiconductor-to-metal phase transition expected in the ultrastrong field regime \cite{Nakamura2021}. As briefly introduced in Section~\ref{sec:RF}, previous radio-frequency AC conductivity measurements using chemical explosive-driven flux compression have already suggested symptoms of this phase transition in magnetization and conductivity behaviors around 400~T.
Figure~\ref{fesidata} summarizes the results of the SRC RF-conductivity measurements conducted at 700~MHz in magnetic fields above 400~T at various temperatures. Initial discharge noises were practically absent in the region of interest, in contrast to measurements performed in STC experiments. Four distinct electronic phase changes were observed successively: first, a positive magnetoresistance, followed by a saturation region ($B_{sh} < B < B_{S-M'}$), a subsequent slow decrease in resistance ($B_{S-M'} < B$), and finally an abrupt drop into metallic conduction---a well-defined phase transition at $B_{S-M} = 267$~T---with increasing magnetic fields. The positive magnetoresistance up to $B_{sh}$ was attributed to hopping-type electron conduction in mid-gap localized states. The observed semiconductor-to-metal phase transition was well explained by the band gap closing into a zero-gap state, driven by the enormous Zeeman shifts attainable only in ultrastrong magnetic fields.

\section{Future Perspectives and Opportunities}
\label{sec:outlook}

\subsection{Evolution of Measurement Techniques for the Megagauss Frontier}
\label{sec:appli}

\begin{figure}[tbp] 
\centering
\includegraphics[width=0.6\columnwidth]{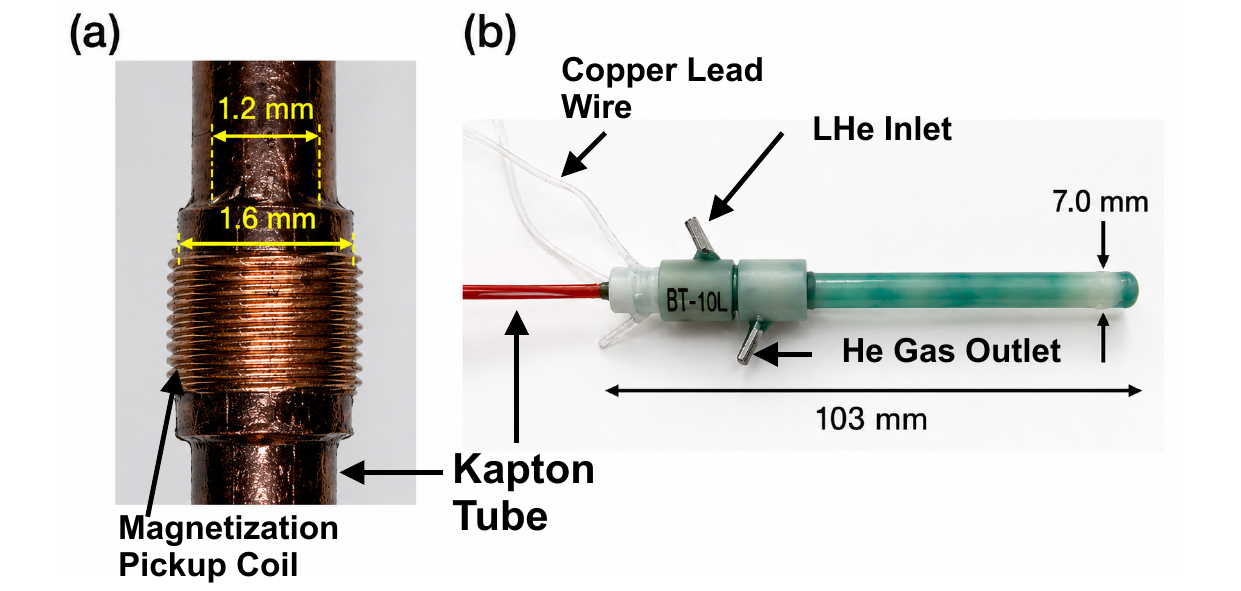} 
\caption{\label{photocoaxial}
(a) Photograph of a coaxial-type compensated magnetization pickup coil wound around Kapton tubes with inner and outer diameters of 1.2~mm and 1.6~mm, respectively. The copper wire is the same as that used in Figure~\ref{fig:comp_coil}. (b) All-FRP (fiber-reinforced plastic) liquid $^4$He cryostat, into which the magnetic pickup coil shown in photograph (a) is inserted. 
[Reproduced from Ref.~\cite{Gen} (2019) with permission from APS.]}
\end{figure}

However, the parallel pair of counter-wound magnetic pickup coils introduced in Sec.~\ref{inductionm} is considered inappropriate for EMFC, where measurements are limited to a single-shot opportunity, unlike the STC case. The coaxial-type compensated magnetization pickup coil shown in Figure~\ref{photocoaxial} is the most promising candidate and is therefore worth pursuing for magnetization measurements in EMFC experiments \cite{Gen,Gen2023PNAS}. 
The coaxial-type coil benefits from the extended axial homogeneity of the EMFC field, enabling higher sensitivity through increased windings. Unlike parallel configurations, the coaxial design results in nearly identical $dB/dt$ induced voltages in both the inner and outer coils. This in-phase induction prevents large voltage gradients between the windings, thereby significantly enhancing the robustness against insulation failure in ultrahigh field environments.

Magnetostriction is closely correlated with magnetization through spin--lattice coupling, which induces deformations of the crystal structure associated with changes in the spin structure induced by magnetic fields. Magnetostriction measurements are indispensable for a deep understanding of the physics underlying magnetization processes in magnetic material. 
One of the classical method for magnetostriction measurement is the capacitance method, but it is mostly applied in steady-state magnetic fields. In pulsed magnetic fields, the strain-gauge method is typically adopted using either resistive gauges or piezoelectric devices. These methods, however, have hardly been applied in destructive short-pulsed magnetic fields due to limited time and space, as well as interference from electromagnetic noise and mechanical vibrations. Owing to recent technological advances in fiber optics, fiber Bragg gratings embedded in optical single-mode fibers can be fabricated with high precision; these are known to be highly effective as sensitive magnetostriction gauges in pulsed magnetic fields \cite{Daou2010rsi}.

This technique was further refined by employing a 100~MHz mode-locked fiber laser, and high-speed, high-resolution magnetostriction measurements were demonstrated in destructive short-pulsed magnetic fields of up to $\sim$150~T using a single-turn coil \cite{IkedaMstric2017}. It was observed that abrupt changes in striction, enhanced by the massive $dB/dt$ in ultrastrong magnetic fields, can evoke artifacts such as substantial signal oscillations that deteriorate system resolution. While the fiber Bragg grating method still has significant room for improvement, it remains an auspicious technique for magnetostriction measurements in the 100--1000~T range provided by EMFC.

Measuring entropy provides additional information necessary for a deeper understanding of quantum phases induced by either external magnetic fields or pressure. In recent decades, magnetocaloric effect and specific heat measurements---which provide direct access to the entropy of materials---have evolved into sophisticated techniques for probing the details of quantum phase transitions in strong pulsed magnetic fields. However, these are generally limited to the magnetic field regions provided by non-destructive long-pulsed magnets. Recent achievements in entropy measurements realized in non-destructive long-pulsed magnetic fields are reviewed in Refs.~\cite{Miyake2020, Kohama2022}. The response time required for temperature measurements has been an impediment to their application in destructive short-pulsed magnetic fields on the order of microseconds above 100~T. A possible approach for such short-pulsed fields is to utilize adiabatic conditions and simply measure the temperature to obtain $T(B)$ curves, which in turn constitute magnetocaloric effect curves. Adiabatic magnetocaloric effect measurements were demonstrated to clarify the phase boundaries of the magnetic field--temperature ($B$--$T$) phase diagram of solid oxygen in long-pulsed magnetic fields \cite{TNomura2017}; this technique could be extended to measurements in ultrastrong short-pulsed magnetic fields.

Ultrasound measurement is an efficient tool for the sensitive detection of sound velocity, reflecting the free energy in materials (both solid and liquid), and has been used to investigate various phase transitions in strong magnetic fields. This technique has been intensively applied in non-destructive strong pulsed magnetic fields to identify evidence of phase transitions that are difficult to detect via magnetocaloric effect or magnetization measurements. For example, this technique proved highly efficient for surveying the spin-nematicity observed around 40~T in the spin-frustrated magnet LiCuVO$_4$ \cite{Gen2019nematic}. The spin-nematic phase transition, like other quantum critical points caused by quadrupole ordering, was shown to couple effectively to ultrasound propagation. Ultrasound measurements using the continuous-wave excitation method were attempted in ultrastrong short-pulsed magnetic fields generated by a single-turn coil megagauss generator \cite{Nomura2021ultrasound}. However, the sound velocity and ultrasound amplitude signals were reliable only in magnetic fields below $\sim$80~T; above this, the signal lines were likely compromised by mechanical vibrations from the shockwave and electromagnetic noise from the exploding coil. In this regard, ultrasound measurement is considered more suitable for EMFC experiments, where the magnetic field pulse profile is more gradual and longer than that of a single-turn coil, and the shockwave front reaches the sample surface only after the peak field has passed.

\subsection{Measurements in Megagauss Field Environment}

While STC system has a nominal maximum field of 300~T, successful cryogenic measurements have been limited to 200~T \cite{Otsuka2018}. This gap of nearly 100~T stems from a critical trade-off between the magnetic field intensity and the available working space: generating 300~T requires a small coil bore of only 3~mm in diameter, which currently cannot accommodate conventional cryostats. Consequently, the development of an ultra-miniaturized cryostat, with an outer diameter small enough to fit within this 3~mm limit, is essential to bridge this gap. Maintaining a stable environment under such extreme spatial and electromagnetic conditions remains a primary challenge.

While the peak magnetic field generated by the STC method is lower than that achieved by EMFC, it offers a significant advantage: the ability to perform repetitive measurements under identical physical environments. This repeatability is a crucial factor in enhancing the reliability and precision of experimental data. Consequently, STC measurements serve as a vital backup, providing high-precision data in the relatively lower-field regime that reinforces the credibility of EMFC results obtained in much higher fields. To fully leverage this advantage, several technical challenges regarding reproducibility must be addressed: \begin{itemize} \item \textbf{Geometrical Reproducibility of the Coil:} Although the STC coil is destroyed in every shot, precise consistency in the shape and dimensions of successive coils is paramount for maintaining field uniformity. \item \textbf{Electrical Contact Reliability:} The quality of the current contact between the collector plates and the STC coil must be perfectly reproducible. Poor contact leads to sparking and fluctuating contact resistance, which alters the current injection profile and destabilizes the field generation. \item \textbf{Durability of the Sample Cryostat:} To maintain a consistent physical environment across multiple shots, the sample cryostat must be designed to withstand repeated explosive shocks. For instance, in magnetization measurements, a set of at least two shots is required to accurately subtract large background signals. This necessitates the use of the exact same magnetic pickup coil under identical conditions. \end{itemize} Therefore, the continuous refinement of ``craftsmanship'' in fabricating bespoke, handmade sample cryostats, alongside the development of robust instrumentation, is essential for achieving the highest standards of precision in megagauss science. 

Regarding EMFC, to enable precision physics at cryogenic temperatures at lowest possible temperature and at higher fields than 600~T, a minimum bore diameter of 6~mm must be preserved until the imploding liner destroys the cryostat. Although a record field of 1200~T was achieved in a 3~mm diameter space \cite{Takeyama1200T} (see Section~\ref{sec:record1200T}), scaling this to a 6~mm bore for 1000~T class measurements requires further technical advancements. These include an increased seed field of 4.5~T (Ref.~\cite{Takeyama1000T})] and a capacitor energy of 5~MJ.
Significant technical challenges remain, particularly regarding the seed-field magnets. Located only 10~cm from the primary coil, these magnets are subjected to intense shockwaves and mechanical forces during the explosion. Future development must focus on optimizing mechanical protection, enhancing electrical insulation against breakdown, and increasing seed-field intensity. Furthermore, while the direct integration of the liner and sample holder onto the copper-lined primary coil (Figure~\ref{emfcsample}, Sec.~\ref{sec:cryo}) ensures precise alignment, it introduces parasitic electromagnetic forces. The rapid $dB/dt$ of the pulsed seed field induces mechanical vibrations in the primary coil, which can destabilize the uniformity of the liner acceleration. 
Overcoming this requires enhancing the mechanical rigidity of both the primary coil and the clamping system to suppress detrimental oscillations triggered by increasing seed-fields, which otherwise compromise the symmetry and stability of the liner implosion.

It is crucial to recognize that while EMFC is inherently associated with drastic structural destruction, our experience underscores that the aftermath of such events is not merely a byproduct, but a vital source of innovation. 
Contrary to the conventional approach of treating the explosion-proof chamber as a ``black box,'' systematic post-experimental analysis---specifically the meticulous cleaning and debris mapping conducted over years of operation---allowed us to identify ``safe zones'' within the explosive environment. 
By accurately delineating the propagation paths of shockwaves and blast pressure, we were able to strategically install high-speed optical monitoring systems and delicate cryogenic infrastructures, such as liquid helium transfer tubes and sample cryostats, directly within the protection proof chamber. 
This optimized configuration was instrumental in achieving the first-ever magnetization measurements at 600~T and 5~K, as well as reaching the landmark field intensity of 1200~T.

The philosophy of ``managing'' rather than merely ``resisting'' explosive energy represents a paradigm shift: it demonstrates that progress is unlocked by treating experimental ``failure'' or destruction not as something to be ignored, but as data to be rigorously analyzed. 
This principle---learning from the wreckage to prepare for the next breakthrough---is a fundamental concept applicable across diverse scientific disciplines. 
True advancement arises when we refuse to overlook the remnants of a failed trial, treating every fragment of debris as a roadmap toward the fulfillment of a higher purpose.

\section{Conclusions}
\label{sec:conclusions}

This review has synthesized the twenty-year trajectory of technical breakthroughs and groundbreaking scientific discoveries that have successfully transformed megagauss field environments into domains of high-precision solid-state metrology. 
In single-turn coil (STC) systems, the~meticulous development of bespoke, ultra-miniaturized sample cryostats, combined with innovative engineering of the blast-shielding structures and protection blocks, has enabled a dramatic evolution in cryogenic experiments. The~reliable and reproducible measurement ceiling has been pushed from the conventional 150~T up to well over 200~T. This robust framework has vastly expanded the repertoire of high-precision metrology over a wide temperature range---extending from cryogenic to high temperatures---enabling highly sophisticated measurements of magneto-optical effects (Faraday rotation and magneto-absorption), cyclotron resonance, magnetization, specific heat, the~magnetocaloric effect, and~ultrasound~propagation. 

These advanced STC capabilities have uniquely unlocked the unambiguous determination of novel quantum phases and microscopic mechanisms in strongly correlated electron systems. Prominent scientific achievements include the comprehensive mapping of magnetic field--temperature ($B$--$T$) phase diagrams, the~determination of successive field-induced phase transitions, and~the elucidation of complex magnetic ordering in frustrated magnets. Furthermore, systematic investigations into the relationship between electronic states and magnetism in high-carrier-density ferromagnetic semiconductors have been established. In~high-temperature superconductors, the~systematic evaluation of upper critical fields has successfully clarified the mechanisms underlying Cooper-pair breaking, culminating in the complete determination of their $B$--$T$ phase diagrams and the first-ever observation of a first-order phase transition, a~phenomenon exceptionally rare in high-temperature superconducting states. In~molecular systems, these configurations led to the landmark discovery of the ultrahigh-field-induced molecular rearrangement and an exotic structural phase ($\theta$-phase) in solid~oxygen.

In electromagnetic flux compression (EMFC) systems, a~parallel paradigm shift has been achieved through rigorous structural and electrical optimization. The~development of advanced primary coils (CL primary coils) and their refined installation modalities, coupled with the optimized design of the high-energy capacitor bank and the entire current transmission line, has significantly enhanced the efficiency, reproducibility, and~controllability of the field-generation process. This stable implementation effectively doubled the attainable peak magnetic field, advancing the frontier from the conventional 600~T regime to a record-breaking 1200~T. This unprecedented field generation clearly delineated the physical limitations of conventional induction-based pickup coil methods, leading to the establishment of highly reliable optical Faraday rotation techniques as the standard for precise field calibration up to 1200~T. 

Furthermore, through strategic innovations both inside and outside the explosion-proof chamber, alongside optimized coil positioning and specialized instrumentation, we successfully shattered the absolute technical barrier that had long plagued multi-megagauss science worldwide. Historically, any attempts at cryogenic measurements under EMFC or the chemical explosive-driven flux compression experiments had been completely blocked below 400~T, with~even room-temperature experiments strictly capped at that limit due to the sheer destructive violence of the implosion. Defying this long-standing impossibility, our meticulous management of the explosive environment culminated in the historic, world-first milestone of high-fidelity physical metrology under unprecedented coexisting extremes: a peak magnetic field of 600~T at a cryogenic temperature of 5~K.

Finally, the~successful development of high-frequency radio-frequency (RF) magneto-conductivity sensing in these extreme environments has enabled high-resolution transport measurements in strongly correlated semiconductors. This approach successfully provided a definitive and unambiguous observation of the semiconductor-to-metal phase transition driven by the giant Zeeman effect, effectively resolving a long-standing historical controversy regarding theoretical predictions in ultrahigh magnetic~fields.

The fundamental open problems that strongly motivate further development into the 1000~T regime lie in exploring quantum and structural regimes where the magnetic energy ceases to be a mere perturbation and instead dominates the intrinsic atomic and molecular architectures. In~this ultrahigh-field domain, the~magnetic field is capable of driving a complete reconfiguration of chemical bonds and breaking strongly bound electron dimers. A~pivotal stepping-stone toward this frontier has already been demonstrated by our pioneering work on tungsten-doped $\text{VO}_2$, where an ultrahigh magnetic field of 500~T was utilized to induce an insulator-to-metal transition via the spin Zeeman effect, effectively dissociating the vanadium dimers and successfully resolving a half-century-long controversy regarding the essential driving force of the metal--insulator transition~\cite{Matsuda2020}. Furthermore, the~1000~T regime serves as the ultimate laboratory to unveil the novel topological and quantum effects described in our introduction, where the magnetic flux directly pierces the electron orbitals. This includes the manifestation of the A-B effect in carbon nanotubes and the induction of distinct asymmetries in the Dirac cones of graphene, transitions that fundamentally alter the underlying electronic~symmetries. 

Ultimately, the~transition from 100~T to 1000~T is not merely a numerical increase in field intensity, but~a fundamental shift in experimental methodology. While the technical hurdles described in this review---ranging from ultra-miniaturized cryogenics to noise-immune optical sensing---are formidable, they are being systematically overcome through the synergy between extreme physical ambition and the disciplined observation of the experimental ``back side.'' By integrating these cutting-edge measurement techniques with the strategic management of explosive energy, the~megagauss science regime is evolving from a frontier of ``survival'' into a domain of precision science. This evolution promises to unveil entirely new physical and chemical quantum states---such as field-induced phase transitions and reconfigured chemical bonds---extending our reach to the grand scale of astrophysics and potentially elucidating the metallic state of hydrogen predicted to exist under the intense magnetic environments of the~cosmos.

\vspace{20pt}

\section*{Acknowledgments}
The author acknowledges his collaborators at the International MegaGauss Science Laboratory (IMGSL), ISSP, the~University of Tokyo (A. Miyata, N. Matsuda, A. Ikeda), with~special thanks to E. Kojima and~D. Nakamura for their intensive assistance. Creative discussions with O. Portugall (LNCMI-Toulouse) are highly appreciated. The~author is also indebted to K. Takekoshi at TERRABYTE Co., Ltd. for insightful discussions regarding computer simulations and for providing an unpublished figure. This article is sincerely dedicated to M. von Ortenberg, G. Kido, N. Miura, and~the late F. Herlach, in~gratitude for their long-standing encouragement throughout the author's research career. 


\vspace{20pt} 
\section*{Abbreviations}

\vspace{12pt}
\noindent
\begin{tabularx}{\textwidth}{@{}l X@{}}
STC       & Single-turn coil \\
EMFC      & Electromagnetic flux compression \\
IMGSL     & International MegaGauss Science Laboratory \\
ISSP      & Institute for Solid State Physics, the University of Tokyo \\
LNCMI     & Laboratoire National des Champs Magnétiques Intenses in Toulouse \\
MC        & Magnetio-cumulative \\
FRP       & Fiber-reinforced plastic \\
LN$_2$    & Liquid nitrogen \\
$B$--$T$  & Magnetic field--temperature (phase diagram) \\
CL        & Copper lined (coil)\\
FR        & Faraday rotation \\
S-C       & self-compensated (induction pickup coil) \\
OD        & Optical density \\
SWCNT     & Single wall carbon nanotube \\
A-B effect & Aharonov--Bohm effect \\
CR        & Cyclotron resonance \\
RF        & Radio frequency \\
TDO       & Tunnel diode oscillator \\
PDO       & Proximity 1231 detector oscillators \\
AC        & Alternating current \\
IC        & Integrated circuit \\
LC        & Inductance--capacitance \\
SRC       & Self-resonant coil \\
VCO       & Voltage-controlled oscillator \\
\end{tabularx}


    \end{document}